\documentclass[aps,pra,twocolumn,superscriptaddress]{revtex4}
\usepackage{graphicx}
\usepackage{amssymb}
\usepackage{comment}
\usepackage{amsmath}
\usepackage[utf8x]{inputenc}
\usepackage[T1]{fontenc}
\usepackage{dsfont}

\usepackage{url}
\usepackage[colorlinks]{hyperref}
\hypersetup{%
    plainpages=true,
    breaklinks=true,
    hypertexnames=false,
    pageanchor=true,
    colorlinks=true,
    linkcolor={blue},
    citecolor={red},
    urlcolor={blue},
    anchorcolor={black}
    }

\usepackage[position=top,caption=false]{subfig}

\newcommand{\ket}[1]{| #1 \rangle}
\newcommand{\bra}[1]{\langle #1 |}

\def\tr{{\rm Tr}}

\def\<{\langle}
\def\>{\rangle}

\begin{document}

\title{Experimental hierarchy and optimal robustness
of quantum correlations of two-qubit states with controllable
white noise}

\author{Kate\v{r}ina Jir\'akov\'a} \email{katerina.jirakova@upol.cz}
\affiliation{Joint Laboratory of Optics of Palack\'y University
and Institute of Physics of Czech Academy of Sciences, 17.
listopadu 12, 779 00 Olomouc, Czech Republic}

\author{Anton\'in \v{C}ernoch} \email{acernoch@fzu.cz}
\affiliation{Institute of Physics of the Czech Academy of
Sciences, Joint Laboratory of Optics of PU and IP AS CR, 17.
listopadu 50A, 772 07 Olomouc, Czech Republic}

\author{Karel Lemr}
\email{k.lemr@upol.cz} \affiliation{Joint Laboratory of Optics of
Palack\'y University and Institute of Physics of Czech Academy of
Sciences, 17. listopadu 12, 779 00 Olomouc, Czech Republic}

\author{Karol Bartkiewicz} \email{karol.bartkiewicz@upol.cz}
\affiliation{Joint Laboratory of Optics of Palack\'y University
and Institute of Physics of Czech Academy of Sciences, 17.
listopadu 12, 779 00 Olomouc, Czech Republic}
\affiliation{Institute of Spintronics and Quantum Information,
Faculty of Physics, Adam Mickiewicz University, 61-614 Pozna\'n,
Poland}

\author{Adam Miranowicz} \email{miran@amu.edu.pl}
\affiliation{Institute of Spintronics and Quantum Information,
Faculty of Physics, Adam Mickiewicz University, 61-614 Pozna\'n,
Poland}

\begin{abstract}
We demonstrate a hierarchy of various classes of quantum
correlations on experimentally prepared two-qubit Werner-like
states with controllable white noise. Werner states, which are
white-noise-affected Bell states, are prototypal examples for
studying such a hierarchy as a function of the amount of white
noise. We experimentally generated Werner states and their
generalizations, i.e., partially entangled pure states affected by
white noise. These states enabled us to study the hierarchy of the
following classes of correlations: separability, entanglement,
steering in three- and two-measurement scenarios, and Bell
nonlocality. We show that the generalized Werner states (GWSs)
reveal fundamentally different aspects of the hierarchy compared
to the Werner states. In particular, we find five different
parameter regimes of the GWSs, including those steerable in a
two-measurement scenario but not violating Bell inequalities. This
regime cannot be observed for the usual Werner states.
Furthermore, we find threshold curves separating different regimes
of the quantum correlations and find the optimal states which
allow for the largest amount of white noise, which does not
destroy their specific quantum correlations (e.g., unsteerable
entanglement). Thus, we could identify the optimal
Bell-nondiagonal GWSs, which are, for this specific meaning, more
robust against white noise compared to the Bell-diagonal GWSs
(i.e., Werner states).
\end{abstract}

\date{\today}

\maketitle


\section{Introduction}

\subsection{Entanglement, steering, and Bell nonlocality}

Quantum entanglement~\cite{Einstein1935} and its generalizations,
i.e. quantum steering~\cite{Schrodinger1935, Schrodinger1936} and
Bell nonlocality~\cite{Bell1964}, are fundamental types of quantum
correlations between spatially separated systems (parties). These
effects reveal the disparity between classical and quantum physics
from a fundamental point of view, but also play a pivotal role in
quantum information and its applications for quantum technologies
of second generation~\cite{HorodeckiReview, BrunnerReview,
CavalcantiReview,UolaReview}. (i) Quantum entanglement (or quantum
inseparability) occurs when the state of one party cannot be
described independently of the state of the other
party~\cite{HorodeckiReview}. (ii) Quantum steering, also referred
to as Einstein-Podolsky-Rosen (EPR) steering, refers to the
ability of one party (say, Alice) to affect the state of the other
party (say, Bob) through the choice of her measurement basis,
which cannot be explained by any local hidden state (LHS)
models~\cite{CavalcantiReview,UolaReview}. Furthermore, (iii)
quantum nonlocality can be defined as the effect detectable by the
violation of the Bell inequality and, thus, which cannot be
explained by any local hidden variable (LHV) models. Here we limit
our interest to the two-qubit Bell inequality in the
Clauser-Horne-Shimony-Holt (CHSH) form~\cite{Clauser1969}. Thus,
we refer to this effect as Bell(-CHSH) nonlocality, having in mind
that quantum nonlocality can also be understood in a much broader
sense~\cite{BrunnerReview}.

The distinction between these effects is fundamental, and their
intuitive operational interpretation can be given from a
measurement perspective, i.e., by referring to their detection
using two types of measuring devices, which can be {perfect or
imperfect from physical and technological points of view, or
trusted or untrusted from a cryptographic perspective, i.e., with
or without prior knowledge about the devices}~\cite{Wiseman2007}.
Specifically, (i) quantum entanglement between two systems can be
detected using trusted devices for both systems, (ii) EPR steering
can be tested by trusted devices for one system and untrusted ones
for the other, and (iii) quantum nonlocality can be detected by
untrusted devices on both sides. Such interpretation has direct
applications for quantum cryptology, including secure
communication. In the same measurement scenarios, Bell nonlocality
implies steering, and steering implies entanglement, but not vice
versa, in general. Indeed, there exist entangled~\cite{Werner1989}
and steerable states which do not violate Bell inequalities, as
well as unsteerable entangled
states~\cite{CavalcantiReview,UolaReview}.

\subsection{Werner states and their experimental generation}

Mixtures of a Bell state and a maximally mixed state (i.e., a
white noise) are prototypal examples of states revealing the
nonequivalence of entanglement and Bell nonlocality, which was
first demonstrated by Werner over 30 years ago~\cite{Werner1989}.
The Werner states were later used to show a hierarchy of criteria
and a hierarchy of some classes of correlations (CC) (which for
short is referred here to as \emph{CC hierarchy}), including
quantum steering (see, e.g., reviews
in~\cite{HorodeckiReview,BrunnerReview,CavalcantiReview,UolaReview}
and the very recent Ref.~\cite{Zhao2020} with references therein).
The effect of white noise on Bell states has also been studied
theoretically to reveal a hierarchy of the following classes of
temporal quantum correlations~\cite{Ku2018hierarchy}: temporal
inseparability~\cite{Fitzsimons2015}, violations of temporal
Bell-CHSH inequalities~\cite{Fritz2010}, and temporal
steering~\cite{Chen2014,Chen2014,Bartkiewicz2016}.

We stress that we only consider von Neumann's projective
measurements in this work. Note that the quantum-correlation
regimes of states assumed for projective measurements are
different from those based on positive-operator-valued measures
(POVMs). However, the same hierarchy relations, as studied here,
still hold assuming POVMs.

Generation of mixed states of discrete photons has been
investigated both
theoretically~\cite{Thew2001PhysRevA,Zhang2004PhysRevA,Wei2005PhysRevA}
and experimentally~\cite{LIMA2008OptComm, Ling2006LasPhys,
Liu2017EurLett, Puentes2006OptLett, White2001PhysRevA,
Zhang2002PhysRevA, Cinelli2004PhysRevA, Caminati2006PhysRevA,
Puentes2007PhysRevA, Aiello2007PhysRevA, Brida2008PhysRevA,
Peters2004PhysRevLett, Barbieri2004PhysRevLett, Kwiat2000Science}.
Temporal decoherence of optical polarization modes in a
birefringent material seems to be a rather widely used technique
in a number of experiments such as those reported in
Refs.~\cite{Kwiat2000Science, White2001PhysRevA}. This technique
has also enabled the experimental generation of maximally
entangled mixed states (MEMSs)~\cite{Munro2001PhysRevA} by Peters
\emph{et al.}~\cite{Peters2004PhysRevLett} and later by Aiello
\emph{et al.}~\cite{Aiello2007PhysRevA}. Recently, Liu \emph{et
al.} incorporated a tunable decoherence
channel~\cite{Jeong2013PhysRevA} to generate the Werner
states~\cite{Liu2017EurLett}. Alternative methods to generate or
simulate temporal decoherence include the generation of mixed
states by exploiting a particular geometry of a spontaneous
parametric down-conversion (SPDC)
source~\cite{LIMA2008OptComm,Brida2008PhysRevA}. Barbieri \emph{et
al.}~\cite{Barbieri2004PhysRevLett} and Cinelli \emph{et
al.}~\cite{Cinelli2004PhysRevA} reported their refined two-photon
sources capable of preparing a broad range of mixed quantum
states, including MEMSs. A highly birefringent material, together
with a wide momentum spectrum of generated photon pairs (resulting
in effective spatial decoherence), was also used as an alternative
method to generate temporal decoherence~\cite{White2001PhysRevA}.
Puentes \emph{et al.} applied wedge depolarizers and bucket
detectors~\cite{Puentes2006OptLett}, and later utilized scattering
in various media~\cite{Puentes2007PhysRevA}. Moreover, Zhang
\emph{et al.} incoherently combined photons generated in two
separate SPDC sources to create mixed quantum
states~\cite{Zhang2002PhysRevA}, while Caminati \emph{et al.}
reported an experiment, where mixed states were generated by
attenuating a high-gain SPDC source~\cite{Caminati2006PhysRevA}.
The idea of using a wide-temporal detection window, such that a
detected state appeared to be mixed, was also implemented in
several experiments~\cite{Ling2006LasPhys, Lemr2016PhysRevA}. It
is also possible to use an experimental setup that can be tuned
(to change properties of generated states) in times shorter than
the measurement integration time~\cite{GAVENDA2005ModPhysLett}.

In this work, we report experimental generation of both Werner
states and their generalizations, i.e., partially entangled pure
states affected by white noise, which we refer to as generalized
Werner states (GWSs). These states were not the focus of the
above-reviewed experiments. Some of the experimental setups cannot
generate these generalized states (e.g.,
Ref.~\cite{Aiello2007PhysRevA}), some could be used after specific
improvements (e.g., Ref.~\cite{Liu2017EurLett}) and the others
might have such capabilities, but these (e.g.,
Ref.~\cite{Zhang2002PhysRevA}) have not been used so far for
demonstrating the CC hierarchy of the Werner states or their
generalizations. In this paper, our experimentally generated and
reconstructed states are applied to reveal a CC hierarchy.


{The remainder of the paper is organized as follows. Two
approaches to study hierarchies of correlations are specified} in
Sec.~\ref{ch:TwoHierarchies}. Measures of quantum correlations of
general two-qubit states are recalled in
Sec.~\ref{ch:Intro_Measures}. These include popular measures of
entanglement, steering, and Bell nonlocality. Moreover, steering
in the two-, three-, and multi-measurement scenarios is explicitly
discussed in Appendices~\ref{App:Steering2}, \ref{App:Steering3},
and \ref{App:SteeringN}, respectively.  In Sec.~\ref{ch:GWS}, we
define GWSs. Because GWSs are a direct generalization of the usual
Werner states based on a Bell state, we refer to them as
Bell-nondiagonal GWSs. Our experiment is described in
Sec.~\ref{ch:Experiment}. We compare various predictions of the
quantum correlations for the theoretical and experimental GWSs
with those for the Werner states in Sec.~\ref{ch:HierarchyAll}. We
also discuss fundamental differences in a CC hierarchy for the
Bell diagonal and nondiagonal GWSs in this section. In
Sec.~\ref{ch:CounterintuitiveResults}, we present our most
counterintuitive theoretical results. Specifically, we show in
Sec.~\ref{ch:CounterintuitiveResults1} that there exist GWSs,
which are steerable in a two-measurement scenario (2MS) but still
admit LHV models. Such a regime cannot be observed for the
standard Werner states. In Sec.~\ref{ch:CounterintuitiveResults2},
we show that some Bell-nondiagonal GWSs are more robust against
white noise than the diagonal GWSs, i.e., the Werner states. In
Sec.~\ref{ch:SteeringN}, we analyze lower and upper bounds on
steering for a large number of measurements. We show better
robustness against white noise of unsteerable entangled
Bell-nondiagonal GWSs compared to the diagonal ones. An example of
a hierarchy of entanglement criteria is discussed in
Appendix~\ref{App:MM} in comparison with the CC hierarchy for the
GWSs. We conclude in Sec.~\ref{ch:Conclusions}.

\section{Two approaches to study a hierarchy of quantum correlations} \label{ch:TwoHierarchies}

{Here we study a CC hierarchy, which is the hierarchy of
\emph{states} with different correlation properties rather than
types of probability distributions, as in the case of certain
research in quantum information. We use the term correlation of a
state by referring to its entanglement, steering, and Bell
nonlocality. For clarity, we recall that: (a) an entangled
(separable) state is a state that cannot (can) be factored into
individual states belonging to separate subspaces; (b) an EPR
steerable (unsteerable) state is the one described by the
statistics, which cannot (can) be reproduced by an LHS model for a
given measurement set (see Sec.~\ref{ch:Intro_Steering} for more
details); and (c) a quantum nonlocal (local) state is the one
described by the statistics, which cannot (can) be reproduced by
an LHV model, which in turn implies the violation (fulfillment) of
a Bell inequality. Since we are focused on analyzing two-qubit
states, the Bell inequalities can be limited to the CHSH
inequality. Furthermore, the steerability of states can be
considered in the limit of an infinite number of measurements, but
it is usually limited to practical resources, including a finite
number of measurements. We focus in this paper on the GWSs, which
are steerable or unsteerable in the 2MSs and 3MS, corresponding,
respectively, to measuring two or three Pauli operators. Thus, we
can consider subclasses of steerable states depending on the
number of performed measurements. In what follows, we study in
detail the hierarchy of correlation classes limited to analyzing
the states which are: (i) separable, (ii) entangled but
unsteerable in 3MSs, (iii) steerable in 3MSs but not in 2MSs, (iv)
2MS-steerable but Bell local, and (v) Bell nonlocal. The hierarchy
is extended in Sec.~\ref{ch:SteeringN} to include the analysis of
the GWSs, which are steerable for a larger number $n$ of
measurements (i.e., $n=136$).}

In general, a hierarchy of quantum correlations can be understood
in several ways including: (i) a hierarchy of conditions (or
criteria) for the observation of a given class of quantum
correlations and (ii) a hierarchy of different classes of quantum
correlations (i.e., a CC hierarchy). This division is also closely
related to experimental demonstrations of a hierarchy by measuring
(nonuniversal or universal) witnesses of quantum correlations
corresponding to performing partial or full quantum state
tomography (QST), respectively.

In this work, we focus on analyzing a CC hierarchy of the GWSs. We
demonstrate different kinds of quantum correlations in question by
performing full QST and then calculating the corresponding
measures on the reconstructed states.

Below we explain the main differences between the two approaches
to study a hierarchy of quantum correlations and explain why a
complete experimental demonstration of the studied CC hierarchy
seems to be very challenging within the present state of the art.

\subsubsection{Hierarchy of criteria for a given class
of quantum correlations} \label{hierarchyI}

Experimental demonstrations of Bell nonlocality via the violations
of the CHSH inequalities have been at the heart of quantum
information since its early days starting from the pioneering
experiments of Aspect \emph{et al.}~\cite{Aspect1982} and, then,
refined in hundreds of experiments, including
significant-loophole-free tests~(see,
e.g.,~\cite{Christensen2013,Giustina2015,Hensen2015} and the
review in~\cite{BrunnerReview} for references). Thus, if one talks
about ``demonstrating'' the nonlocality of a quantum state, one
would normally expect to see a violation of a Bell inequality,
rather than QST.

However, this approach usually reveals only a hierarchy of
criteria (i.e., either sufficient or necessary conditions) for the
observation of a specific class of quantum correlations. This is
because it is usually based on measuring nonuniversal witnesses of
quantum correlations by testing the violation of specific
inequalities. Note that nonuniversal witnesses correspond usually
to \emph{sufficient but not necessary conditions} of a specific
quantum (temporal or spatial) correlation effect. Thus, such a
witness can usually be determined \emph{without} a complete QST.

Within this hierarchy approach, one can analyze a hierarchy of,
e.g., different Bell inequalities or even the Bell-CHSH
inequalities but for different angles of polarizers in a
description of Bell nonlocality; specifically, by choosing
different angles $\phi_1$, $\phi_2$, $\phi'_1$, and $\phi_2'$ as
described in Eq.~(\ref{CHSH}). By having \emph{a priori}
information about a given generated state, one can choose optimal
angles of the polarizers to maximize the violation of the
Bell-CHSH inequalities, and, thus, to be able to quantify Bell
nonlocality (i.e., to determine a nonlocality measure) for the
state. However, without knowing \emph{a priori} a given state, one
has to measure many copies of the state at different angles of the
polarizers, to find their optimal rotation. Such scanning of the
angles corresponds to complete or partial QST.

The hierarchy of criteria has also been studied based on the
matrices of the moments of, e.g., the annihilation and creation
operators of bosonic or fermionic states of any dimension. Indeed,
a number of works demonstrated: (i) a hierarchy of sufficient
conditions for observing entanglement (i.e., entanglement
witnesses), which include the conditions based on the
Shchukin-Vogel criterion~\cite{Shchukin2005,Miranowicz2006}, which
are related to the Peres-Horodecki criterion and its generalized
versions using positive maps beyond partial
transpose~\cite{Miranowicz2009}; (ii) a hierarchy of sufficient
conditions for observing quantum steering (i.e., steering
witnesses)~\cite{Kogias2015}; (iii) a hierarchy of necessary
conditions for revealing Bell nonlocality (i.e., nonlocality
requirements)~\cite{Navascus2007}; and (iv) a hierarchy of
sufficient conditions for observing spatial~\cite{Richter2002} and
spatio-temporal~\cite{Vogel2008,Miranowicz2010} nonclassicality
(i.e., nonclassicality witnesses).

An illustrative detailed example of a hierarchy of entanglement
criteria is discussed in Appendix~\ref{App:MM}.

Note that the upper and lower bounds of measures of quantum
correlations, which correspond to their sufficient and necessary
conditions, can be determined using such a hierarchy of matrices
of moments without a complete QST. However, for an unknown state,
to make these bounds tight to a true measure, one needs to
increase the number of moments to be detected. This in turn leads
to a partial moment-based QST, which approaches more and more a
complete QST as explained in Appendix~\ref{App:Wunsche}.

In conclusion, this approach, in general, enables a direct but
partial demonstration of a hierarchy, which is discussed below.

\subsubsection{Hierarchy of various classes of correlations}
\label{hierarchyCC}

A hierarchy of various classes of correlations can be revealed by
their measures or by the conditions, which are both necessary and
sufficient for their observation. It should be stressed that we
are focused on demonstrating such a CC hierarchy in this paper.

Indeed, experimental methods for a complete demonstration of a CC
hierarchy can be based on experimentally reconstructed density
matrices (in the case of standard single-time spatial
correlations) or the Choi-Jamiolkowski matrices (in the case of
temporal correlations) for a given system via quantum state or
process tomographies, respectively. This approach enables the
calculation of necessary and sufficient conditions for observing
and quantifying the amount of any class of quantum temporal or
spatial correlations for a given state or process.

Experimental demonstration of such a CC hierarchy has usually been
done using a complete QST, although it can also be done with an
incomplete QST, {as discussed in
Appendix~\ref{App:IncompleteTomography}.}

Here, we apply an indirect approach based on experimental
detecting and reconstructing states via a full QST, and only then
calculating their correlation measures on the reconstructed
states. This approach has important fundamental and experimental
advantages, which include the following (in addition to the
above-mentioned ones):

(i) We want to test the above-mentioned five classes of quantum
correlations on the same footing (preferably using the same setup)
based on either complete or incomplete tomography. However, it is
seen that we can determine experimentally the Horodecki
nonlocality measure without QST, but detecting the negativity and
the steerable weights (or, equivalently, steering robustness) can
be done effectively only via a complete QST.

(ii) We want to use the same experimental states for testing
different quantum properties. The problem is that we do not have a
perfect control of, especially, the mixing parameter determining
the amount of white noise in a pure state. Thus, we cannot
generate the same GWSs even using the same setup. Such a state
generation would be even more demanding using different setups for
testing different classes of correlations. However, this is
feasible using a full QST to reconstruct a state, which is only
then numerically studied for its quantum correlations.

\section{Measures of quantum correlations of general two-qubit
states} \label{ch:Intro_Measures}

As a part of our introduction, we shortly recall standard measures
of quantum correlations for general two-qubit states $\rho$, which
can be written in the Bloch representation as follows:
\begin{equation}
\rho  =  \frac{1}{4}\Big(I\otimes
I+\boldsymbol{u}\cdot\boldsymbol{\sigma}\otimes
I+I\otimes\boldsymbol{v}\cdot\boldsymbol{\sigma}+\!\!\!\sum
\limits_{n,m=1}^{3}T_{nm}\,\sigma _{n}\otimes \sigma _{m}\Big),
\label{rhoGeneral}
\end{equation}
where $u_{i}=\mathrm{Tr[}\rho(\sigma_{i}\otimes I)]$ and
$v_{i}=\mathrm{Tr[}\rho(I\otimes\sigma_{i})]$ are the elements of
the Bloch vectors $\boldsymbol{u}=[u_{1},u_{2},u_{3}]$ and
$\boldsymbol{v}=[v_{1},v_{2},v_{3}]$ of the first and second
qubits, respectively, and $I$ is the single-qubit identity
operator. Moreover, the correlation matrix elements
$T_{ij}=\mathrm{Tr}[\rho(\sigma_{i}\otimes\sigma_{j})]$ and
$\boldsymbol{\sigma}=[\sigma_{1},\sigma_{2},\sigma_{3}]\equiv
[X,Y,Z]$ are expressed via the Pauli matrices.

\subsection{Entanglement measures} \label{ch:Intro_Entanglement}

Here, we recall the standard definitions and physical meaning of
the two most popular measures of two-qubit entanglement, i.e., the
negativity and concurrence, which are in the following sections
compared with the measures of steering and Bell nonlocality.

The negativity is defined as ${N}({\rho})=\max \{0,-2\mu
_{\min}\}$, where $\mu_{\min}=\min{\rm eig}(\rho^{\Gamma})$ and
$\rho^{\Gamma}$ denotes a partial transpose of $\rho$. It was
first introduced in Ref.~\cite{Zyczkowski1998} as a quantitative
version of the Peres-Horodecki entanglement
criterion~\cite{Horodecki1996negativity}. The two-qubit negativity
(or, more directly, the logarithmic negativity $\log_2[{N}
({\rho})+1]$) has various quantum-information interpretations.
Specifically: (i) it is a measure of the entanglement cost under
operations preserving the positivity of partial transpose for
two-qubit systems~\cite{Audenaert2003,Ishizaka2004}, (ii)~it gives
an upper bound of distillable entanglement~\cite{HorodeckiReview},
and (iii) it determines a dimensionality of entanglement, i.e.,
the number of the degrees of freedom of entangled
subsystems~\cite{Eltschka2013}.

The Wootters concurrence~\cite{Wootters1998}, which is
monotonically related to the entanglement of formation, is given
by $C({\rho})=\max \{0,2\lambda_{\max}-\sum_j\lambda_j\}$, where
$\lambda^2 _{j} = \mathrm{eig}[{\rho }({\sigma }_{2}\otimes
{\sigma }_{2}){\rho}^{\ast }({ \sigma }_{2}\otimes {\sigma
}_{2})]_j$, with $\sigma_2$ denoting the Pauli $Y$-operator, and
$\lambda_{\max}=\max_j\lambda_j$.

Note that both measures have been applied in quantifying not only
entanglement but also, e.g., nonclassicality (quantumness) of
single-qubit (or single-qudit)
states~\cite{Asboth2005,Adam2015a,Adam2015b}. These two related
measures reach unity for the Bell states and vanish for separable
states. For the brevity of our presentation, we plotted the negativity
as the only entanglement measure.

These entanglement measures of various two-qubit states have been
typically determined experimentally only indirectly, based on a
full QST, which is also the case in this work. Note that an
experimental universal test of entanglement without a complete QST
was proposed in Ref.~\cite{Bartkiewicz2015} (see
Appendix~\ref{App:IncompleteTomography}). This test is a necessary
and sufficient criterion of two-photon polarization entanglement.
It is based on measuring a collective universal witness of
Ref.~\cite{Augusiak2008}, which gives tight lower and upper bounds
for the negativity and concurrence, and can be used as an
entanglement measure on its own. However, since its
quantum-information interpretation and applications are limited,
we prefer to use the standard entanglement measures, even if they
are determined indirectly using experimental density matrices.

\subsection{Steerable weight}
\label{ch:Intro_Steering}

The steerable weight~\cite{Skrzypczyk2014} and the
steering robustness~\cite{Piani2015} are arguably the most
popular measures of EPR
steering~\cite{CavalcantiReview,UolaReview,Ku2018}. They can be
applied for quantifying not only standard spatial steering, but
also (after a minor modification) to quantify
temporal~\cite{Chen2014,Chen2016,Bartkiewicz2016,Ku2016,Ku2018hierarchy}
and spatio-temporal~\cite{Chen2017} steering.

An intuitive and general idea behind the steerable weight,
according to Skrzypczyk \emph{et al.}~\cite{Skrzypczyk2014}, is
based on the decomposition of a given assemblage of Alice,
$\sigma_{a|x}$, into its steerable ($\sigma_{a|x}^{{\rm S}}$) and
unsteerable ($\sigma_{a|x}^{{\rm US}}$) parts, for the values of
$a$ and $x$ specified in Appendices~\ref{App:Steering3} and
\ref{App:Steering2}, i.e.,
\begin{eqnarray}
  \sigma_{a|x} &=& \mu\sigma_{a|x}^{{\rm US}} + (1-\mu) \sigma_{a|x}^{{\rm
  S}},
\label{decom}
\end{eqnarray}
for $\mu\in[0,1]$. {Note that the unsteerable assemblages
$\sigma_{a|x}^{{\rm US}}$ can be created via classical strategies,
and a model based on $\sigma_{a|x}^{{\rm US}}$ can be referred to
as an LHS model.} The steerable weight $S=1-\mu^*$ is defined as
the maximum amount of unsteerable assemblage $\sigma_{a|x}^{{\rm
US}}$ necessary to reproduce Alice's assemblage $\sigma_{a|x}$.
This general definition can be formulated as solutions of
semidefinite programs (SDPs) as demonstrated in
Refs.~\cite{Skrzypczyk2014,CavalcantiReview}, and are given
explicitly in {Appendices~\ref{App:Steering3} and
\ref{App:Steering2} for the 3MS and 2MS, respectively. Moreover,
sufficient and necessary conditions for observing steering in
multi-measurement scenarios are discussed in
Appendix~\ref{App:SteeringN}.}

{The LHS models are relevant to quantum steering as
follows~\cite{Wiseman2007}: A given state $\rho$ is referred to as
quantum (EPR) \emph{unsteerable} (in the communication from Alice
to Bob) for Alice's measurement set $\{M_{a|x}\}$ if one can find
a variable $\lambda$ allowing for the following Bell local
decomposition~\cite{CavalcantiReview,UolaReview}
\begin{equation}
\label{LHS} p(ab|xy) \;=\; \int d\lambda\; \pi(\lambda) \; p_A
(a|x,\lambda ) \; \tr ( M_{b|y} \sigma_\lambda),
\end{equation}
where $\sigma_\lambda$ is the local (hidden) quantum state of Bob
and $p_A (a|x,\lambda )$ is Alice's response distribution.
Otherwise a given state for the measurement set $\{M_{a|x}\}$ is
referred to as quantum (EPR) \emph{steerable}, i.e., when its
statistics cannot by reproduced by an LHS model. Note that
Eq.~(\ref{LHV}), which defines a Bell local state, reduces in the
special case to Eq.~(\ref{LHS}) by setting $p_B (b|y,\lambda)=\tr
( M_{b|y} \sigma_\lambda)$. It is usually assumed that Bob's
measurements $M_{b|y}$ enable a complete QST of his qubit. The
collection of Bob's states $\sigma_{a|x} = \tr_A (M_{a|x} \otimes
\openone \; \rho), $ conditioned on Alice's measurements, is
called an assemblage.}

{The steerable weight and, equivalently, the steering robustness
of Ref.~\cite{Piani2015} are defined via necessary and sufficient
conditions for quantum information characterization of quantum
steering in the specified measurement scenarios. Thus, a spatially
separated two-qubit state $\rho$ is referred to as steerable (or,
more precisely, $S_n$-steerable) in the discussed $n$-measurement
scenarios if there exists a set of measurements such that the
steerable weight is nonvanishing, $S_n(\rho)>0$. Otherwise, it is
referred to as unsteerable (or $S_n$-unsteerable).}

{The question arises why our interest is focused on analyzing
steering in two- and three- measurement scenarios only, except in
Sec.~\ref{ch:SteeringN} and Appendix~\ref{App:SteeringN}. In
principle, one could also consider steering in the limit of an
infinite number (of the types) of measurements. However, this
would require knowing universal criteria (i.e., which are both
sufficient and necessary) for detecting this type of steering.
Unfortunately, such universal criteria are not known for the GWSs.
Note that the upper and lower bounds for steering have only been
calculated numerically so far for large but still finite numbers
$n$ of measurements (i.e., at most for $n=136$, as shown in
Fig.~\ref{fig8}(a) based on the results of
Refs.~\cite{Hirsch2016,Fillettaz2018}). Moreover, our analysis of
steering includes not only criteria but also steering measures, as
shown in Figs.~\ref{fig:theoryWS}--\ref{fig:ExpGWS}.
Unfortunately, the calculations of the steerable weight and the
steering robustness are much more involved beyond 3MS. Finally, we
remark that our experimentally generated states are not exactly
GWSs, so the calculations of their measures or even universal
criteria of steering beyond the 3MS are even more complicated
compared to those for the perfect GWSs.}

\subsection{Horodecki measure of Bell nonlocality}
\label{ch:Intro_Nonlocality}

Here we recall the Horodecki measure~\cite{Horodecki1995,
Horodecki1996} of quantum nonlocality for two-qubit states.

Note that quantum nonlocality is usually studied and interpreted
in the context of Bell inequalities (including the CHSH
inequality) and, then, it is often referred to as Bell(-CHSH)
nonlocality~\cite{BrunnerReview}. A Bell inequality violation
(BIV) demonstrates the impossibility of any LHV models to fully
reproduce quantum mechanical predictions~\cite{Bell1964}. For
convenience, we use the terms BIV and Bell(-CHSH) nonlocality
interchangeably, in the context of our two-qubit experiments. Note
that BIV implies a violation of local realism. So, BIV can, in
principle, be explained by \emph{nonlocal} (non)realistic
theories, but also by \emph{local} nonrealistic ones. Moreover,
quantum nonlocality can be defined without referring to BIV. And
such (generalized) quantum nonlocality can occur without quantum
entanglement in, e.g., three qubits or two qutrits (three-level
systems)~\cite{Bennett1999}. Thus, it should be stressed that, in
general, neither BIV implies quantum nonlocality nor quantum
nonlocality implies BIV (see, e.g.,
Refs.~\cite{Popescu1994,BrunnerReview}).

The Horodecki measure of Bell nonlocality for a two-qubit state
$\rho$ quantifies the amount of the maximal violation of the
Bell-CHSH inequality~\cite{Clauser1969},
\begin{equation}
|\langle {\cal B}\rangle_{}|=|{\cal E}(\phi_1,\phi_2)+{\cal
E}(\phi_1',\phi_2)+{\cal E}(\phi_1,\phi_2')-{\cal
E}(\phi_1',\phi_2')|\leq 2, \label{CHSH}
\end{equation}
which is given in terms of the Bell-CHSH operator ${\mathcal
B}=\boldsymbol{a}\cdot \boldsymbol{\sigma }\otimes (\boldsymbol{
b}+\boldsymbol{b}^{\prime })\cdot \boldsymbol{\sigma
}+\boldsymbol{a}^{\prime }\cdot \boldsymbol{\sigma }\otimes
(\boldsymbol{b}-\boldsymbol{b}^{\prime })\cdot \boldsymbol{\sigma
}.$ Furthermore, $\phi_i$ and $\phi_i'$ are two dichotomic
variables of the $i$th ($i=1,2$) qubit corresponding to the
rotations of a polarizer in typical optical implementations; while
${\cal E}(\phi_1,\phi_2)$ is the expectation value of the joint
measurement of $\phi_1$ and $\phi_2$, and, analogously, for the
other expectation values. For a given two-qubit state $\rho$, the
expected value of the Bell-CHSH operator ${\mathcal B}$, which is
maximized over real-valued three-dimensional unit vectors
$\boldsymbol{a},\,\boldsymbol{a}^{\prime },\,\boldsymbol{b},$ and
$\boldsymbol{b}^{\prime }$,
reads~\cite{Horodecki1995,Horodecki1996}:
\begin{equation}
\max_{{\mathcal B}}\,\mathrm{Tr}\,(\rho \,{\mathcal B}_{
\mathrm{CHSH}})=2\,\sqrt{M(\rho )}, \label{MaxBellOp}
\end{equation}
where $M(\rho )=\max_{j<k}\;\{h_{j}+h_{k}\}\leq 2,$ and $h_{j}$
$(\,j=1,2,3)$ are the eigenvalues of $U=T^{T}\,T$, which is the
real symmetric matrix constructed from the correlation matrix $T$
(and its transpose $T^{T}$) defined below Eq.~(\ref{rhoGeneral}).
Thus, the Bell-CHSH inequality is violated if
$M(\rho)>1$~\cite{Horodecki1995,Horodecki1996}. To quantify the
degree of BIV and Bell nonlocality we apply the
parameter~\cite{Adam2004}:
\begin{equation}
B(\rho )\equiv \sqrt{ \max \,[0,\,M(\rho )-1]}. \label{BIV}
\end{equation}
Note that this nonlocality measure is exactly equal to the
concurrence and negativity for two-qubit \emph{pure} states. For a
given state $\rho$, the Bell-CHSH inequality in Eq.~(\ref{CHSH})
is satisfied if and only if $B(\rho)=0$. If $B(\rho)=1$ then the
inequality is maximally violated, which is the case for Bell
states. We refer to $B(\rho )$ as a Bell nonlocality measure.

In this work, we refer to Bell nonlocal and local states in the
following meaning. Usually, a spatially separated state is
referred to as Bell local if local measurements and classical
communication can generate a correlation admitting an LHV
model~\cite{Bell1964,BrunnerReview}. Otherwise, the state is
referred to as Bell nonlocal.

{More specifically, an LHS model can be introduced by considering
two distant observers (Alice and Bob) who share an entangled
two-qubit state $\rho$. Assume that Alice (Bob) performs a set of
measurements $\{M_{a|x}\}$ ($\{M_{b|y}\}$) satisfying
$M_{a|x},M_{b|y}\geq 0$ and $\sum_a M_{a|x}=\sum_b M_{b|y} =
\openone$, where $x,y$ label measurements and $a,b$ are their
outcomes. The resulting statistics $p(ab|xy) = \tr ( M_{a|x}
\otimes M_{b|y}\, \rho )$ are referred to as Bell \emph{local}
(with respect to the measurement sets $\{M_{a|x} \}$ and
$\{M_{b|y}\}$) if they allow for a Bell local
decomposition~\cite{BrunnerReview}:
\begin{equation}
\label{LHV} p(ab|xy) \;=\; \int d\lambda\; \pi(\lambda) \; p_A
(a|x,\lambda ) \; p_B (b|y,\lambda),
\end{equation}
where $\lambda$ is a shared local hidden variable distributed with
density $\pi(\lambda)$, while $p_A (a|x,\lambda )$ and $p_B
(b|y,\lambda)$ are local response distributions. Thus, a state is
called Bell \emph{local} if it can be reproduced by an LHV model
with properly chosen $\{\lambda, \pi(\lambda), p_A(a|x,\lambda ),
p_B(b|y,\lambda)\}$. Otherwise, the state is referred to as Bell
\emph{nonlocal} for the measurement sets $\{M_{a|x} \}$ and
$\{M_{b|y}\}$. This Bell nonlocality can be witnessed by the
violation of a Bell inequality, which reduces to testing the
Bell-CHSH inequality in the case of two-qubit states.  So, in
terms of the Horodecki measure, a given two-qubit state is Bell
local (nonlocal) if and only if $B(\rho)=0$ ($B(\rho)>0$).}

The Horodecki measure of Bell nonlocality can be determined
\emph{without} a complete QST, which was experimentally
demonstrated in an entanglement-swapping device
in~\cite{Bartkiewicz2017} (see
Appendix~\ref{App:IncompleteTomography}). However, here, we apply
a full QST for determining $\rho_{\rm GW}^E$ and then calculating
$B(\rho_{\rm GW}^E)$.

Note that various alternative approaches to quantifying nonlocality
have been proposed~\cite{BrunnerReview}. These include a
nonlocality measure introduced by Elitzur \emph{et al.} in
Ref.~\cite{Elitzur1992}, which can be interpreted as a fraction of
a given ensemble that cannot be expressed via local correlations.
Thus, this quantifier has been referred to as a fraction of
nonlocality~\cite{Barrett2006,Aolita2012}.

\section{Generalized Werner states and their experimental generation}

\subsection{Werner states and their generalizations}
\label{ch:GWS}

In this work, we focus on comparing quantum correlations of
experimental states, which are special cases of those in
Eq.~(\ref{rhoGeneral}). Specifically, we directly generated
{Werner(-like) states (also referred to as isotropic states or
Bell states with white noise)}~\cite{Werner1989}:
\begin{eqnarray}
  \rho_{\rm W} &=& p\ket{\phi^{+}}\bra{\phi^{+}}+\frac{1-p}{4} I\otimes I,
\label{WS}
\end{eqnarray}
which are mixtures of any Bell state [say,
$\ket{\phi^{+}}=(\ket{00}+\ket{11})/\sqrt{2}$)] and the maximally
mixed state for various values of the mixing parameter
$p\in[0,1]$. {Note that the original definition of the Werner
state is based on the singlet state~\cite{Werner1989}, instead of
$\ket{\phi^{+}}$. However, this local change does not effect
measures of entanglement, steering, and nonlocality.} Thus, the
state defined in Eq.~(\ref{WS}) is also often referred to as a
Werner state~(see, e.g,
Refs.~\cite{Munro2001PhysRevA,Ghosh2001,Wei2003,Adam2004,Fillettaz2018}).
This terminology is used in this paper.

We are also interested in {partially entangled states with the
white noise, which we call GWSs,} which are obtained from
Eq.~(\ref{WS}) by replacing $\ket{\phi^{+}}$ by a general
two-qubit pure state, $\ket{\phi_q}=\sqrt{q}\ket{00}
+\sqrt{1-q}\ket{11}$ with the superposition coefficient
$q\in[0,1]$. Thus, the GWSs can be defined as
\begin{eqnarray}
  \rho_{\rm GW}(p,q) &=& p\ket{\phi_q}\bra{\phi_q}+\frac{1-p}{4} I\otimes I.
\label{GWS}
\end{eqnarray}
These states for $q=1/2$ can be referred to as the Bell-diagonal
GWSs corresponding to the Werner states $\rho_{\rm W}(p)$, which
are diagonal in the Bell basis. While for $q\neq 1/2$, we refer to
them as the Bell-nondiagonal GWSs.

{These states have been generated by us in the experimental setup
described below.}

\subsection{{Experimental setup}} \label{ch:Experiment}

\begin{figure}
\centering
\hfill\includegraphics[scale=1.]{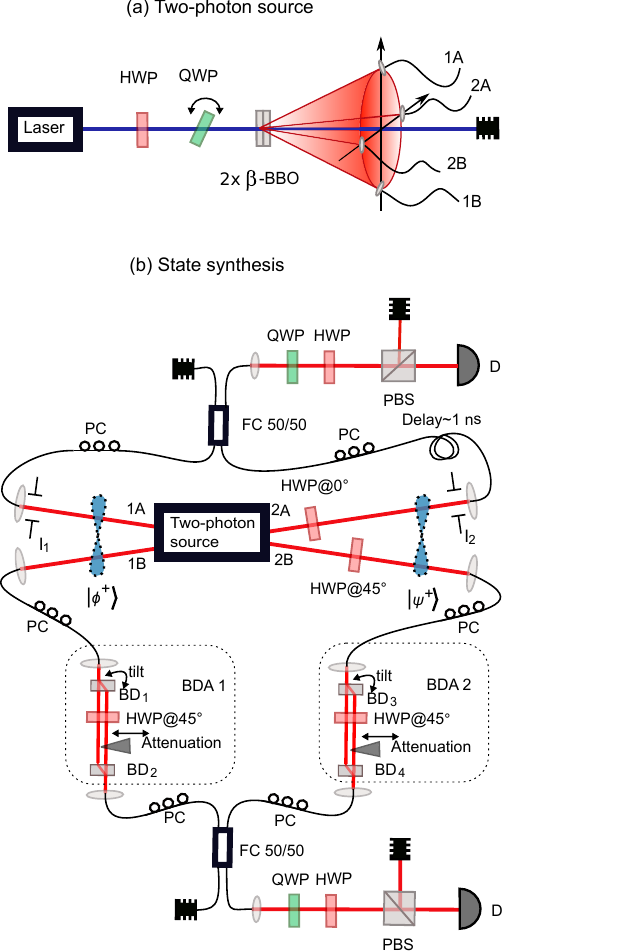}\hspace*{\fill}
\caption{Our experimental setups for (a) photon-pairs generation
and (b) state synthesis. 1A and 1B (2A and 2B) stand for photons
propagating in vertical (horizontal) planes, BD -- beam displacer,
BDA -- beam displacer assembly, D -- detector, FC -- fiber
coupler, HWP -- half-wave plate, PBS -- polarization beam
splitter, PC -- polarization controller, QWP -- quarter-wave
plate, $\mathrm{I}_1, \mathrm{I}_2$ -- irises 1, 2 and $\beta$-BBO
stands for nonlinear crystal ($\beta$-barium borate).}
\label{fig:ExpSetup}
\end{figure}

Here we describe our experimental setup, which is designed to be
as much versatile as possible, being capable of generating a broad
class of mixed quantum states in the form of
\begin{equation}
{\rho} =
\begin{pmatrix}
A & 0 & 0 & E \\
0 & B & F & 0\\
0 & F^* & C & 0\\
E^* & 0 & 0 & D
\end{pmatrix}\,.\label{eq:SetupGeneral}
\end{equation}
This class of states includes (i) the Werner~\cite{Werner1989} and
Werner-like states, (ii) the Horodecki states, which are mixtures
of a Bell state and a separable state orthogonal to
it~\cite{Horodecki1996negativity}, (iii) Bell-diagonal states
(including the Werner states), and (iv) various types of MEMSs,
e.g., those defined in~\cite{Munro2001PhysRevA}. Moreover, the
capabilities of our method are not limited to the Werner or
Horodecki states based on a ``balanced'' Bell state, but also
allow for (v) their generalized forms based on unbalanced
entangled states $\sqrt{1-q} \ket{00} + \sqrt{q} \ket{11}$ for any
$q\in[0,1]$ instead of considering only $q=1/2$.

In this work, we focus on experimentally generating the Werner
states and GWSs, which are prepared on a platform of quantum
linear optics using the experimental setup depicted in
Fig.~\ref{fig:ExpSetup}. Qubits are encoded into polarization
states of single photons. The process of type-I SPDC occurring in
a cascade of two nonlinear $\beta$-BBO crystals, serves as a
source of entangled photons~\cite{Kwiat1999PRA}. When pumped by a
beam at a wavelength of $\lambda = 355$~nm the source generates
two polarization-entangled photons in mutually different spatial
modes at $\lambda = 710$~nm [Fig.~\ref{fig:ExpSetup}(a)]. The
state of these photons can be expressed in the form
$\ket{\phi^{+}} = ( \ket{00} +\ket{11})/\sqrt{2}$, where $\ket{0}$
and $\ket{1}$ denote horizontally ($H$) and vertically ($V$)
polarized photons, respectively. Due to the geometry of type-I
SPDC, photons are generated in symmetrically opposite directions
on the surface of a cone with its axis coincidental with the pump
beam. We choose to couple photon pairs propagating in the vertical
and horizontal planes, denoting them by (1A,1B) and (2A,2B),
respectively [see Fig.~\ref{fig:ExpSetup}(a)]. Assuming that only
two-photons are generated (so higher-photon-number processes are
negligible), these photons are simultaneously in either fmodes
{(1A,1B)} or (2A,2B). Employing a half-wave plate (HWP) at
45$^\circ$ in the {2B mode}, the state $\ket{\phi^{+}}$ is
transformed into the Bell state $\ket{\psi^{+}} = (\ket{01}
+\ket{10})/\sqrt{2}$. Thus, we obtained states spanning the two
subspaces $\ket{\phi^{+}}$ and $\ket{\psi^{+}}$.

Our goal is to prepare the Werner states and their generalizations
for various values of the mixing parameter $p$. The main idea
behind the design of our setup is to decrease temporal coherence
of the states $\ket{\phi^{+}}$ (in the modes 1A and 1B) and
$\ket{\psi^{+}}$ (in the modes 2A and 2B) using beam displacer
assemblies (BDAs). A BDA consists of a pair of beam displacers
(BDs) with an HWP inserted between them. This allows us to split
and subsequently rejoin the horizontal and vertical components of
a photon polarization state. By introducing a difference in the
propagation time between these two components (which is done by
tilting one BD) we can achieve their mutual phase difference (by
fine tilting) and tunable distinguishability (by coarse tilting).
Polarization-sensitive losses can easily be implemented by
partially blocking one of the polarization paths. Subsequently,
the modes (1A,2A) and (1B,2B) are incoherently mixed in fiber
couplers.

First, the subspace $\ket{\phi^+}$ is adjusted while arms 2A and
2B (belonging to the subspace $\ket{\psi^+}$) are blocked. By
means of the polarization-sensitive losses in BDA${_1}$, we
regulated the intensity ratio of the matrix elements $A$ and $D$
[see Eq.~(\ref{eq:SetupGeneral})] in the computational basis,
i.e., $\ket{00}$ and $\ket{11}$ (or $\ket{HH}$ and $\ket{VV}$ in
the polarization terms). The ratio accounts for
\begin{equation}
R_{\textrm{A,D}} = \frac{4pq+1-p}{4p(1-q)+1-p}\,,
\end{equation}
where $p$ and $q$ are both tuned parameters. The next step
consists of tuning the decoherence by observing coincidence counts
in the projections onto $\ket{++}$ and $\ket{+-}$, where
$\ket{\pm} = (\ket{0} \pm \ket{1})/\sqrt{2}$ stand for diagonal
and anti-diagonal polarization states, respectively. We find such
a coarse tilt of BD${_1}$, so that the visibility accounts for
\begin{equation}
\nu = \frac{2E}{A+D},
\end{equation}
while the phase is set by fine tuning, the tilt using a piezo
actuator, which minimizes the signal in the $\ket{+-}$ projection
by setting the effective value of $E$ to be real.

Second, when adjusting the subspace $\ket{\psi^+}$, in turn, the
arms 1A and 1B are blocked. In analogy with the adjustment of the
$\ket{\phi^+}$ subspace, the same two steps are performed. This
time, however, the target intensity ratio $R_{\textrm{B,C}}$ is
equal to 1 because $B = C$. The coarse tilt of BD${_3}$ needs to
be sufficient to decrease the coherence of the state completely
since $\nu = 0$ resulting in $F = 0$. The phase becomes
meaningless.

Finally, all arms are unblocked and both subspaces are balanced to
adjust the ratio between the matrix elements $A$ and $B$. While
having the projection onto $\ket{00}$ and $\ket{01}$, the required
ratio is
\begin{equation}
R_{\textrm{A,B}} = \frac{4pq+1-p}{1-p}\,.
\end{equation}
For this purpose, we partially closed the irises in the 1A and 2A
couplers, which are depicted in Fig.~\ref{fig:ExpSetup}(b) by
labels I${_1}$ and I${_2}$, respectively.

After all the adjustments are implemented, the measurement itself
is carried out and it consists of a standard full
QST~\cite{Halenkova2012ApplOpt}. Polarization projection is
performed on both photons utilizing a set of quarter- and
half-wave plates, as well as polarizers and single-photon
detectors. Coincidence detections within $2$-ns window are
detected under all 36 two-fold combinations of single-photon
projections onto the basis states: $\ket{0}$, $\ket{1}$,
$\ket{+}$, $\ket{-}$, and $(\ket{0} \pm i\ket{1}/\sqrt{2}$, where
the latter states are the right- and left-hand circularly
polarized states, respectively. Density matrices are estimated via
a maximum likelihood
method~\cite{Banaszek1999,James2001,Hradil2004,Sanderson2016,Sanderson2018}.

Because of experimental imperfections, the setup produces states
with the $p$ and $q$ parameters slightly different from those
targeted by the above-described procedure. To observe a better
agreement with theoretical predictions, we estimate the
best-fitting parameters $p_\mathrm{est}$ and $q_\mathrm{est}$ by
finding such a $\rho_{\rm GW}(p_\mathrm{est},q_\mathrm{est})$ that
its fidelity with the experimentally reconstructed density matrix
is maximized. We find that the deviations of the estimated value
of the mixing parameter $p_\mathrm{est}$ from the value of $p$,
which is set with a limit precision in our experiment, are on the
average are equal to 0.01 for the Werner states and 0.03 for the
GWSs. For the estimated value of the superposition parameter
$q_\mathrm{est}$, the observed parameter deviations from a given
value of $q$ equal on average to 0.02. The maximal deviation
values are 0.03 for both Werner states and GWSs. Note that the
superposition parameter $q$ is manually set by an HWP in the
source part of the setup shown in Fig.~\ref{fig:ExpSetup}(a).
Experimental data as well as the estimated density matrices are
provided in the Supplementary Material~\cite{supplement}.

The states $\rho_{\rm W}$ and $\rho_{\rm GW}$ can also be
expressed by Eq.~(\ref{eq:SetupGeneral}) with $F = 0$. In this
matrix, the subspace spanned by the states $\ket{\phi_q}$ for
$q\in[0,1]$ is represented by the elements $A, D, E,$ and $E^*$,
while the corresponding subspace for the white-noise term
corresponds to only diagonal elements ($A,B,C,D$). For the reasons
specified below, we set, in our experiments, the superposition
coefficient at $q=0.9$, in addition to $q=0.5$.

Note that it is irrelevant to replace $\ket{\phi_q}$ by a
four-term superposition state
$\ket{\phi_{abcd}}=a\ket{00}+b\ket{01}+c\ket{10}+d\ket{11}$ at
least in the analysis of nonclassical correlations. This is
because $\ket{\phi_{abcd}}$ can be reduced to $\ket{\phi_q}$
solely by local rotations, so the studied two-qubit quantum
correlations are unchanged. As mentioned above, the GWSs are not
diagonal in the Bell basis, except $q=0,1/2,1$. This property
greatly complicates analytical calculations of correlation
measures. So, for the Bell-nondiagonal GWSs, we present analytical
formulas of the entanglement and nonlocality measures only,
contrary to the corresponding results for the Werner states, which
include also our formulas for the steerable weights.

\begin{figure}
 \hspace*{-8mm} \#1 \hspace*{5mm} \#2 \hspace*{12mm} \#3 \hspace*{12mm} \#5
\hfill\includegraphics[scale=1.]{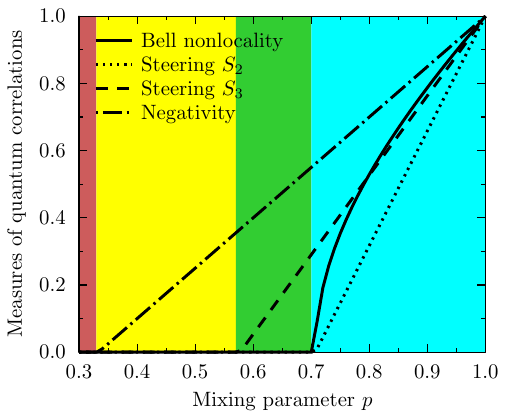}\hspace*{\fill}
\caption{Four correlation regimes of the Werner states
corresponding to those listed in Table~\ref{table1}. Note that
regime \#4 is missing. Theoretical plots for the negativity $N$
(or, equivalently, the concurrence $C$), the steerable weights
$S_2$ and $S_3$, and the Bell nonlocality measure $B$ as a
function of the mixing parameter $p$. Exact definitions of the
depicted quantum correlation measures are given in
Sec.~\ref{ch:HierarchyAll}.} \label{fig:theoryWS}
\end{figure}

\begin{figure}
\centering
 \hspace*{12mm} \#1 \hspace*{15mm} \#2 \hspace*{11mm} \#3 \#4 \hspace*{5mm} \#5
\hfill\includegraphics[scale=1.]{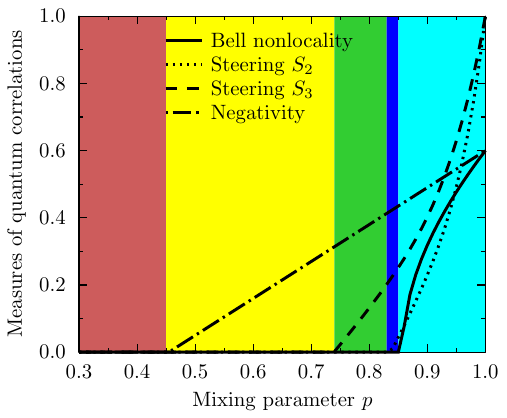}\hspace*{\fill}
\caption{Five correlation regimes of the GWSs corresponding to the
regimes in Table~\ref{table2}. Curves are analogous to those in
Fig.~\ref{fig:theoryWS}, but for the superposition parameter
$q=0.9$ or, equivalently, $q=0.1$.} \label{fig:theoryGWS}
\end{figure}

We begin our detailed comparative analysis by presenting various
theoretical relations between chosen correlation measures for the
Werner states and GWSs showed in Figs.~\ref{fig:theoryWS}
and~\ref{fig:theoryGWS}, respectively. These curves show the
negativity $N$ (or equivalently the concurrence $C$), the
steerable weights $S_2$ and $S_3$, and the Bell nonlocality
measure $B$ as a function of the mixing parameter $p$. Each
colored region starts, where a given correlation measure becomes
non-zero with an increasing value of the mixing parameter $p$. We
refer to these regions as quantum correlation regimes, which are
also listed in Tables~\ref{table1} and~\ref{table2}.

\begin{figure}
\hspace*{-20mm} (a)  \hspace*{35mm} (b)  \\ \vspace*{-1mm}
\centering 
{\includegraphics[width=.5\columnwidth]{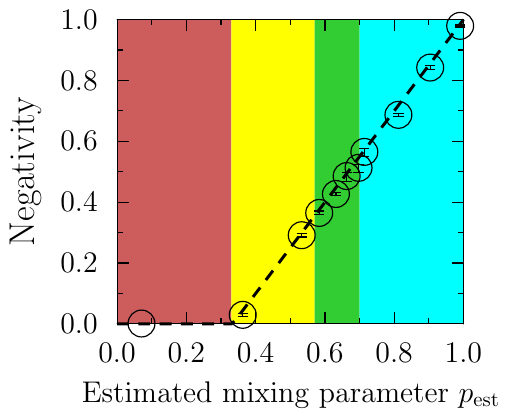}} \hspace*{-4mm}
{\includegraphics[width=.5\columnwidth]{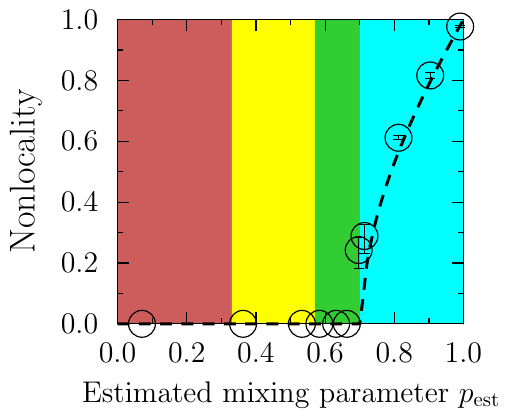}}
\\
\vspace*{-1mm} \hspace*{-20mm} (c)  \hspace*{35mm} (d) \\ \vspace*{-1mm}
{\includegraphics[width=.52\columnwidth]{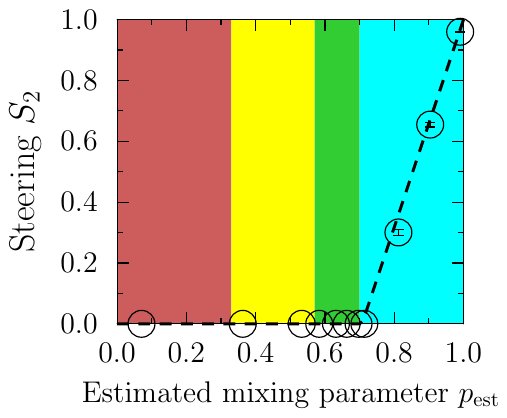}}
\hspace*{-6mm}
{\includegraphics[width=.52\columnwidth]{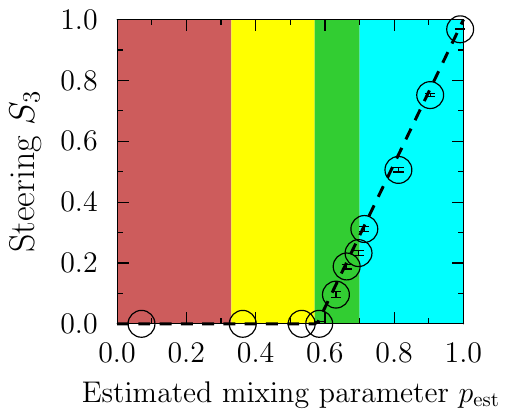}}
\caption{Quantum correlations for the theoretical and experimental
Werner states as a function of the estimated mixing parameter
$p_{\rm est}$: (a) the negativity $N$, (b) the Bell nonlocality
measure $B$, and the steerable weights (c) $S_2$ and (d) $S_3$.}
\label{fig:ExpWS}
\end{figure}
\begin{figure}
\hspace*{-20mm} (a)  \hspace*{35mm} (b)  \\ \vspace*{-1mm}
\centering 
{\includegraphics[width=.5\columnwidth]{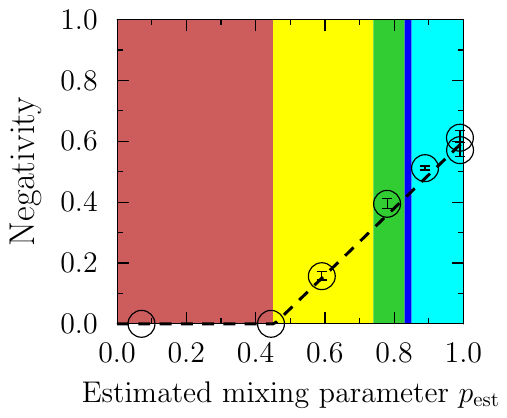}}
\hspace*{-4mm} 
{\includegraphics[width=.5\columnwidth]{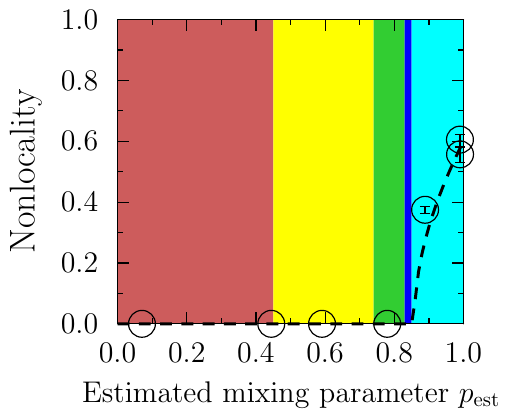}}
\\
\vspace*{-1mm} \hspace*{-20mm} (c)  \hspace*{35mm} (d) \\ \vspace*{-1mm}
{\includegraphics[width=.52\columnwidth]{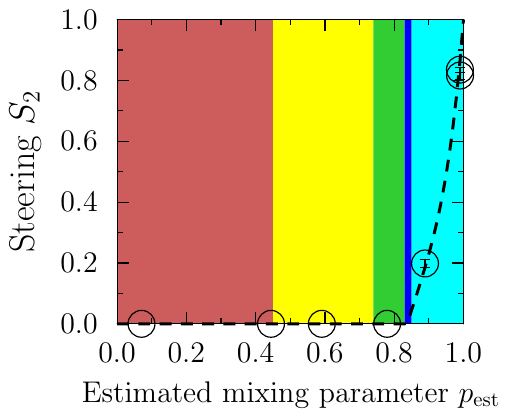}}
\hspace*{-6mm} 
{\includegraphics[width=.52\columnwidth]{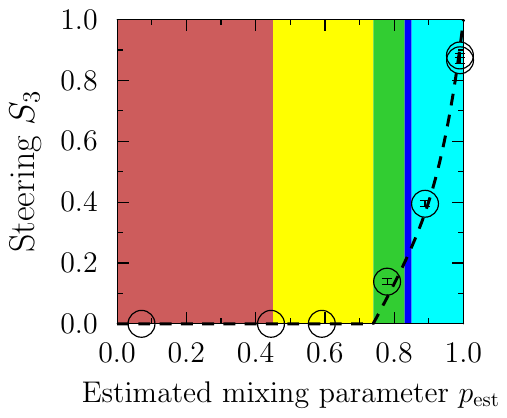}} \caption{Same
as in Fig.~\ref{fig:ExpWS}, but for the GWSs for the estimated
superposition coefficient $q_{\rm est}\approx q=0.9$ (see the text
for details).}\label{fig:ExpGWS}
\end{figure}

\begin{table*}
\caption{Hierarchy of classes of correlations for the Werner
states, $\rho_{\rm W}(p)$, depending on the mixing parameter~$p$.
The four regimes of vanishing or nonvanishing different classes of
quantum correlations correspond to the regimes shown in
Figs.~\ref{fig:theoryWS} and~\ref{fig:ExpWS}.} \label{table1}
\begin{tabular}{c c c c c c c}
    \hline\hline
     Regime   & $B$         & $S_2\equiv S_2^{ij}$ & $S_3$    & $N$         & $p$ & Experiment \\
    \hline
        \#1 & $B=0$ & $S_2=0$ & $S_3=0$ & $N=0$ & $p\in[0,\frac13]$  & direct \\
        \#2 & $B=0$ & $S_2=0$ & $S_3=0$ & $N>0$ & $p\in(\frac13,\frac1{\sqrt{3}}]$ & direct \\
        \#3 & $B=0$ & $S_2=0$ & $S_3>0$ & $N>0$ & $p\in(\frac1{\sqrt{3}},\frac1{\sqrt{2}}]$ & direct \\
        \#4 & $B=0$ & $S_2>0$ & $S_3>0$ & $N>0$ & $p\in \emptyset $ & impossible \\
        \#5 & $B>0$ & $S_2>0$ & $S_3>0$ & $N>0$ & $p\in(\frac1{\sqrt{2}},1]$ & direct \\
    \hline
    \hline
\end{tabular}
\end{table*}

\begin{table*}[t]
\caption{Hierarchy of classes of correlations exhibited by the
Bell-nondiagonal GWSs, $\rho_{\rm GW}(p,q)$, for different values
of the mixing parameter $p$ and a fixed value of the superposition
parameter at $q=0.9$ or, equivalently, $q=0.1$. This table lists
the five regimes shown in Figs.~\ref{fig:theoryGWS},
\ref{fig:ExpGWS}, and~\ref{fig6}(a). The threshold values read:
$p'_N\equiv p_N(q)=5/11=0.45(45)$ and $p'_B\equiv
p_B(q)=5/\sqrt{32}=0.8574\cdots$, are given by Eqs.~(\ref{pN})
and~(\ref{pB}) for $q=0.9$ (or 0.1), respectively, while
$p'_{S_3}\equiv p_{S_3}(q)=0.7390\cdots$ and $p'_{S_2}\equiv
p_{S_2}(q)=0.8370\cdots$ are obtained numerically. The term
\emph{hybrid} experiment refers to averaging of two directly
generated experimental states according to Eq.~(\ref{GWS085}). 2MS
and 3MS stand for two- and three-measurement scenarios,
respectively.} \label{table2}
\begin{tabular}{c c c c c c c c c c}
    \hline\hline
    Regime & states & $B$ & $S_2^{XY}$ & $S_2\equiv S_2^{XZ}=S_2^{YZ}$ & $S_3$ & $N$ & $p$ & Experiment \\
    \hline
        \#1 & separable & $B=0$ & $S^{XY}_2=0$ & $S_2=0$ & $S_3=0$& $N=0$ & $p\in[0,p'_N]$  & direct \\
        \#2 & 3MS-unsteerable but entangled & $B=0$ & $S^{XY}_2=0$ & $S_2=0$ & $S_3=0$& $N>0$ & $p\in(p'_N,p'_{S_3}]$ & direct \\
        \#3 & steerable in 3MS but not in 2MS & $B=0$ & $S^{XY}_2=0$ & $S_2=0$ & $S_3>0$& $N>0$ & $p\in(p'_{S_3},p'_{S_2}]$ & direct \\
        \#4 & Bell local but 2MS-steerable & $B=0$ & $S^{XY}_2=0$ & $S_2>0$ & $S_3>0$& $N>0$ & $p\in(p'_{S_2},p'_B]$ & hybrid \\
        \#5 & Bell nonlocal & $B>0$ & $S^{XY}_2>0$ & $S_2>0$ & $S_3>0$& $N>0$ & $p\in(p'_B,1]$ & direct \\
    \hline
    \hline
\end{tabular}
\end{table*}
 \begin{figure}[t]
\begin{center}
\subfloat[CC hierarchy]
{\includegraphics[width=.52\columnwidth]{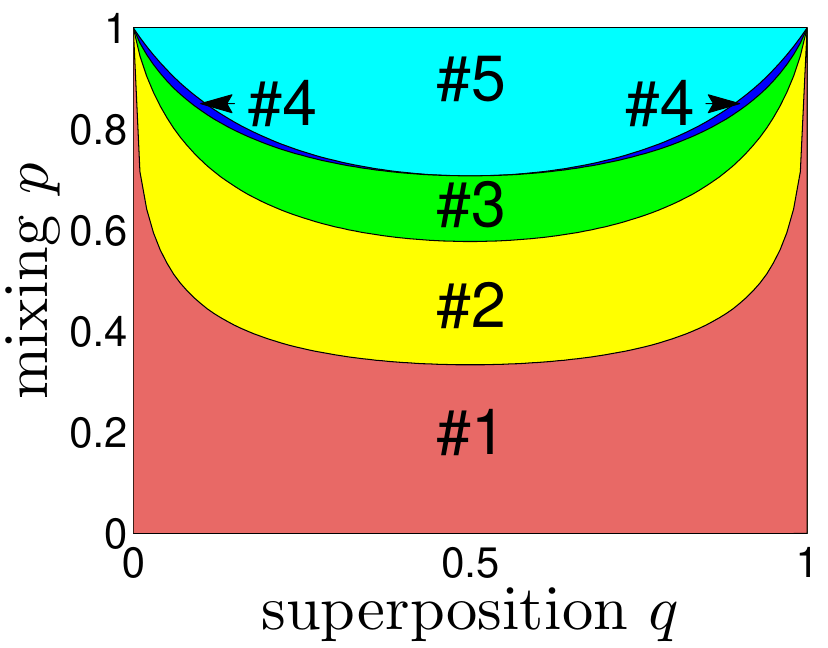}}\\
\subfloat[Optimal robustness]
{\includegraphics[width=\columnwidth]{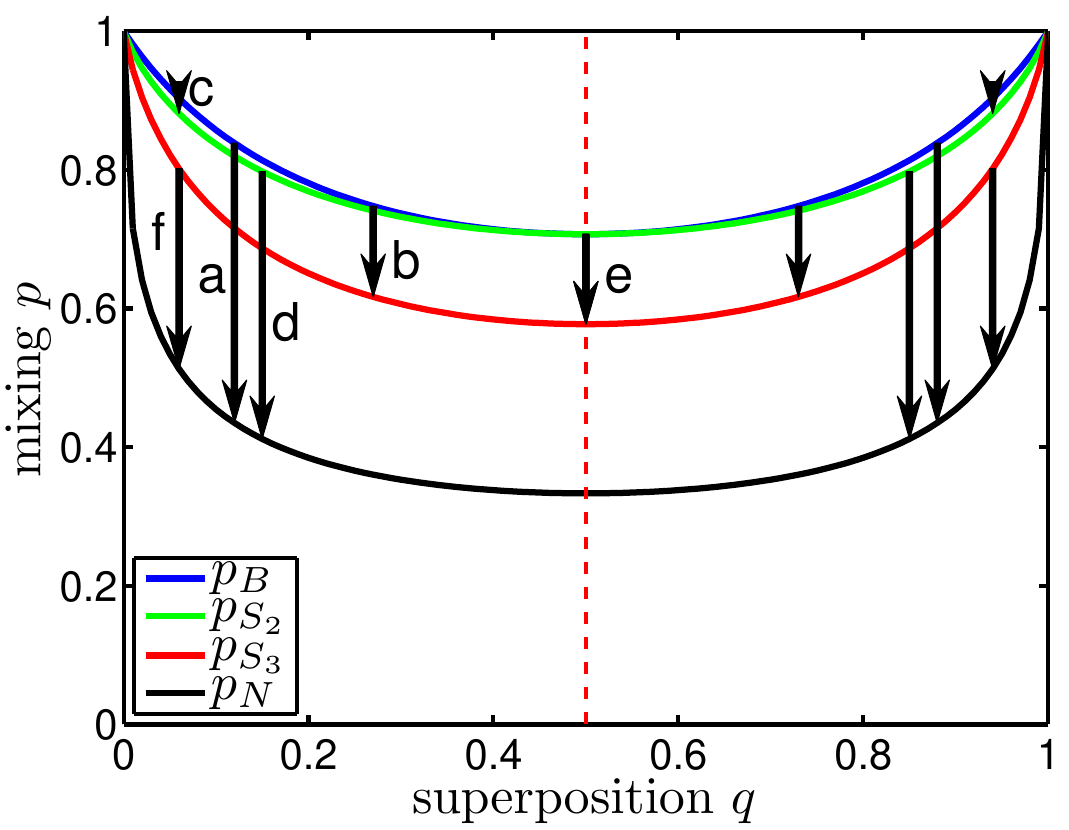}}
\caption{Threshold mixing parameters $p_i(q)$ versus the
superposition parameter $q$ for the GWSs. (a) The threshold curves
separate the five regimes in the hierarchy of the classes of
quantum correlations, which are listed in Table~\ref{table2}.
(b)~Transitions between various curves, requiring the largest
amount of white noise, are indicated by arrows. It can be seen
that the only arrow $e$ for the Werner states (i.e., the GWSs at
$q=1/2$) is marked for the transition between the curves
$p_{S_2}(q)$ and $p_{S_3}(q)$. All the other arrows are plotted at
$q\neq 1/2$. This explains the meaning of enhanced robustness of
the Bell-nondiagonal GWSs against white noise compared to that of
the Werner states. The locations at $q_{\rm opt}$ and lengths of
the labeled arrows are listed in Table~\ref{table3}. The
unlabelled arrows are located at $q'_{\rm opt}=1-q_{\rm opt}$.}
 \label{fig6}
 \end{center}
 \end{figure}

\section{Hierarchy of the classes of correlations for Werner-like states} \label{ch:HierarchyAll}

\subsection{Entanglement of Werner-like states}
\label{ch:GWS_Entanglement}

It is well known that, for the Werner states, the concurrence and
negativity, which were defined in
Sec.~\ref{ch:Intro_Entanglement}, are equal to each other and are
given by a linear function of the mixing parameter $p$, i.e.,
\begin{eqnarray}
  N(\rho_{\rm W}) = C(\rho_{\rm W}) &=& \max\left[0,(3p-1)/2\right],
\label{N_WS}
\end{eqnarray}
as shown in Fig.~\ref{fig:theoryWS} by the dot-dashed curve. The
good agreement of the negativities calculated for the theoretical
and experimental Werner states is shown in
Fig.~\ref{fig:ExpWS}(a).

We find that the negativity and concurrence for the GWSs read:
\begin{equation}
  N(\rho_{\rm GW}) = C(\rho_{\rm GW}) = \max\left\{0,\tfrac{1}{2} \left[p (1+4 \sqrt{x})-1\right]\right\},
\label{N_GWS}
\end{equation}
with $x=q(1-q)$, which is plotted in Fig.~\ref{fig:theoryGWS} by
the dot-dashed curve. Figure~\ref{fig:ExpGWS}(a) demonstrates the
good fit of the negativities calculated for the theoretical and
experimental GWSs for different values of the superposition
parameter $q$. Note that not only $N(\rho_{\rm W})$, but also
$N(\rho_{\rm GW})$ is a linear function of the mixing parameter
$p$ for a fixed value of the superposition coefficient $q$. In a
special case for a pure state $\ket{\phi_q}$ (i.e., when $p=1$),
Eq.~(\ref{N_GWS}) reduces to
$N(\ket{\phi_q})=C(\ket{\phi_q})=2\sqrt{q(1-q)}$.

Equation~(\ref{N_GWS}) vanishes for $p\in[0,p_N(q)]$ at the
threshold value given by
\begin{equation}
  p_N(q)=1/\big[1+4 \sqrt{q(1-q)}\big],
  \label{pN}
\end{equation}
which is plotted in Fig.~\ref{fig6}. It can be seen that
$N[\rho_{\rm W}(p)]>0$ if $p>1/3$ and $N[\rho_{\rm GW}(p,0.9)]>0$
if $p>p'_N=5/11$. These threshold values are below compared with
those for the other measures of quantum correlations and also
marked in Figs.~\ref{fig6} and~\ref{fig7}.

Note that $N(\rho_{\rm GW}) = C(\rho_{\rm GW})$ should hold for
the ideal GWSs, including the Werner states. However, our
experimental GWSs do not exactly satisfy this condition. Thus, we
calculate both measures, because their difference shows how much
our experimental states deviate from the ideal Werner states.
These deviations quantify also the precision of our measurements.
Specifically, the observed experimental differences between the
negativity and concurrence are on the average 0.02\% for the
Werner states and 0.06\% for the GWSs. Thus, on the scale of
Figs.~\ref{fig:ExpWS}(a) and~\ref{fig:ExpGWS}(a) one could not see
any differences between $N(\rho^E_{\rm GW})$ and $C(\rho^E_{\rm
GW})$.


\subsection{Steering of Werner-like states in the three-measurement scenario}
\label{ch:GWS_Steering3}

Steering in a 3MS on Alice's side can be quantified by the
steerable weight $S_3$ of Ref.~\cite{Skrzypczyk2014}, as defined
as an SDP in Appendix~\ref{App:Steering3}. We find that this
steerable weight $S_3$ for the Werner states is a \emph{linear}
function of the mixing parameter $p$, specifically,
\begin{equation}
  S_3(\rho_{\rm W}) = \max\left(0,\frac{\sqrt{3}p-1}{\sqrt{3}-1}\right),
  \label{S3_WS}
\end{equation}
which means that the state $\rho_{\rm W}(p)$ is steerable in the
3MS if $p>1/\sqrt{3}$ [see Fig.~\ref{fig:ExpWS}(d) and
Table~\ref{table1}]. By contrast to this, the steerable weight
$S_3$ for the GWSs is a \emph{nonlinear} function of the mixing
parameter $p$ for $q\neq 1/2$. This is shown for $q=0.9$ in
Fig.~\ref{fig:ExpGWS}(d). It can be seen that these GWSs are
steerable for $p > p_{S_3}=0.7390$ (see also Table~\ref{table2}).
This means that $\rho_{\rm GW}(p,0.9)$ is steerable for a much
shorter range of the mixing parameter $p$ than that for $\rho_{\rm
W}(p)\equiv \rho_{\rm GW}(p,1/2)$. Figures~\ref{fig:ExpWS}(d)
and~\ref{fig:ExpGWS}(d) show the weight $S_3$ for our experimental
states compared to those for the theoretical states. These results
show the good agreement of the theory with our experimental
results.

\subsection{Steering of Werner-like states in two-measurement scenarios}
\label{ch:GWS_Steering2}

To quantify steering of the Werner states and GWSs in 2MSs on the
Alice side, we apply the steerable weights $S_2^{ij}$ of
Ref.~\cite{Skrzypczyk2014} defined in
Appendix~\ref{App:Steering2}.

We find that the weights $S_2^{ij}$ for the Werner states, are
equal to each other, $S_2(\rho_{\rm W})\equiv S_2^{XY}(\rho_{\rm
W})=S_2^{XZ}(\rho_{\rm W})=S_2^{YZ}(\rho_{\rm W})$, being a
\emph{linear} function of the mixing parameter $p$,
i.e.,\begin{equation}
  S_2(\rho_{\rm W}) = \max\left(0,\frac{\sqrt{2}p-1}{\sqrt{2}-1}\right).
  \label{S2_WS}
\end{equation}
This implies the steerability of the states in the 2MS if
$p>1/\sqrt{2}$ [see Fig.~\ref{fig:ExpWS}(c) and
Table~\ref{table1}]. However, the steerable weights for the GWSs
become much more complicated. We find that $S_2^{XY}(\rho_{\rm
GW})\le S_2^{XZ}(\rho_{\rm GW})=S_2^{YZ}(\rho_{\rm GW})\equiv
S_2(\rho_{\rm GW})$, and there exist two threshold values
$p'_{S_2}$ and $p'_B$, as listed in Table~\ref{table2}.
Specifically, (i) $S_2^{XZ}(\rho_{\rm GW})=S_2^{YZ}(\rho_{\rm
GW})>0$ if $p>p'_{S_2}=0.8370\cdots$ and (ii) $S_2^{XY}(\rho_{\rm
GW})>0$ if $p>p'_B=5/\sqrt{32}$, which is the same threshold
parameter as that for the Bell nonlocality measure $B>0$, as
discussed above. Furthermore, the dependence of
$S_2^{ij}(\rho_{\rm GW})$ on the mixing parameter $p$ becomes
nonlinear for $q\neq 1/2$. Different values of the threshold
parameters for $p'_B$ and $p'_{S_2}$ imply the occurrence of
region \#4 for the GWSs, which is shown in
Figs.~\ref{fig:theoryGWS}, \ref{fig6}(a), and \ref{fig7}(c), and
explained in detail in Sec.~\ref{ch:CounterintuitiveResults1}.

\subsection{Nonlocality of Werner-like states} \label{ch:GWS_Nonlocality}

To estimate the degree of quantum nonlocality or, equivalently, to
quantify the violation of the Bell-CHSH inequality for two-qubit
states~\cite{Clauser1969}, we use the Horodecki
measure~\cite{Horodecki1995, Horodecki1996}, which is as defined
in Sec.~\ref{ch:Intro_Nonlocality}.

The Bell nonlocality measure for the Werner states reads as
\begin{equation}
  B(\rho_{\rm W}) = \sqrt{\max(0,2p^2-1)},
  \label{B_WS}
\end{equation}
which instantly implies a standard result that the Werner states
are nonlocal if $p>1/\sqrt{2}$ (see also Table~\ref{table1}).
However, if $p\in (1/3,1/\sqrt{2})$, the Werner states are
entangled without BIV (admitting an LHV model), as already
demonstrated by Werner in 1989 in~\cite{Werner1989}.

We find that the Bell nonlocality measure for the GWSs is given by
\begin{equation}
  B(\rho_{\rm GW}) = \max\left\{0,\sqrt{p^2 [1+4
  q(1-q)]-1}\right\}.
\label{B_GWS}
\end{equation}
Note that for pure states ($p=1$), Eq.~(\ref{B_GWS}) reduces to
the standard result
$B(\ket{\phi_q})=N(\ket{\phi_q})=2\sqrt{q(1-q)}$. It can be seen
that $B(\rho_{\rm GW})$ is zero for the values of the mixing
parameter in the range $p\in[0,p_B(q)]$ with the threshold value
given by
\begin{equation}
  p_B(q)=1/\sqrt{1+4q(1-q)},
  \label{pB}
\end{equation}
which is plotted in Fig.~\ref{fig6}. For the diagonal GWSs (with
$q=1/2$), we can reproduce the well-known threshold value
$p_B(1/2)=1/\sqrt{2}$ for the Werner states. In another special
case for $q=0.9$, which is set in our experiments, we have the
threshold value $p'_B\equiv p_B(q=0.9)=p_B(q=0.1)=5/\sqrt{32}$.
Thus, the GWSs $\rho_{\rm GW}(p,0.9)$ for $p\in
(p'_N,p'_B)=(5/11,5/\sqrt{32})$ are entangled without Bell
nonlocality, which occurs for a wider range of the mixing
parameter $p$ compared to that for the Werner states, i.e.,
$p'_B-p'_N\approx 0.4029 > 1/\sqrt{2}-1/3\approx 0.3738$, as it is
explained in detail in Sec.~\ref{ch:CounterintuitiveResults2}.

In Fig.~\ref{fig:ExpWS}(b), we plot $B(\rho_{\rm W})$ in
comparison to the numerically calculated $B(\rho^E_W)$ for the
experimental Werner states $\rho^E_W(p)$ for various values of the
mixing parameter $p$ and fixed $q=0.9$. Analogous results for the
Bell nonlocality measure $B(\rho_{\rm GW})$ for the GWSs generated
experimentally, $\rho^E_{\rm GW}(p;q=0.9)$, are shown in
Fig.~\ref{fig:ExpGWS}(b) in comparison to those for the ideal
GWSs, $\rho_{\rm GW}(p;q=0.9)$. Note that $B(\rho_{\rm GW})>0$ if
$p>p'_B$ (see also Table~\ref{table2}) assuming $q=0.9$ or $0.1$,
which is clearly larger than the corresponding threshold value
$1/\sqrt{2}$ for the Werner states. Both Figs.~\ref{fig:ExpWS}(b)
and~\ref{fig:ExpGWS}(b) show a relatively good agreement of our
experimental results compared to the corresponding theoretical
predictions. More details about the accuracy of our experimental
results are given in Sec.~\ref{ch:Experiment}.

\section{Counterintuitive results} \label{ch:CounterintuitiveResults}

Here we present, arguably, the most interesting theoretical
results of our paper for the states generated experimentally
(either directly or in a hybrid way).

\subsection{Steerability $S_2$ without Bell nonlocality}
\label{ch:CounterintuitiveResults1}

\begin{figure}
\begin{center}
\subfloat[entangled states without nonlocality:  regimes \#2,3,4]{\includegraphics[width=0.49\columnwidth]{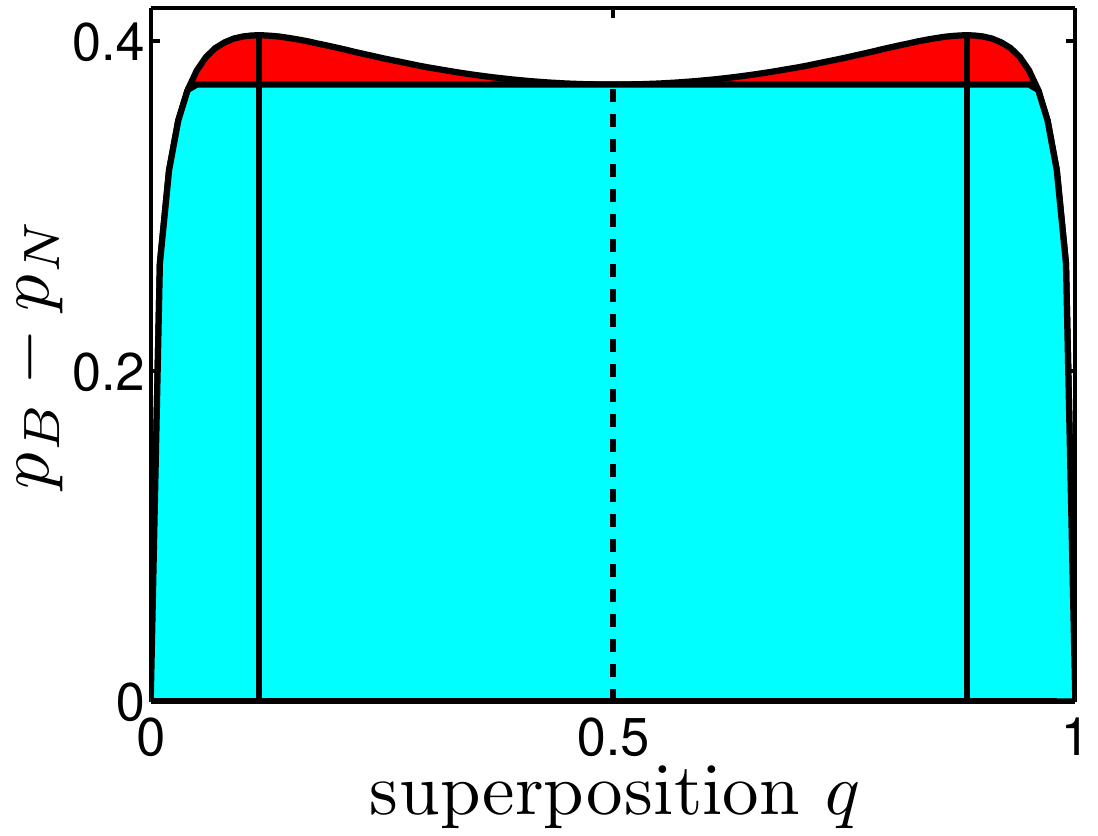}}\hspace*{0pt} %
\subfloat[3MS-steerable states without nonlocality: reg. \#3,4] {\includegraphics[width=0.49\columnwidth]{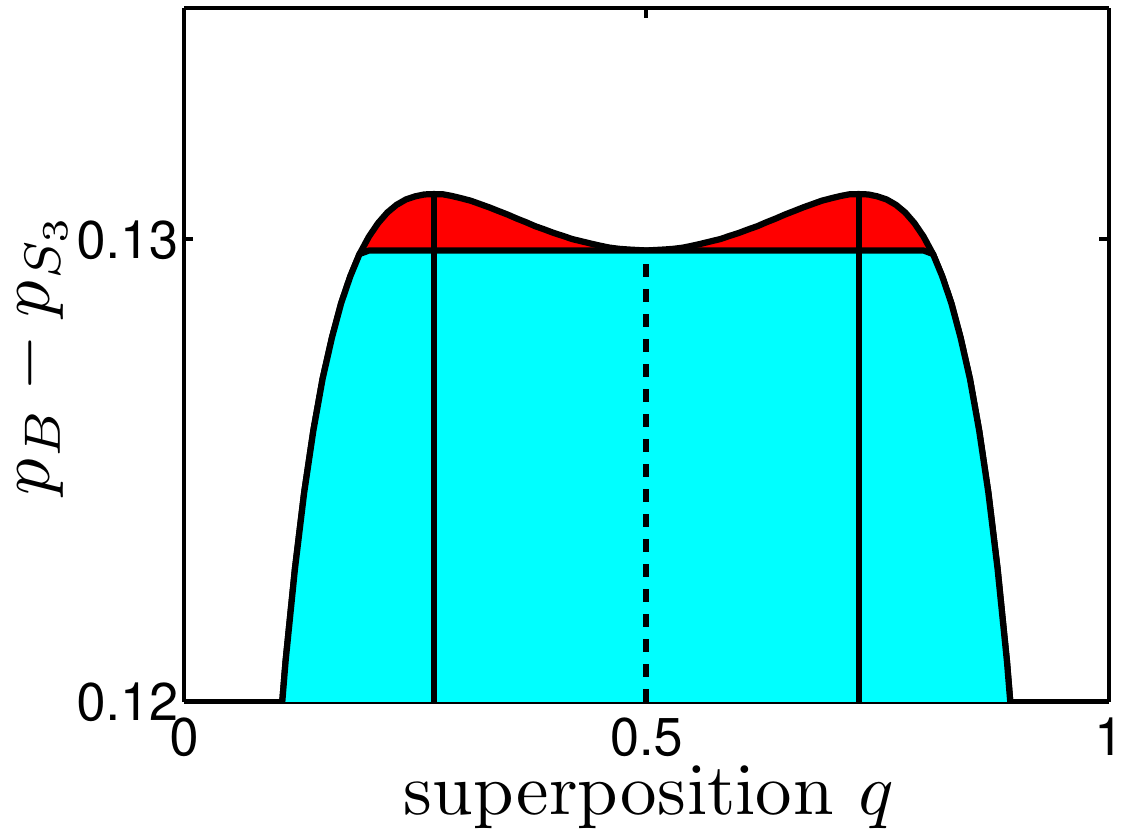}}\hspace*{0pt}\\ %
\subfloat[2MS-steerable states without nonlocality: regime \#4]{\includegraphics[width=0.49\columnwidth]{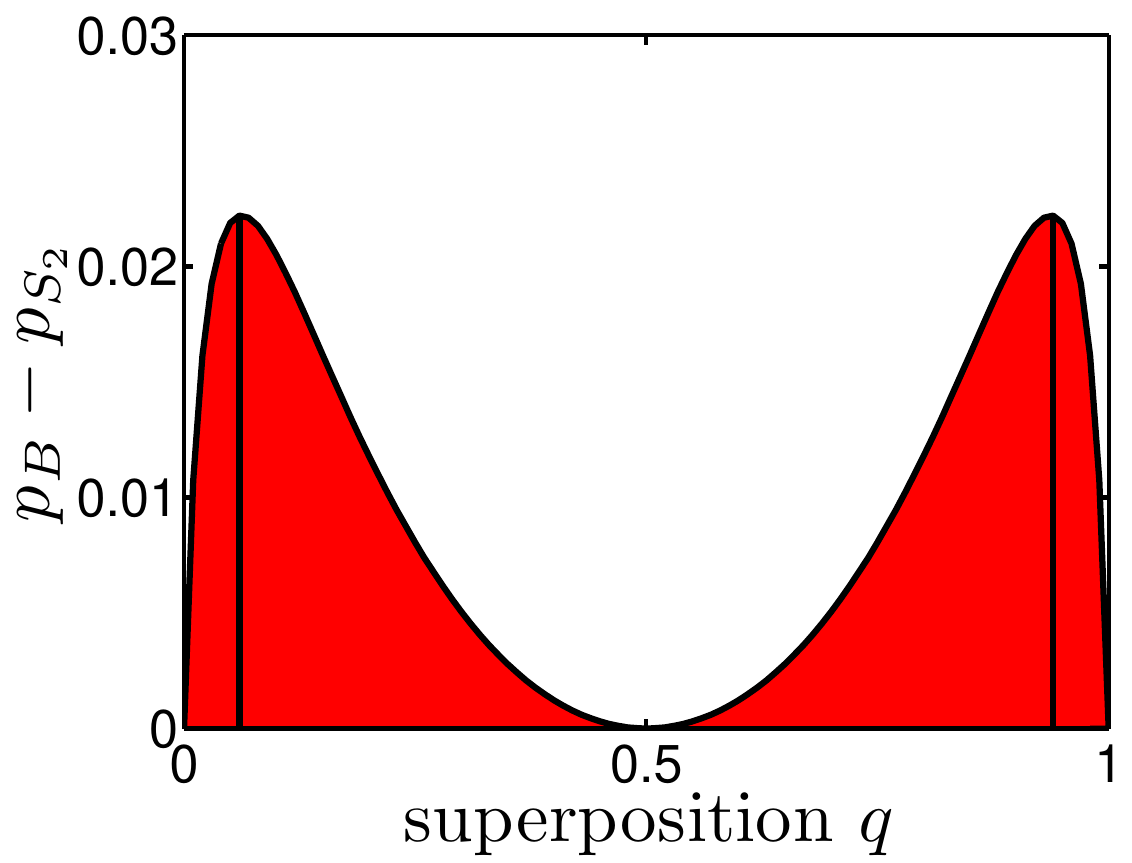}}\hspace*{0pt} %
\subfloat[2MS-unsteerable entangled states:  regimes \#2,3] {\includegraphics[width=0.49\columnwidth]{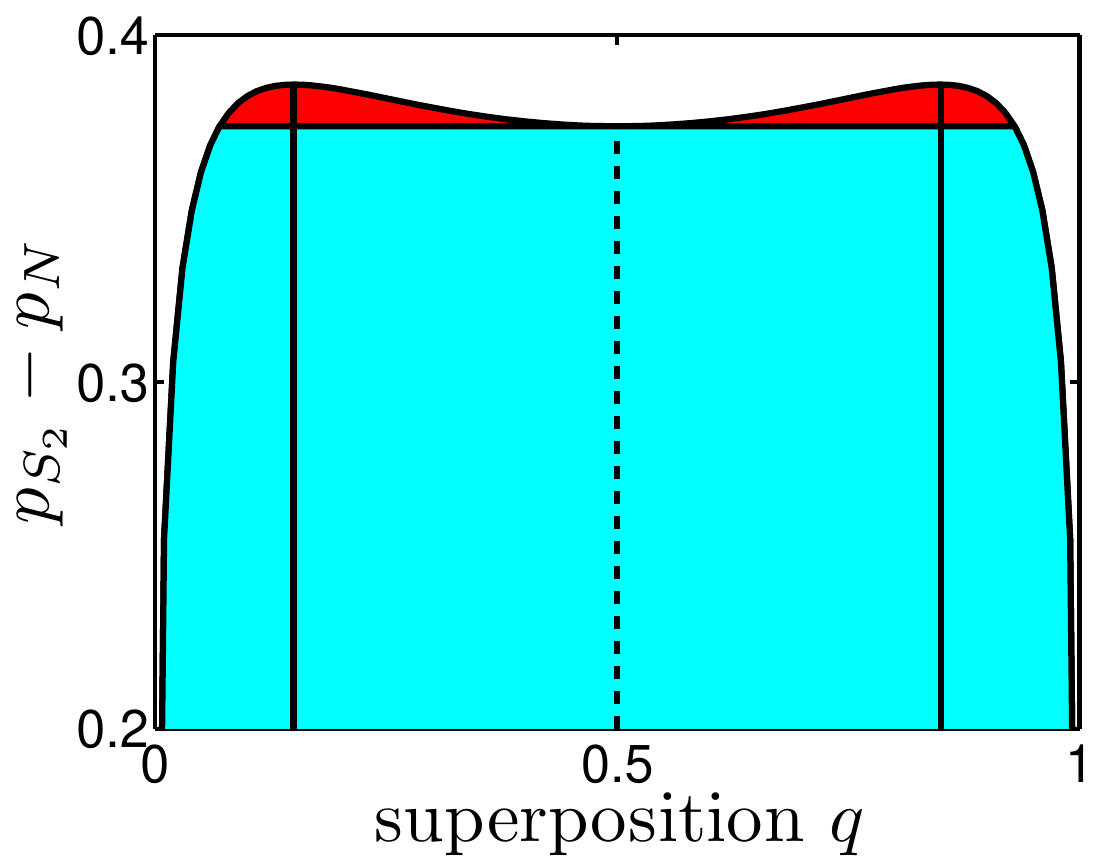}}\hspace*{0pt}\\ %
\subfloat[steerable states in 3MS but not in 2MS: regime \#3]{\includegraphics[width=0.49\columnwidth]{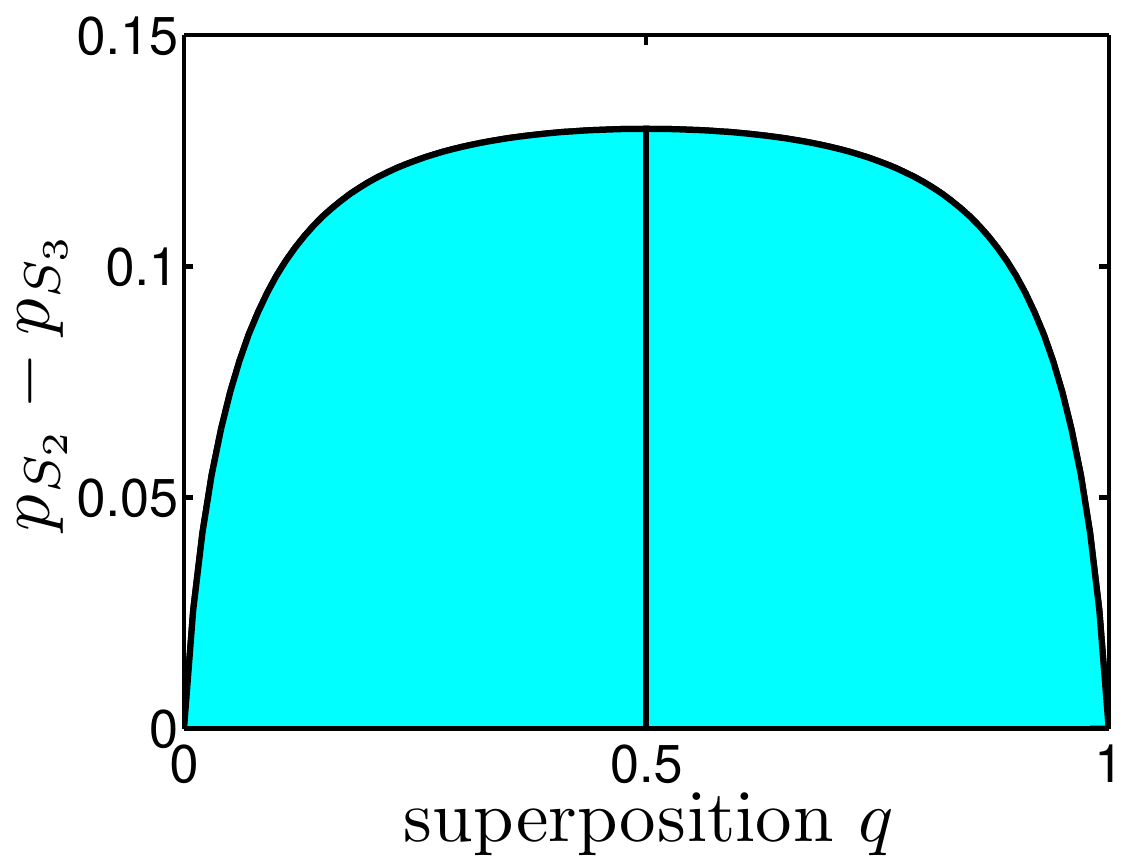}}\hspace*{0pt} %
\subfloat[3MS-unsteerable entangled states: regime \#2] {\includegraphics[width=0.49\columnwidth]{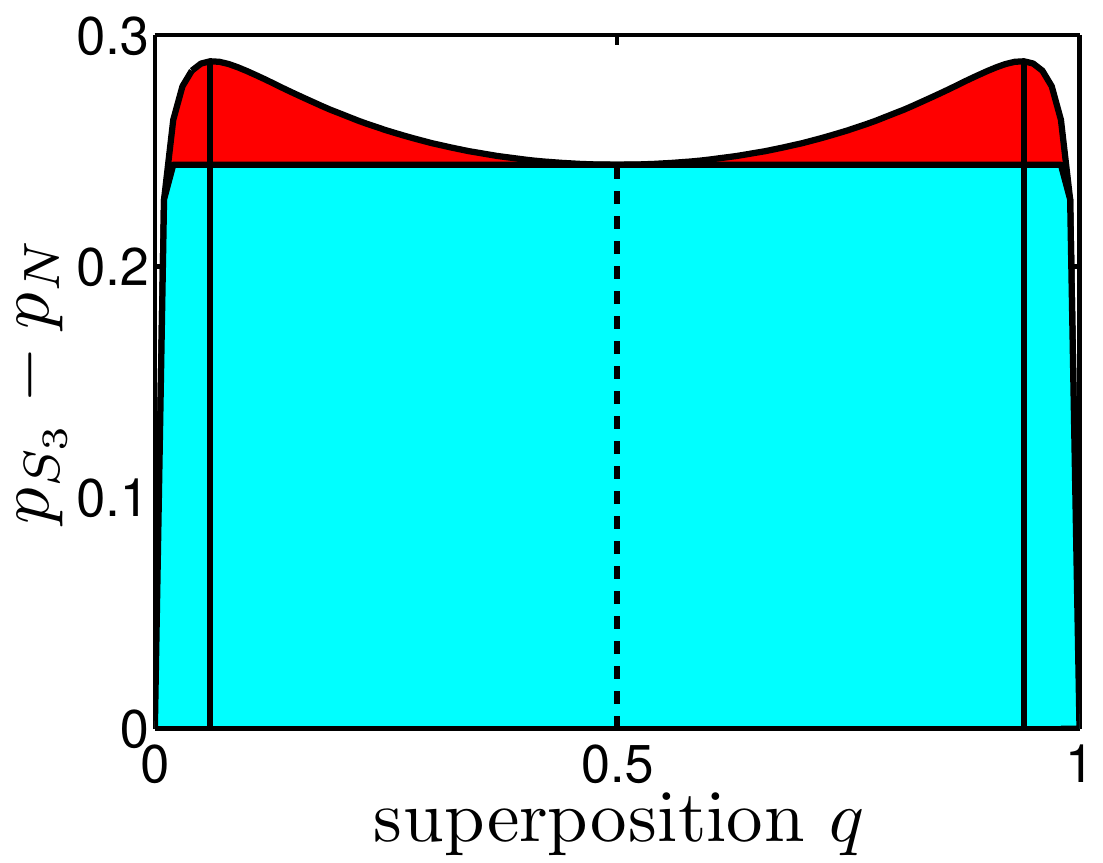}}\hspace*{0pt}\\ %
\caption{Differences $\Delta_{ij}(q)=p_i(q)-p_j(q)$ of the
threshold mixing parameters versus the superposition parameter $q$
for the GWSs corresponding to the transitions shown by the arrows
in Fig.~\ref{fig6}. The red-colored regions show explicitly the
improved robustness against the white noise of the
Bell-nondiagonal GWSs compared to the diagonal ones in the Bell
basis (i.e., the standard Werner states), except the case shown in
panel (e). Combined red and cyan regions correspond to the regimes
indicated in the captions of all these panels and those listed in
Table~\ref{table2}.}
 \label{fig7}
 \end{center}
 \end{figure}

Here we show that Bell-nondiagonal GWSs are steerable in 2MSs on
Alice's side but still admit an LHV model. So, the existence of
such quantum correlations cannot be revealed by the violation of
the Bell-CHSH inequality. The GWSs exhibiting the
$S_2$-steerability without Bell nonlocality correspond to the
regime \#4 in Table~\ref{table2} and are shown in
Figs.~\ref{fig:theoryGWS}, \ref{fig6}(a), and \ref{fig7}(c).

Our analytical and numerical results clearly demonstrate that the
regime \#4 cannot be observed for the Werner states, for which
$p_B(1/2)=p_{S_2}(1/2)$ holds, as seen in Fig.~\ref{fig:theoryWS}.
However, this degeneracy is broken for the GWSs with $q\neq
0,1/2,1$.

We find this result interesting, although the amount of the
required white noise destroying the correlations is small [i.e.,
$\max_q\Delta_{B,S_2}(q)=0.023$] compared to all the other cases
shown in Fig.~\ref{fig7}, except panel \ref{fig7}(e).

Moreover, regime \#4 can be observed for the mixing parameter $p$
limited to a very narrow range $[p'_{S_2},p'_B]\approx
[0.837,0.857]$ assuming $q=0.9$ (or, equivalently, 0.1), as shown
in Figs.~\ref{fig:ExpGWS}(b) and \ref{fig:ExpGWS}(c). We have
experimentally generated the GWSs for $p=0.8$ and $p=0.9$, but
unfortunately they are outside the desired range
$[p'_{S_2},p'_B]$.

To solve this problem, we recall that mixtures of any two GWSs,
say $\rho_{\rm GW}(p_1,q)$ and $\rho_{\rm GW}(p_2,q)$ for a fixed
value of $q$, are also GWSs. Specifically,\begin{equation}
  \rho^E_{\rm GW}(p,q) = \frac{p_2-p}{p_2-p_1} \rho_{\rm GW}(p_1,q)+
  \frac{p-p_1}{p_2-p_1}\rho_{\rm GW}(p_2,q).
  \label{GWSmix}
\end{equation}
Thus, we can use this property to produce (or simulate) a GWS,
which was not measured directly in our experiment, e.g.,
\begin{equation}
  \rho_{\rm GW}(0.85,q)
  = \frac{1}{2} [\rho^E_{\rm GW}(0.8,q)+\rho^E_{\rm GW}(0.9,q)],
  \label{GWS085}
\end{equation}
simply by balanced post-measurement numerical mixing of the two
experimental GWSs, $\rho^E_{\rm GW}(p,q)$ for $p=0.8$ and $0.9$
assuming $q=0.9$. We refer to this method as a \emph{hybrid}
experimental generation, as written in Table~\ref{table2} for the
regime \#4. By contrast to this regime, we have \emph{directly}
generated experimental states in all other regimes listed in
Tables~\ref{table1} and~\ref{table2}. Moreover, all the states
plotted in our figures correspond to those \emph{directly}
experimentally generated without using any post-measurement
numerical mixing.

{Our prediction of the existence of states in regime \#4 is a
surprising result and our experiment just confirms it. This
prediction seems to be especially counterintuitive} in the context
of the Girdhar-Cavalcanti theorem that ``All two-qubit states that
are steerable via CHSH-type correlations are Bell nonlocal''
in~\cite{Girdhar2016} (see also
Refs.~\cite{Cavalcanti2015,Costa2016}), which seemingly implies
the impossibility of generating states in this regime. However,
the Girdhar-Cavalcanti theorem is valid in a 2-2 measurement
scenario only, i.e., for ``a scenario employing only correlations
between two arbitrary dichotomic measurements on each
party''~\cite{Girdhar2016}. Our steering measures $S_2$ and $S_3$
refer to 2-3 and 3-3 measurement scenarios, respectively. Indeed
we always assume a full tomography on Bob's side corresponding to
the measurement of the three Stokes parameters:
$\langle\sigma_x\rangle$, $\langle\sigma_y\rangle$, and
$\langle\sigma_z\rangle$. While the projective measurements on the
Alice's side can be limited to 2MS or 3MS. It can be seen that our
and Girdhar and Cavalcanti's steering results refer to different
measurement scenarios. Thus, the observation of regime \#4 in our
steering scenarios does not imply the violation of the
Girdhar-Cavalcanti theorem.

\begin{table*}
\caption{Transitions between the threshold values of different
correlations of the GWSs for the optimal superposition parameter
$q_{\rm opt},$ which maximizes the white-noise robustness,
$\Delta_{if}(q_{\rm opt}) = p_i(q_{\rm opt})-p_f(q_{\rm opt})$ for
$i\neq f\in\{N,S_2,S_3,N\}$. These transitions correspond to the
arrows shown in Fig.~\ref{fig6}(b), and the length of a given
arrow is given by $\Delta_{if}(q_{\rm opt})$. The parameter $p_i$
($p_f$) is the threshold value of the mixing parameter $p$ for the
initial (final) class of correlations, or, equivalently, the
position of the beginning (end) of the corresponding arrow. The
last column shows the relative robustness with respect to the
standard Werner states (i.e., the GWS for $q=1/2$). Note that, for
every $q_{\rm opt}$, there is a second optimal value of the
superposition parameter, $q'_{\rm opt}=1-q_{\rm opt}$, exhibiting
the same quantum correlation properties. }
\begin{ruledtabular}
\begin{tabular}{lcccccc}
 Transition  &   $q_{\rm opt}$ &  $p_i(q_{\rm opt})$& $p_f(q_{\rm opt})$ & $\Delta_{if}(q_{\rm opt})$ &  $\Delta_{if}(\tfrac12)$ &  $\Delta_{if}(q_{\rm opt})-\Delta_{if}(\tfrac12)$     \\ \hline
(a)  $p_B\to p_{N}$       & 0.1170 &  $p_B=0.8412$     &  $p_{N}=0.4375$ & 0.4037  & 0.3738    &  0.0299   \\
(b)  $p_B\to p_{S_3}$     & 0.2692 &  $p_B=0.7481$     &  $p_{S_3}=0.6171$ & 0.1310  & 0.1298   &  0.0012  \\
(c)  $p_B\to p_{S_2}$     & 0.0625 &  $p_B=0.9001$     &  $p_{S_2}=0.8779$  & 0.0222 & 0 &  0.0222   \\
(d)  $p_{S_2}\to p_{N}$   & 0.1508 &  $p_{S_2}=0.7971$ &  $p_{N}=0.4113$ & 0.3858  & 0.3738    &  0.0120   \\
(e)  $p_{S_2}\to p_{S_3}$ & 0.5000 &  $p_{S_2}=0.7071$ &  $p_{S_3}=0.5774$ & 0.1298  & 0.1298   & 0 \\
(f)  $p_{S_3}\to p_{N}$   & 0.0630 &  $p_{S_3}=0.7959$ &  $p_{N}=0.5071$ & 0.2888  & 0.2440   &  0.0448   \\
\label{table3}
\end{tabular}
\end{ruledtabular}
\end{table*}
\subsection{Increased robustness against white noise of
Bell-nondiagonal generalized Werner states}
\label{ch:CounterintuitiveResults2}

Even a quick analysis of Figs.~\ref{fig6}(b) and~\ref{fig7}, and
Table~\ref{table3} shows one of the main theoretical results of
this paper, i.e., increased robustness against the white noise of
Bell-nondiagonal GWSs compared to the standard (Bell-diagonal)
Werner states. Below we give a more intuitive and detailed
explanation of this result.

We recall that Bell diagonal (nondiagonal) GWSs are the maximally
(partially) entangled states affected by white noise. Let us
analyze the amount of white noise (i.e., $1-p$), which is
necessary to make the transition of a GWS from one threshold
value, say $p_i(q)$, to another (final) value, $p_f(q)$, for a
given value of the superposition parameter $q$. Thus, the required
white noise can be quantified by
\begin{equation}
  \Delta_{if}(q)\equiv p_i(q)-p_f(q),
  \label{Delta}
\end{equation}
for $i\neq f\in\{N,S_3,S_2,B\}$, which is plotted in
Fig.~\ref{fig7} and numerically given in Table~\ref{table3}.

For example, let us consider the maximally entangled Werner state
admitting an LHV model, i.e. $\rho_{\rm W}(p_B)$. Our question is
about the minimum amount of the white noise, which should be added
to this state to make it separable, i.e., $\rho_{\rm W}(p_N)$. The
answer is $\Delta_{BN}(q=1/2)=1/\sqrt{2}-1/3\approx 0.3738$. we
find that, in the case of the GWSs, the minimum amount of the
white noise to convert the maximally entangled GWS $\rho_{\rm
GW}[p_B(q),q]$, admitting an LHV model, to the closest separable
state $\rho_{\rm GW}[p_N(q),q]$ can be larger than that for the
Werner states, $\Delta_{BN}(q)> \Delta_{BN}(1/2)$, for some values
of the superposition parameter $q$ corresponding to the red
regions in Fig.~\ref{fig7}(a). Assuming that $q=0.9$ (as set in
our experiments), we obtain $\Delta_{BN}(0.9)=0.4029> 0.3738$.
Actually, the largest value
$\max_q\Delta_{BN}(q)=\Delta_{BN}(q')=0.4037$ can be achieved for
$q'=0.8829$ and $1-q'$, which can be calculated by solving the
following sixth-order equation $(1+4x^2)^3=x^2(1+4x)^4$ with
$x=\sqrt{q'(1-q')}$.

The same conclusion, about higher robustness of the
Bell-nondiagonal GWSs against white noise compared to that of the
Werner states, can also be drawn for other transitions indicated
by the arrows in Figs.~\ref{fig6}(b) and~\ref{fig7}, and also
listed in Table~\ref{table3}. The only exception is observed for
the transition corresponding to $\Delta_{S_2,S_3}(q)$, which
reaches the largest value for the Werner states,  as shown in
Fig.~\ref{fig7}(e).

More white noise should be added to a Bell state to reach any
threshold value $p_j$ compared to that for any partially entangled
state, because $1-p_j(1/2)> 1-p_j(q)$ for $q\neq 1/2$ and
$j\in\{N,S_3,S_2,B\}$, i.e.,  the amount of the white noise
destroying completely any quantum properties of the states,
including entanglement, steering, and nonlocality. So, in that
sense, the Werner states are more robust against white noise than
the nondiagonal GWSs. However, by choosing proper reference states
or proper transitions, one can draw the opposite conclusion, as we
have demonstrated in this section and it is clearly visualized in
Figs.~\ref{fig6}(b) and~\ref{fig7}.

 \begin{figure}
\begin{center}
\subfloat[extended CC hierarchy]{\includegraphics[width=0.49\columnwidth]{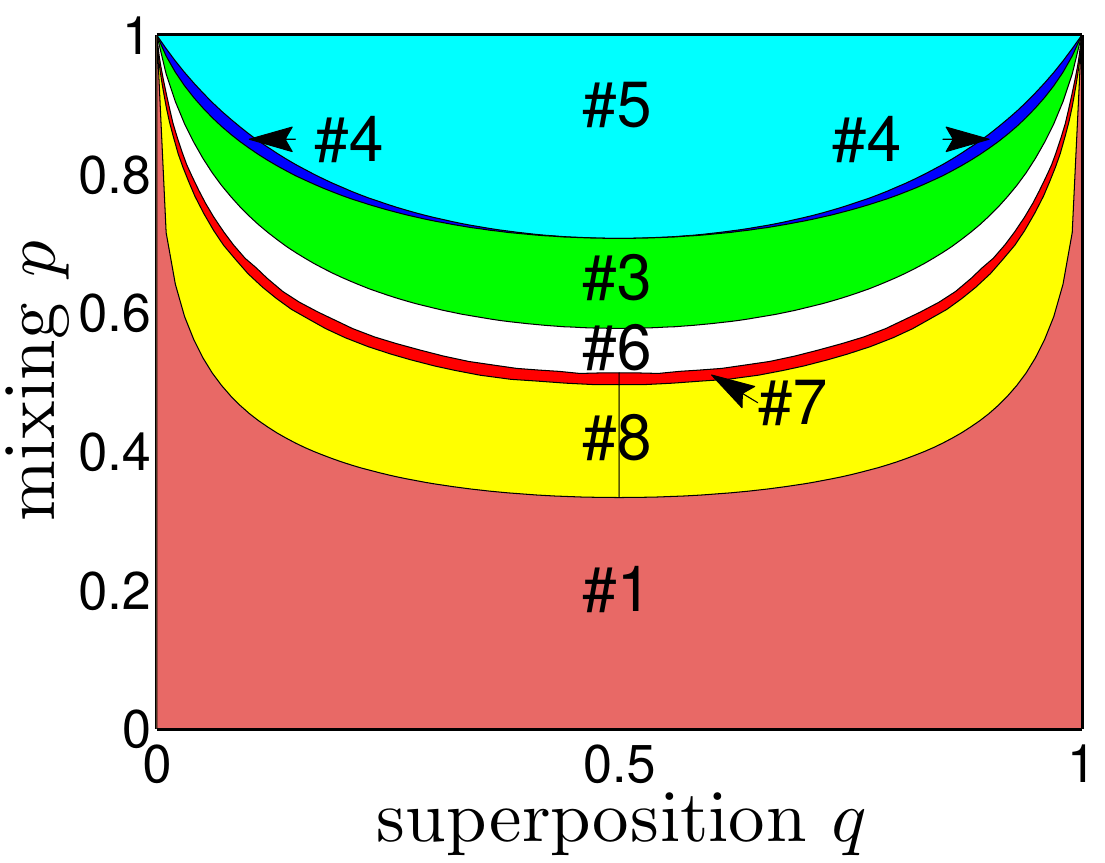}}\hspace*{0pt} \\ %
\subfloat[supremum of unsteerable entangled states: reg. \#7,8]{\includegraphics[width=0.49\columnwidth]{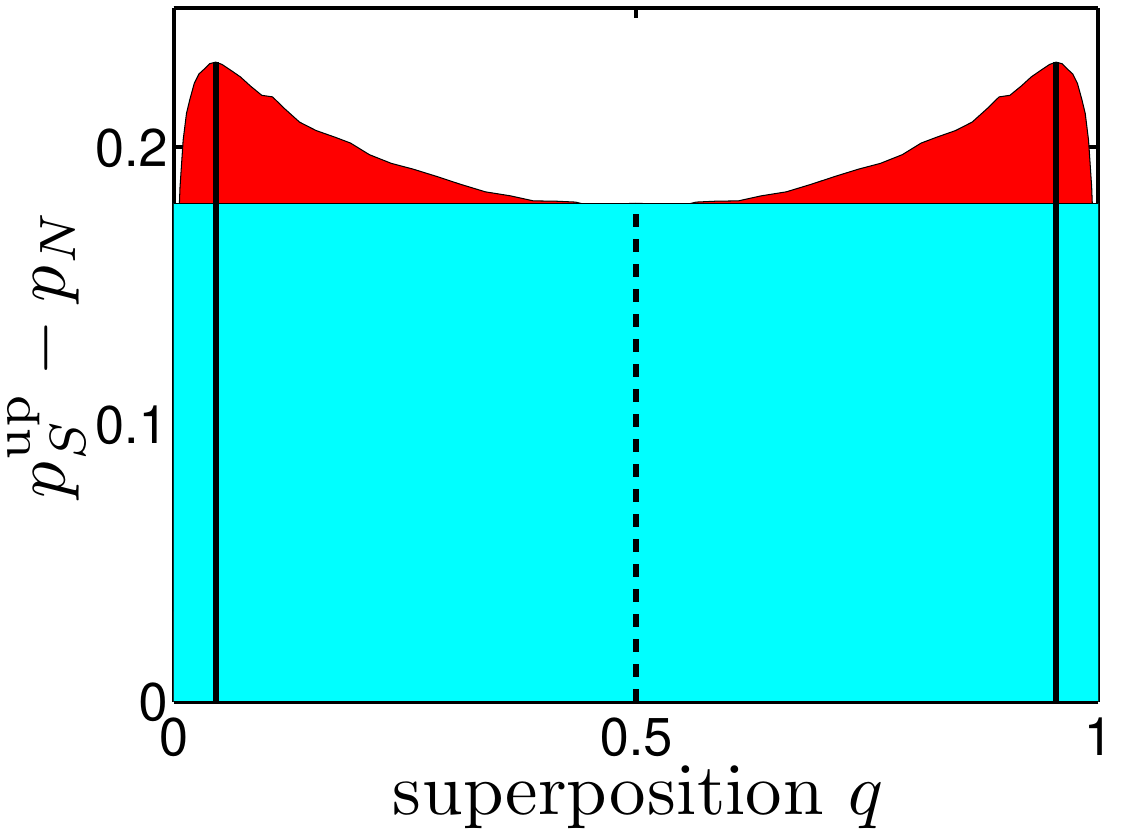}}\hspace*{0pt} %
\subfloat[unsteerable entangled states: regime \#8] {\includegraphics[width=0.49\columnwidth]{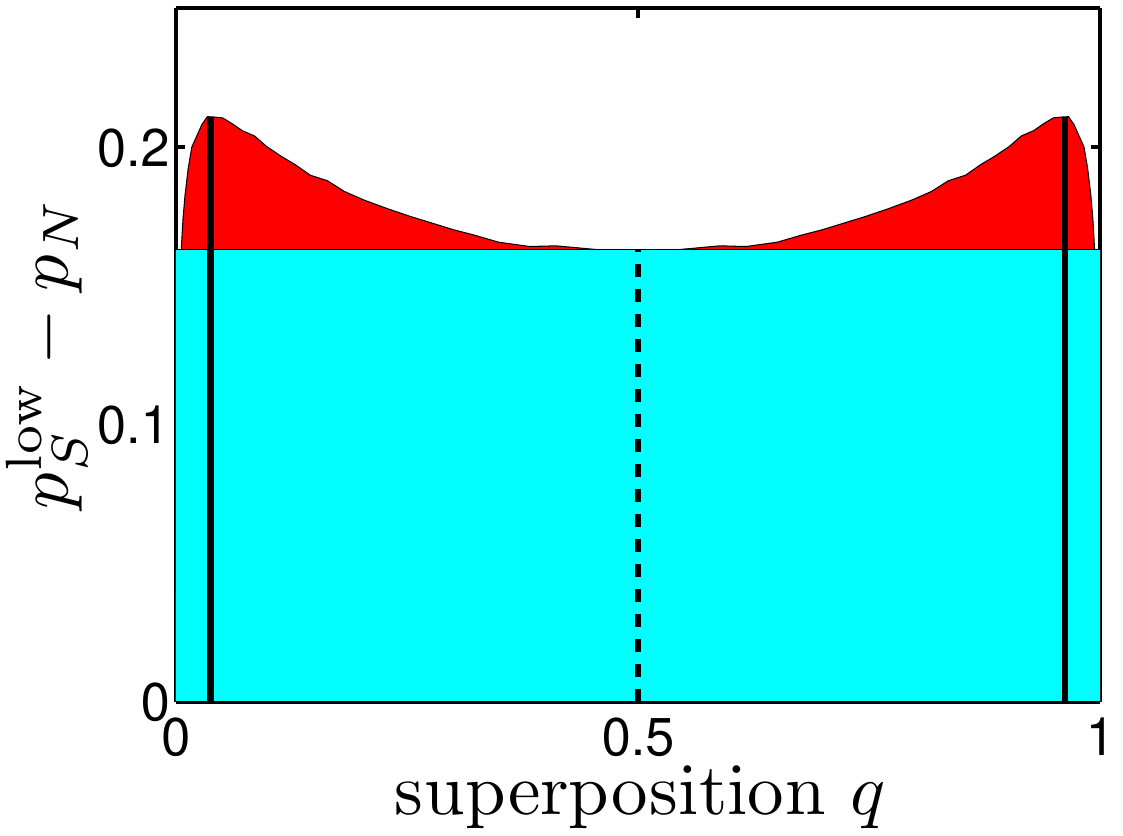}}\hspace*{0pt} %
\caption{(a) Same as in Fig.~\ref{fig6}(a) but with additional
regions (regimes) \#6, \#7, and \#8 of steerability in the limit
of a large number of measurements. Differences (b) $p_S^{\rm
up}-p_N$ and (c) $p_S^{\rm low}-p_N$, where $p_N$ is given by
Eq.~(\ref{pN}). The curve $p_S^{\rm up}$ is the border between the
regimes \#6 and \#7, which corresponds to a sufficient condition
for steerability of Ref.~\cite{CavalcantiReview}, while the curve
$p_S^{\rm low}$ is the border between regimes \#7 and \#8, which
corresponds to a sufficient condition of unsteerability based on
the algorithm and numerical data of
Refs.~\cite{Hirsch2016,Fillettaz2018} assuming 136 projective
measurements. Panels (b,c) show, analogously to those in
Fig.~\ref{fig7}, that the optimal robustness of steering assuming
a large number of measurements compared to the entanglement of the
GWSs is observed for the Bell nondiagonal GWSs with the
superposition parameter $q\neq 1/2$ (as denoted by blue solid
lines).}
 \label{fig8}
 \end{center}
 \end{figure}

\subsection{Increased robustness of {steering for a larger number of measurements}}
\label{ch:SteeringN}

Our paper is focused on analyzing steering in only the two- and
three-measurement scenarios. Nevertheless, in
Appendix~\ref{App:SteeringN}, we also discuss steering in
multi-measurement scenarios including the case of steering in the
limit of an infinite number of types of available measurements.

Specifically, we analyze lower and upper bounds on {steering for a
much larger number $n$ of measurements (even $n=136$)}. We
demonstrate that entangled GWSs, which are unsteerable for a very
large (or, in principle, infinite) number of measurements, can be
more robust against white noise if they are nondiagonal in the
Bell-state basis compared to the diagonal ones (i.e., the Werner
states).

First, we recall that, while the analyzed entanglement measures
reveal the property of a given state independent of its
measurements, the measures for steerability and Bell nonlocality
additionally depend on the available measurements. Thus, one can
raise the following question: (i) Is a larger spread
(corresponding to higher robustness) between different classes of
correlations in the GWSs an artifact stemming from the fact that
the considered steering and Bell-nonlocality measures perform
better on less entangled states? This question can also be
rephrased: (ii) Can one expect to find the same robustness
behavior for some tight bounds for Bell nonlocal states and
steerable states taking into account any measurement scenario?

We answer these questions by calculating tight upper ($p_S^{\rm
up}$) and lower ($p_S^{\rm low}$) {bounds on steering for the GWSs
for a large number of measurements. These numerical bounds
strongly suggest that the hierarchy also holds for an arbitrary
number of measurements. Indeed,} similar analysis can be performed
for Bell nonlocality of the GWSs, as discussed
in~\cite{Hirsch2016}, to show that the Horodecki measure fully
describes the nonlocality in two-qubit states with no restriction
on the number of measurements.

Two bounds on multi-measurement steering are shown in
Fig.~\ref{fig8}. Specifically, the upper bound $p_S^{\rm up}$,
which corresponds to the border curve between regimes \#6 and \#7
in Fig.~\ref{fig8}(a), is a sufficient condition for the
steerability of the GWSs. This bound was obtained numerically in
Refs.~\cite{Hirsch2016,Fillettaz2018} from a criterion of
Ref.~\cite{CavalcantiReview} using an SDP technique for 13
measurements on the Bloch sphere. Furthermore, the lower bound
$p_S^{\rm low}$, which is shown by the curve between regimes \#7
and \#8, corresponds to a sufficient condition of the
unsteerability of the GWSs based on the algorithm of
Refs.~\cite{Hirsch2016,Fillettaz2018} for constructing LHS models
assuming 136 projective optimal (or  almost optimal) measurements
corresponding to the fourth level of their algorithm. The curves
for both $p_S^{\rm low}$ and $p_S^{\rm up}$ are plotted using the
numerical data of Ref.~\cite{Fillettaz2018}. Thus, any GWS above
the $p_S^{\rm up}$ curve in Fig.~\ref{fig8}(a) is steerable, while
any state below the $p_S^{\rm low}$ curve is unsteerable. The
unsteerability of some of the states in regime \#7 (lying close to
the border curve $p_S^{\rm low}$) can be tested by applying the
algorithm of Refs.~\cite{Hirsch2016,Fillettaz2018} for higher
levels, which corresponds to analyzing a larger number of
measurements ($n\gg 136$). However, it is unclear whether any GWSs
lying inside regime \#7 can be steerable in the limit of
$n\rightarrow\infty$.

Figure~\ref{fig8}(a) shows that by including the criteria for
{steering in multi-measurement scenarios,} in addition to $S_2$
and $S_3$, one can study a CC hierarchy, which is more refined
than that in Fig.~\ref{fig6}(a). Note that regime \#2 in
Fig.~\ref{fig6}(a) corresponds to the sum of regimes \#6, \#7, and
\#8 shown in Fig.~\ref{fig8}(a).

To answer the questions raised above, we plotted the differences
$p_S^{\rm up}-p_N$ and $p_S^{\rm low}-p_N$ in Figs.~\ref{fig8}(b)
and \ref{fig8}(c), respectively. Both figures are quite similar
and show that the optimal robustness against noise is observed for
the Bell \emph{nondiagonal} GWSs with the superposition parameter
$q\neq 1/2$ (denoted by black solid lines). Thus, even without
knowing the exact threshold values between the steerability and
unsteerability of the GWSs in the limit of an infinite number of
measurements, one can conclude that the predicted optimal
robustness is \emph{not} an artifact at least for the cases shown
in Figs.~\ref{fig8}(b) and \ref{fig8}(c). This is the answer to
question (i). Concerning question (ii), the robustness behavior is
different for different $p_i=p_S^{\rm up},p_S^{\rm
low},p_{S_2},p_{S_3}$. Indeed, the optimal values of the
superposition parameter $q$ maximizing $p_{i}-p_N$ depend on $i$.
However, this property does not weaken our conclusion about higher
robustness against white noise of some Bell nondiagonal GWSs
compared to that of the Werner states.

\section{Conclusions} \label{ch:Conclusions}

The main purpose of this work was to analyze a CC hierarchy of
theoretical and experimental Werner states and their
generalization, i.e., the Bell-nondiagonal GWSs. We recall that
the considered GWSs are the mixtures of partially entangled
two-qubit pure states (not only of a Bell state) and the maximally
mixed state (white noise). We have shown that the Bell-nondiagonal
GWSs exhibit a more refined CC hierarchy compared to that of the
Bell-diagonal GWSs, i.e., the Werner states.

By tuning the mixing and superposition parameters of the GWSs, we
have experimentally generated and tomographically reconstructed
such GWSs, which reveal the hierarchy of the following classes of
correlations: \#1 separability, \#2 entanglement without
steerability in 3MS, \#3 steerability in the 3MS but not steerable
in the 2MS, \#4 steerability in the 2MS without violating the
Bell-CHSH inequality (so admitting LHV models), and \#5 Bell
nonlocality, which cannot be explained within LHV models. Note
that the case of steering is a little more subtle since the
measures assume specific measurements. Thus, we have also analyzed
a sufficient condition for unsteerability assuming a very large
number (i.e., 136) of measurements.

In particular, we found five different parameter regimes of the
GWSs, including the states, which are steerable in a 2MS without
violating Bell inequalities, and, thus, corresponding to the
regime \#4. This is a counterintuitive result, especially when
compared with the Girdhar-Cavalcanti theorem~\cite{Girdhar2016},
which states that: ``All two-qubit states that are steerable via
CHSH-type correlations are Bell nonlocal''~\cite{Girdhar2016}. In
Sec.~\ref{ch:CounterintuitiveResults1}, we have explained why the
observation of regime \#4 in our steering scenarios does not imply
the violation of the Girdhar-Cavalcanti theorem. We also
demonstrated that regime \#4 cannot be observed for the usual
Werner states.

Moreover, we have shown that the robustness against white noise
for, e.g., steerable states admitting LHV models can be stronger
for some Bell-nondiagonal GWSs than that for the diagonal GWSs
(i.e., the Werner states). This can be achieved by properly
choosing the value of the superposition coefficient $q$, as shown
in Figs.~\ref{fig6}(b) and~\ref{fig7}. Thus, we addressed the
problem of optimal robustness of states against white noise.
Specifically, we analyzed threshold values (curves) separating the
five regimes of quantum correlations. Then we could find optimal
transitions between various curves corresponding to the largest
amount of white noise or, in other words, to the largest spread in
the hierarchy. Thus, we discovered the optimal Bell-nondiagonal
GWSs, which are more robust against white noise than the Werner
states.

Furthermore, we considered lower and upper bounds on {steering in
multi-measurement scenarios}. Again we demonstrated better
robustness against white noise of some unsteerable entangled
Bell-nondiagonal GWSs compared to the diagonal ones. Thus, such
enhanced robustness is not limited only to the two- and
three-measurement steering scenarios; it can also be observed for
steering in the limit of a large number of measurements.

Possible applications of the discovered optimal robustness against
white noise can be found for quantum cryptography. For instance,
imagine that legitimate users of some secure quantum
communications system want to use steering (or entanglement) such
that it should not be detected by the violations of Bell
inequalities by others. Thus, assuming that the communication is
via a depolarizing channel, it is convenient to use partially
steerable (or partially entangled) states, which are Bell local
and are the most robust against white noise. Such optimal states
are indicated by arrows in Fig.~\ref{fig6}(b).

Our study of the hierarchy of the classes of spatial quantum
correlations can be generalized to analyze a hierarchy of their
temporal or spatio-temporal analogs. Indeed, the concepts of
spatial and temporal quantum correlations are closely related.
Formally, it is enough to replace two-qubit measurements for
testing spatial correlations by measurements on a single qubit,
followed by transmission through a channel, to reveal temporal
correlations, as explained in the example of spatial and temporal
steering in Ref.~\cite{Chen2016}. Thus, many of the results
discussed here for spatial correlations can also be generalized to
temporal correlations. We explicitly indicated such relations in
various sections of this paper. Analyses of CC hierarchies of
temporal correlations can lead to a deeper understanding of, e.g.,
quantum causality~\cite{Brukner2014} or enable designing new types
of quantum cryptosystems and finding new methods of breaking the
standard ones.

We believe that analyzing such CC hierarchies is interesting
concerning both fundamental aspects of quantum mechanics and
possible cryptographic applications for, e.g., secure
communication, secure information retrieval, and zero-knowledge
proofs of (quantum) identity.

\section*{Acknowledgements}

We acknowledge helpful discussions with Huan-Yu Ku. K.J., K.B.,
and K.L. acknowledge financial support from the Czech Science
Foundation through Project No. 20-17765S. The authors also
acknowledge project CZ.02.1.01/0.0/0.0/16\_019/0000754 of the
Ministry of Education, Youth, and Sports of the Czech Republic.
K.J. also acknowledges the Palack\'y University internal grant No.
IGA-PrF-2021-004. A.M. was supported by the Polish National
Science Centre (NCN) under the Maestro Grant No.
DEC-2019/34/A/ST2/00081.

\appendix

\section{Universal detection of quantum correlations
without full quantum state tomography}
\label{App:IncompleteTomography}

{In this work, we determined quantum correlations from
experimentally generated and reconstructed states using a full
QST. Here we address the question of universal detection of
quantum correlations \emph{without} full QST.}

(a) \emph{Universal detection of an entanglement measure without
QST.} The first experimental universal detection of standard
two-qubit entanglement without full QST was proposed in
Ref.~\cite{Bartkiewicz2015} (see also~\cite{Bartkiewicz2015b})
based on the universal witness of Ref.~\cite{Augusiak2008}. This
method was later improved in Ref.~\cite{Bartkiewicz2017b} to show
theoretically a direct experimental method for determining the
negativity of a general two-qubit state based on 11 measurements
performed on multiple copies of the state using Hong-Ou-Mandel
interference. None of these methods of universal entanglement
detection without a full state tomography was demonstrated
experimentally yet because of the complexity of such setups and
low probability of required multiple coincidences. Note that an
experimental detection, without a complete tomography, of the
fully entangled fraction of Bennett \emph{et
al.}~\cite{Bennett1996} was demonstrated by us in
Ref.~\cite{Bartkiewicz2017}. Unfortunately, the fully entangled
fraction is \emph{not} a universal entanglement witness in
general, so it usually only gives  a sufficient (but not
necessary) condition of entanglement.

(b) \emph{Universal detection of a steering measure without QST.}
Such methods have not been implemented or even proposed for the
steering robustness and the steerable weight. The calculations of
these popular steering measures for general states are based on
numerical optimization (using semidefinite programs). Thus, in
general, these measures could only be determined experimentally
for tomographically reconstructed states or processes, as it was
done in dozens of experimental works (see the reviews
in~\cite{CavalcantiReview,UolaReview} and references therein). Of
course, there are many experiments demonstrating quantum steering
via nonuniversal witnesses (to reveal a hierarchy of criteria),
i.e., by observing the violations of steering
inequalities~\cite{CavalcantiReview, UolaReview}. We note that
measures of steering (e.g., that proposed for a 2MS and a 3MS in
Ref.~\cite{Costa2016}), which are based on the maximal violation
of well-established steering inequalities can be measured without
a complete QST. For example, the optimal violation of the
Cavalcanti-Jones-Wiseman-Reid inequality~\cite{Cavalcanti2009}
can, in principle, be experimentally demonstrated with polarized
photons without scanning all the angles of polarizers. This can be
done, as we anticipate, in systems similar to those demonstrating
the Horodecki measure of Bell nonlocality~\cite{Bartkiewicz2017}.

(c) \emph{Universal detection of a nonlocality measure without
QST.} The Horodecki measure~\cite{Horodecki1995, Horodecki1996})
of Bell-CHSH nonlocality of two-qubit states can indeed be
measured without a full QST, but, to our knowledge, it was first
determined experimentally only recently in our
experiment~\cite{Bartkiewicz2017} without scanning the angles of
the polarizers to obtain an optimal value of the angles maximizing
the violation of the Bell-CHSH inequality for an unknown two-qubit
state. To demonstrate the power of this method, we have
implemented an entanglement-swapping device. To our knowledge, no
other experimental universal detections of a nonlocality measure
(without scanning the polarization angles or without \emph{a
priori} information about a given generated state) have been
reported yet.

\section{Steerable weight in a three-measurement scenario}
\label{App:Steering3}

Here we consider two-qubit EPR steering in a 3MS, when Alice
performs the measurements of the three Pauli operators:
  $X = \ket{+}\bra{+}-\ket{-}\bra{-}$,
  $Y = \ket{R}\bra{R}-\ket{L}\bra{L}$,
  $Z = \ket{0}\bra{0}-\ket{1}\bra{1}$,
of qubits encoded in the polarization states of photons, as in our
experiment. Thus, these measurements are just the projections onto
the Pauli-operator eigenstates
$\ket{\pm}=(\ket{0}\pm\ket{1})/\sqrt{2}$,
$\ket{R}=(\ket{0}+i\ket{1})/\sqrt{2}$,
$\ket{L}=(\ket{0}-i\ket{1})/\sqrt{2}$, $\ket{0}$, and $\ket{1}$,
which correspond to the diagonal, antidiagonal, right-circular,
left-circular, horizontal, and vertical polarization states,
respectively. These measurements of Alice generate unnormalized
states $\sigma_{a|x}$ of Bob for $x=X,Y,Z$ assuming measured
eigenvalues $a=\pm 1$. By defining $f(\ket{m})
=\tr_A[(\ket{m}\bra{m}\otimes I)\rho]$, the six possible
unnormalized Bob states $\sigma_{a|x}$ read as:
\begin{eqnarray}
  \sigma_{+1|X} &=&f(\ket{+}), \quad   \sigma_{-1|X} =f(\ket{-}),
\nonumber \\
  \sigma_{+1|Y} &=&f(\ket{R}), \quad   \sigma_{-1|Y} =f(\ket{L}),
\nonumber \\
  \sigma_{+1|Z} &=&f(\ket{0}), \quad   \sigma_{-1|Z} =f(\ket{1}).
\label{sigma_ax}
\end{eqnarray}
Alice, after performing her measurements, holds a classical random
variable $\lambda\equiv[x,y,z]=
[\bra{x}X\ket{x},\bra{y}Y\ket{y},\bra{z}Z\ket{z}]$, with hereafter
$x,y,z=\pm 1$. Thus, the variable $\lambda$ can take the values
$\lambda_1=[-1,-1,-1]$, $\lambda_{2}=[-1,-1,1]$, ..., and
$\lambda_{8}=[1,1,1]$. The unsteerable assemblage
$\sigma_{a|x}^{{\rm US}}$, can now be expressed as: $\sigma_{\pm
1|X}^{{\rm US}} = \sum_{y,z} \sigma_{\pm 1,y,z}$, $\sigma_{\pm
1|Y}^{{\rm US}} = \sum_{x,z} \sigma_{x,\pm 1,z}$, and $\sigma_{\pm
1|Z}^{{\rm US}} = \sum_{x,y} \sigma_{x,y,\pm 1}$, where
$\sigma_{\lambda}\equiv\sigma_{xyz}$ are the states held by Bob.

The steerable weight $S_3$ in our 3MS can be given by the solution
of the SDP:
\begin{equation}
S_3= 1 -\max \tr \sum_{x,y} \sigma_{xyz},
  \label{S3}
\end{equation}
such that  $\sigma_{xyz}\ge 0$ and
\begin{eqnarray}
  \sigma_{\pm 1|X}&-& \sum_{y,z} \sigma_{\pm1,y,z}\;\ge0,
  \nonumber \\
  \sigma_{\pm 1|Y}&-& \sum_{x,z} \sigma_{x,\pm1,z}\;\ge0,
  \nonumber \\
   \sigma_{\pm 1|Z}&-& \sum_{x,y} \sigma_{x,y,\pm1}\;\ge0.
  \label{S3cond}
\end{eqnarray}

\section{Steerable weight in two-measurement scenarios}
\label{App:Steering2}

The above approach can be simplified when analyzing EPR steering
in 2MSs, i.e., when Alice is performing the measurements of only
two Pauli operators ($XY$, $XZ$, and $YZ$). Thus, one can consider
the following three measures:

(i) For the steerable weight $S_2^{XY}$ for the measurements of
$X$ and $Y$, the corresponding unsteerable assemblage
$\sigma_{a|x}^{{\rm US}}$ can be expressed as $\sigma_{\pm
1|X}^{{\rm US}} = \sum_{y} \sigma_{\pm 1,y}$ and $\sigma_{\pm
1|Y}^{{\rm US}} = \sum_{x} \sigma_{x,\pm 1}$, where
$\sigma_{\lambda}\equiv\sigma_{xy}$ are the states held by Bob.
Then the corresponding steerable weight $S_2^{XY}$ can be
calculated as the solution of the SDP:
\begin{equation}
S_2^{XY} = 1 -\max \tr \sum_{x,y} \sigma_{xy},
  \label{S_XY}
\end{equation}
under the constraints: $\sigma_{xy}\ge 0$ and
\begin{eqnarray}
  \sigma_{\pm 1|X} -  \sum_{y} \sigma_{\pm1,y}\ge0,
  \quad
  \sigma_{\pm 1|Y} -  \sum_{x} \sigma_{x,\pm1}\ge0.
  \label{S_XYcon}
\end{eqnarray}
(ii) The steerable weight $S_2^{XZ}$, based on Alice's
measurements of the Pauli operators $X$ and $Z$, is given by:
\begin{equation}
S_2^{XZ} = 1 -\max \tr \sum_{x,z} \sigma_{xz},
  \label{S_XZ}
\end{equation}
such that $\sigma_{xz}\ge 0$ and
\begin{eqnarray}
   \sigma_{\pm 1|X}-  \sum_{z} \sigma_{\pm1,z}\ge0,
  \quad
   \sigma_{\pm 1|Z}-  \sum_{x} \sigma_{x,\pm1}\ge0.
  \label{S_XZcond}
\end{eqnarray}
(iii) The steerable weight $S_2^{YZ}$ corresponding to measuring
the Pauli operators $Y$ and $Z$, can be calculated as
\begin{equation}
S_2^{YZ} = 1 -\max \tr \sum_{y,z} \sigma_{y,z},
  \label{S_YZ}
\end{equation}
under the conditions $\sigma_{yz}\ge 0$, and
\begin{eqnarray}
  \sigma_{\pm 1|Y}- \sum_{z} \sigma_{\pm1,z}\ge0,
  \quad
  \sigma_{\pm 1|Z}-  \sum_{y} \sigma_{y,\pm1}\,\ge0.
  \label{S_YZconc}
\end{eqnarray}

The optimized 2MS steerable weight ($S_2$) can be given as the
maximum value of the steerable weights for specific measurement
choices, i.e.,
\begin{equation}
  S_2=\max(S_2^{XY},S_2^{XZ},S_2^{YZ}).
  \label{S2}
\end{equation}
This definition of $S_2$ can directly be applied to symmetric
states, including the Werner states and GWSs. However, for
non-symmetric states (including some of our experimental density
matrices), the optimal projectors can be found numerically by
maximizing the steerable weight over unitary transformations for
any two Pauli operators. In our experiments and theoretical
analysis, we apply only single Pauli operators (rather than their
linear combinations) and then optimize them over their unitary
transformations. Thus, we obtain the steerable weights, which are
optimized over von Neumann's projection-valued measures, instead
of the most general case of POVMs. Note that the required
optimization over POVMs is more demanding both experimentally and
theoretically, and it is not applied in this work. We find that,
on the scale of Figs.~\ref{fig:ExpWS}(c) and~\ref{fig:ExpGWS}(c),
no differences can be seen for $S_2$ if it is calculated by the
optimized projectors and by applying directly Eq.~(\ref{S2}) for
any of the measured states.

Note that, in this approach to determine $S_2$, we are limiting
the number of the types of measurements on Alice's side, but a
full QST is always assumed on Bob's side corresponding to
measuring all the Pauli operators. Thus, the steerable weight
$S_3$ corresponds to a 3-3 measurement scenario, i.e.,  three
types of measurements on Alice and Bob sides (assuming that the
efficiency of detectors is known), while the steerable weight
$S_2^{ij}$ (for the specific choice of two Pauli operators)
corresponds to a 2-3 scenario, i.e., based on two types of
measurements on Alice's side and three -- on Bob's side.

All these steerable weights in the two- and three-measurement
scenarios can be efficiently numerically calculated as solutions
of the described semidefinite programs using standard numerical
packages for convex optimization. Our numerical programs are based
on the software for disciplined convex programming of
Ref.~\cite{CVX}. The steerable weights in our work are calculated
using experimental density matrices, which are reconstructed using
a full quantum tomography.

\section{Steerability in multi-measurement scenarios} \label{App:SteeringN}

A related question arises about the steerability using a larger
number $n$ of types of measurements on Alice's side, especially in
the limit of an infinite number of measurements. The algorithms of
Refs.~\cite{Cavalcanti2016, Hirsch2016, Fillettaz2018} for
constructing LHS models can be applied to arbitrary entangled
states and, thus, can be used for finding numerically a sufficient
condition of unsteerability (i.e., a lower bound on steerability)
based on a given number of projective measurements. Note that for
the GWSs, such a lower bound on steerability was determined up to
$n=136$ measurements in Ref.~\cite{Fillettaz2018}. For
convenience, we consider here a steering lower bound $p_S^{\rm
low}(n)$, which can be numerically determined by the protocols of
Refs.~\cite{Hirsch2016,Fillettaz2018} for a given number $n$ of
measurements. We also consider a steering upper bound $p_S^{\rm
up}(n)$, being a sufficient condition for steerability, based on
an SDP technique of Ref.~\cite{CavalcantiReview} (see
also~\cite{Fillettaz2018}) assuming specifically 13 measurements
on the Bloch sphere.

The algorithm of Refs.~\cite{Hirsch2016, Fillettaz2018} has
already been applied to the steerability of the Bell-diagonal
states (including the Werner states) and GWSs (there referred to
as partially entangled states with white noise). Sufficient
conditions of unsteerability, corresponding to $n=6$, 16, 46, and
136 types of measurements, were found for four levels of the
algorithm~\cite{Fillettaz2018}. These results can enable
calculating $p_S^{\rm low}(n)$. Note that each type of measurement
is characterized by a Bloch vector, and all such vectors form a
polyhedron on the Bloch sphere.

It is quite challenging to numerically calculate the lower bound
$p_S^{\rm low}(n)$ of {steerability in multi-measurement
scenarios,} even for the next layer of the protocol of Fillettaz
\emph{et al.}~\cite{Fillettaz2018} (corresponding to the number of
measurements greater than $136$) because of the problem, which is
closely related to the ``curse of dimensionality''. Indeed, the
number of deterministic strategies to be checked numerically grows
exponentially with the number of measurements. The results should
also be optimized for the orientation of the polyhedra, otherwise
the results differ significantly, as explicitly shown in
Ref.~\cite{Hirsch2016}.

The ranges of the allowed values of the mixing ($p$) and
superposition ($q$) parameters in $\rho_{\rm GW}(p,q)$, for which
the GWSs are steerable, increase with the number of measurements
$n$. Thus, finding numerically a solution to these steering
problems could, in principle, enable us to analyze a more refined
hierarchy of the classes of steerability as a function of the
number of measurements such that a given state is steerable using
a given number of measurements, but unsteerable using a smaller
number of measurements. However, an experimental demonstration of
such a refined hierarchy is quite challenging as explained below.

Clearly, a direct experimental demonstration that a given state is
indeed unsteerable based on 136 types of measurements is extremely
demanding using linear optics. However, even theoretical
demonstration of such a refined hierarchy of the classes of
multi-measurement steerability for tomographically reconstructed
experimental states is quite challenging. These problems include
the following:

\emph{First problem.} We recall that our experimental GWSs,
$\rho_{\rm GW}^E(p,q)$, have a high Bures fidelity $F$ compared to
the theoretical optimal GWSs, $\rho_{\rm GW}(p_{\rm opt},q_{\rm
opt})$, on the average are equal to 0.97. Nevertheless, $\rho_{\rm
GW}^E(p,q)$ and $\rho_{\rm GW}(p_{\rm opt},q_{\rm opt})$ can still
have very different steering properties, such that one of the
states is steerable and the other is unsteerable in the same
$n$-measurement scenario, especially for $n>3$.

Note that all the examples of multi-measurement steerability,
based on the protocols of Refs.~\cite{Cavalcanti2016, Hirsch2016,
Fillettaz2018}, were numerically tested only for highly-symmetric
states (including the Werner states and GWSs). Unfortunately, our
experimental states $\rho_{\rm GW}^E$ have usually a broken
symmetry compared to that of the theoretical GWSs, $\rho_{\rm
GW}$. So, the calculation of the steerability of $\rho_{\rm GW}^E$
in the 2MS and 3MS is sometimes much more time-consuming and less
precise. This is even the case for calculating the steerable
weight and steering robustness using standard packages in the 2MS.
For example, the calculations of these two steering measures for
$\rho_{\rm GW}$ take at most a few seconds on a standard PC, while
those for the generated $\rho_{\rm GW}^E$ require sometimes dozens
of minutes assuming the same precision in both cases. These
numerical problems grow very fast with the increasing number $n$
of measurements.

\emph{Second problem.} Our experimental tuning of the parameters
$p$ and $q$ for the GWSs is not fine enough, as explained in
greater detail in Sec.~\ref{ch:Experiment}. Note that the ranges
of the parameters $p$ and $q$ of the GWSs are very small such that
a given GWS is steerable with $n+1$ measurements and unsteerable
with $n$ measurements for $n>3$. Our experimental tuning of $p$
and $q$ was good enough to directly generate states in regime \#3
corresponding to $S_3>0$ and $S_2=0$. However, we were not able to
\emph{directly} generate experimentally GWSs belonging to
different regimes of steerability for a larger number $n$. Note
that even our experimental GWS in regime \#4, corresponding to
$S_2>0$ and $B=0$, was not generated directly. Indeed, we obtained
it in a hybrid way, i.e., by numerically mixing experimental
states belonging to other regimes, as explained in
Sec.~\ref{ch:CounterintuitiveResults1}.

\emph{Third problem.} It is numerically very challenging to check
whether a given $\rho_{\rm GW}^E$ is $n$-measurement steerable and
$(n-1)$-measurement unsteerable, which is crucial in experimental
demonstrating such a refined hierarchy of the steerability classes
for multi-measurement scenarios. Specifically, if we numerically
obtain ${S}_{n}(\rho_{\rm GW}^E)\sim 10^{-12}$, which is the
precision of our numerical calculation of the steering measures,
it is quite biased to decide whether this state $\rho_{\rm GW}^E$
is indeed steerable or not. With the increasing number $n$ of
measurements, the numerically estimated ${S}_{n}(\rho_{\rm GW}^E)$
become less and less precise. So, the question arises how to
correctly classify the steerability of a given experimental state
in the hierarchy of the classes of steerability in various
multi-measurement scenarios.

\emph{Fourth problem.} The border between the steerable and
unsteerable theoretical GWSs is not precisely determined in the
limit of an infinite number $n$ of measurements on Alice's side.
Indeed, the border corresponds to region \#7 in Fig.~\ref{fig8}(a)
spanned by the curves $p_S^{\rm low}$ and $p_S^{\rm up}$.
Estimating $p_S^{\rm low}$ for our experimental imperfect GWSs,
$\rho_{\rm GW}^E$, is even more demanding because $\rho_{\rm
GW}^E$ usually exhibits a broken symmetry compared to that of the
ideal GWSs $\rho_{\rm GW}$.

Thus, for these numerical and experimental reasons, we have
decided to analyze in detail the steerability of our experimental
states for the two simplest types of measurement scenarios only.
We believe that this is good enough to show the hierarchy of some
classes of correlations (including steerability in 2MS and 3MS)
for experimental states.

\section{Hierarchy of entanglement criteria}\label{App:MM}

\subsection{Hierarchy of the Shchukin-Vogel entanglement criteria}

Here we briefly recall the Shchukin-Vogel entanglement criteria
for the universal detection of distillable entanglement via the
matrices of moments of the annihilation and creation
operators~\cite{Shchukin2005}. This approach, in principle, does
not require a full QST, so it is an alternative to the approach
applied in our experiment using QST. We indicate some advantages
and drawbacks of this approach for detecting two-qubit
entanglement.

The Shchukin-Vogel criteria are based on the Hermitian matrices of
moments for a given two-mode state $\rho$, which are defined as
\begin{eqnarray}
  {\cal M}^{\rm org}_N=\left[
\begin{array}{cccc}
M_{11} & M_{12} & ... & M_{1N} \\
M_{21} & M_{22} & ... & M_{2N} \\
... & ... & ... & ... \\
M_{N1} & M_{N2} & ... & M_{NN}
\end{array}
\right],
\end{eqnarray}
where $ M_{ij}= \<({a}^{\dagger i_{2}}{a}^{i_{1}}{b}^{\dagger
i_{4}}{b}^{i_{3}}) ({a}^{\dagger j_{1}}{a}^{j_{2}}{b} ^{\dagger
j_{3}}{b}^{j_{4}})\>$ are  the moments of the annihilation ($a,b$)
and creation ($a^\dagger,b^\dagger$) operators of two modes of
arbitrary dimension. Here, $i,j$ label multi-indices, e.g.,
$(i_1,i_2,i_3,i_4)$. These moments can be detected experimentally
(at least for not too high powers) using, e.g., the setup based on
homodyne detection as described by Shchukin and
Vogel~\cite{Shchukin2005exp}. A partially transposed matrix of
moments can be obtained from ${\cal M}^{\rm org}_N$ as
\begin{eqnarray}
M_{ij}^\Gamma&=&\< ({a}^{\dagger i_{2}}{a} ^{i_{1}}{a}^{\dagger
j_{1}}{a}^{j_{2}})({b}^{\dagger i_{4}}{b }^{i_{3}}{b}^{\dagger
j_{3}}{b}^{j_{4}})\>^\Gamma \nonumber \\
&=&\< ({a}^{\dagger i_{2}}{a} ^{i_{1}}{a}^{\dagger
j_{1}}{a}^{j_{2}})({b}^{\dagger i_{4}}{b }^{i_{3}}{b}^{\dagger
j_{3}}{b}^{j_{4}})^\dagger\> \nonumber \\
&=&\< ({a}^{\dagger i_{2}}{a}^{i_{1}}{a}^{\dagger j_{1}}
{a}^{j_{2}})({b}^{\dagger j_{4}}{b}^{j_{3}}{b}^{\dagger
i_{3}}{b}^{i_{4}})\>,
\end{eqnarray}
where the superscript $\Gamma$ denotes partial transposition
applied here for the second mode. This relation between ${\cal
M}^{\rm org}_N$ and ${\cal M}^{\Gamma}_N$ is a key observation of
Ref.~\cite{Shchukin2005}.  Let ${\cal M}_{N,(r_1, r_2, \cdots,
r_n)}$ denote the $n \times n$ submatrix of  ${\cal M}_N$ having
${\cal M}_{r_i ,r_j}$ elements. The Shchukin-Vogel criteria are
based on the following Sylvester's theorem~\cite{Miranowicz2006}:
${\cal M}_N$ is \emph{positive semidefinite} if and only if all
its \emph{principal minors} are nonnegative, i.e., $\det {\cal
M}_{N,(r_1, r_2, \cdots,r_n)}\ge 0$. Thus, the Shchukin-Vogel
criteria correspond to the positive-partial-transposition-based
Peres-Horodecki criterion, but formulated in terms of the matrix
moments as~\cite{Shchukin2005,Miranowicz2006}:
\begin{eqnarray}
\rho {\rm ~is~PPT}&\Leftrightarrow& \forall N,\forall \{r_k\}:
\quad
\det {\cal M}_{N,(r_1, r_2, \cdots, r_n)}^\Gamma \ge 0,\nonumber \\
\rho {\rm ~is~NPT}&\Leftrightarrow& \exists N, \; \exists \{r_k\}:
\quad \det {\cal M}_{N,(r_1, r_2, \cdots, r_n)}^\Gamma <0,\nonumber \\
\end{eqnarray}
where $1\le r_1< r_2<\cdots < r_{n}\le N$, ${n}=1,2,\cdots,N$, and
PPT (NPT) stands for positive (nonpositive) states under partial
transposition. Many popular entanglement criteria can be derived
from the Shchukin-Vogel
criteria~\cite{Shchukin2005,Miranowicz2010}, including the
Hillery-Zubairy inequalities, which below are recalled and applied
to the GWSs.

 \begin{figure}[t]
\begin{center}
\subfloat[] {\includegraphics[width=.5\columnwidth]{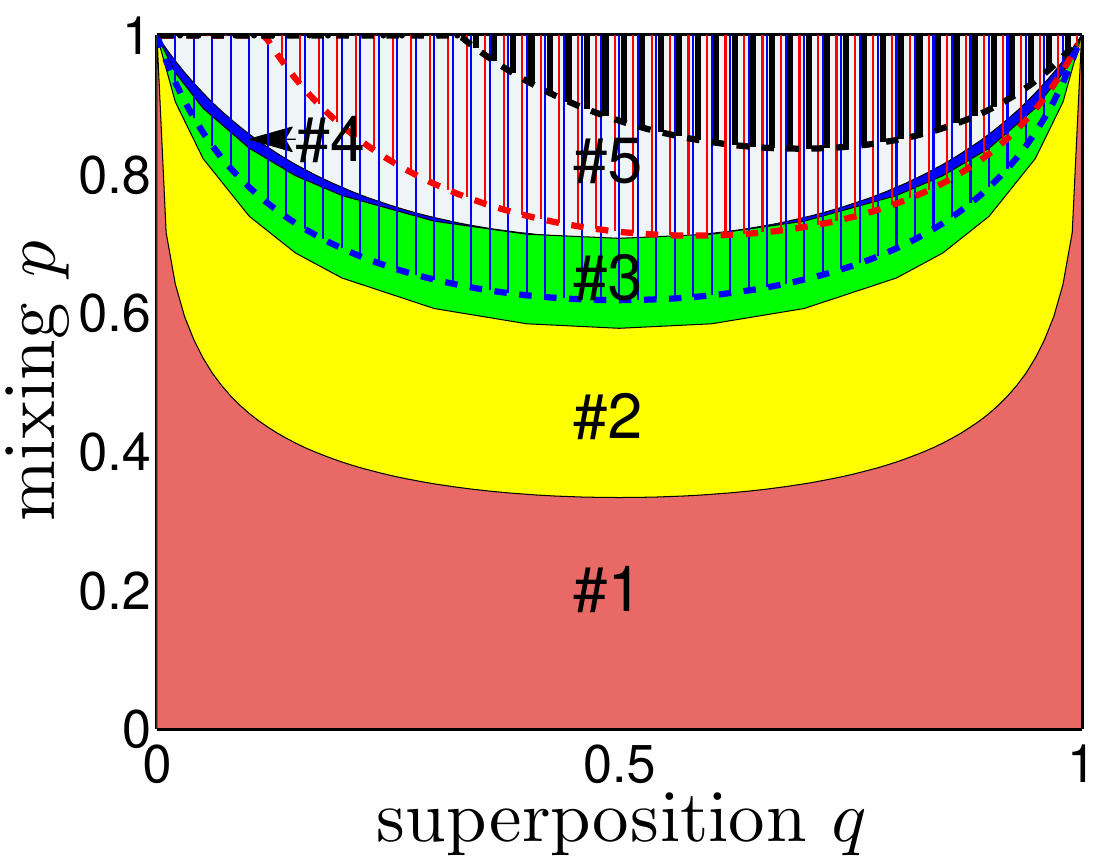}}
\subfloat[] {\includegraphics[width=.5\columnwidth]{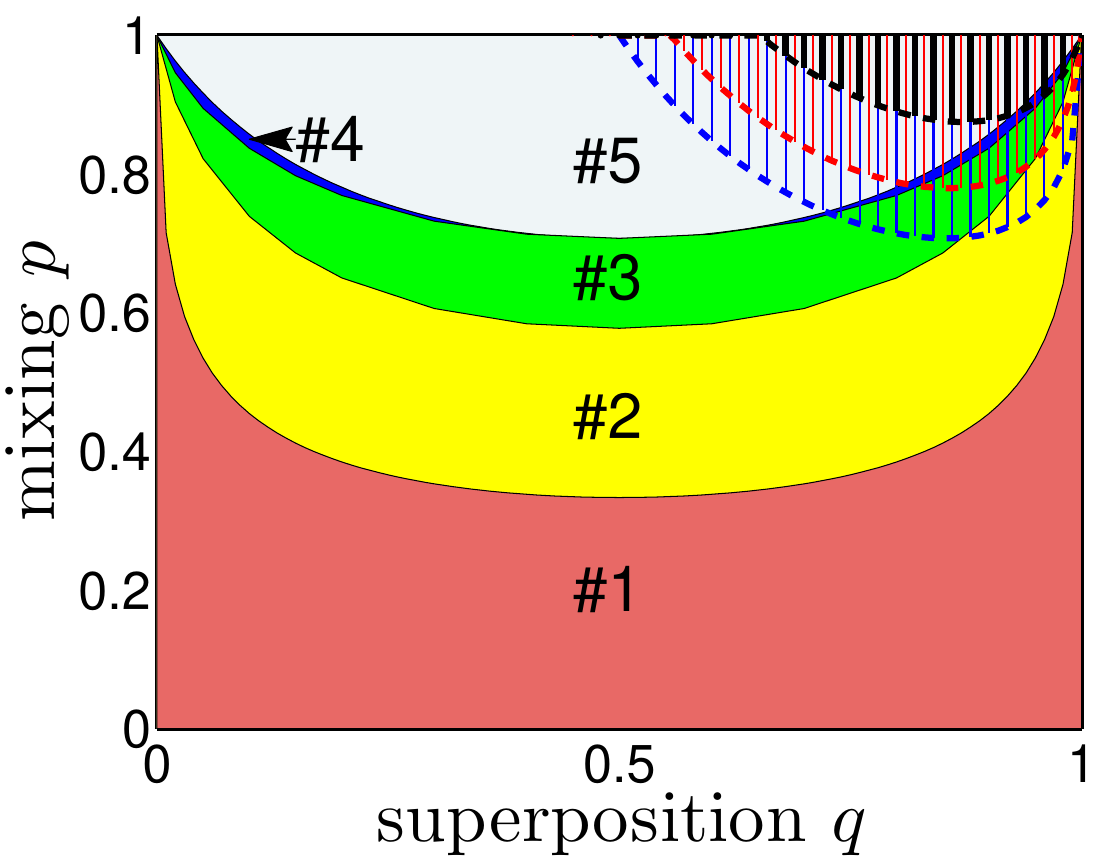}}
\caption{Hierarchy of criteria versus the CC hierarchy for the
GWSs. Specifically, the criterion hierarchy is based on different
nonuniversal witnesses for a given class of quantum correlation,
while the CC hierarchy reveals different types of correlations
determined by their measures or universal witnesses. This is shown
here by the example of nonuniversal entanglement witnesses using
the (a) first and (b) second HZ witnesses. The color regions
reveal the CC hierarchy, as in Fig.~\ref{fig6}(a), while the areas
filled with parallel lines show the criterion hierarchy. The
latter areas determine the allowed values of the mixing parameter
$p$ and the superposition parameter $q$ for the locally rotated
GWSs, $\rho_{\phi}(p,q),$ for which entanglement can be revealed
by the corresponding HZ witnesses: (a)~$\bar H_1(\rho_{\phi})$ for
$\phi=\pi$ (area filled with blue lines), $\phi=0.8\pi$ (red-line
area), and $\phi=0.7\pi$ (black-line area); and (b) $\bar
H_2(\rho_{\phi})$ for $\phi=0$ (blue-line-filled area), $\phi=0.2$
(red-line area), and $\phi=0.3$ (black-line-filled area). For
$\phi=\pi/2$ neither of the HZ witnesses can detect the
entanglement of the GWSs. The dashed curves are obtained from the
analytical formulas in Eqs.~(\ref{p_H2}), (\ref{p_H1}),
(\ref{p_H1phi}), and (\ref{p_H2phi}).}
 \label{fig:HZ}
 \end{center}
  \end{figure}
\subsection{Hierarchy of the Hillery-Zubairy entanglement criteria}

The Hillery-Zubairy (HZ) entanglement criteria for nonuniversal
detection of two-mode entanglement~read as~\cite{Hillery2006}:
\begin{eqnarray}
  H_1(\rho)&\equiv & \langle n_1n_2\rangle -|\langle ab^{\dag }\rangle |^{2}\
<0, \label{Hillery1} \\
H_1(\rho)&\equiv & \langle n_{1}\rangle \langle n_{2}\rangle
-|\langle ab\rangle |^{2}<0, \label{Hillery2}
\end{eqnarray}
where $n_1=a^\dagger a$ and $n_2=b^\dagger b.$ Thus, if
$H_1(\rho)<0$ or $H_2(\rho)<0$ then $\rho$ is entangled. The
criteria are simple and useful witnesses of entanglement, and have
already been experimentally tested in a number of setups (see,
e.g.,~\cite{Gomes2009}). These two criteria can be derived from
the Shchukin-Vogel criteria by calculating
\begin{eqnarray}
  H_n(\rho)=\det {\cal M}_{n}^{\Gamma}
\label{N1}
\end{eqnarray}
for
\begin{eqnarray}
{\cal M}_1^{\text{org}}=\left[
\begin{array}{cc}
1 & \langle ab\rangle  \\
\langle a^{\dag }b^{\dag }\rangle  & \langle n_1n_2\rangle
\end{array}
\right] ,\; {\cal M}_1^{\Gamma }=\left[
\begin{array}{cc}
1 & \langle ab^{\dag }\rangle  \\
\langle a^{\dag }b\rangle  & \langle n_1n_2\rangle
\end{array}
\right], \label{Hillery1a}
\end{eqnarray}
and
\begin{eqnarray}
{\cal M}_2^{\text{org}}=\left[
\begin{array}{cc}
\langle n_{1}\rangle  & \langle a^{\dag }b\rangle  \\
\langle ab^{\dag }\rangle  & \langle n_{2}\rangle
\end{array}
\right] ,\; {\cal M}_2^{\Gamma }=\left[
\begin{array}{cc}
\langle n_{1}\rangle  & \langle a^{\dag }b^{\dag }\rangle  \\
\langle ab\rangle  & \langle n_{2}\rangle
\end{array}
\right],
 \label{Hillery2a}
\end{eqnarray}
respectively. To analyze the HZ criteria on the same footing as
the discussed measures of quantum correlations, one can redefine
$H_n$ to be the HZ witnesses,
\begin{eqnarray}
  \bar H_n &=& \max \left\{0,-H_n\right\}.
\label{barH}
\end{eqnarray}

Let us now analyze in detail the hierarchy and effectiveness of
these criteria in detecting the entanglement of the GWSs compared
to the true measures of entanglement and other correlations.

We find the HZ witnesses for the original GWSs:
\begin{eqnarray}
  \bar H_1(\rho_{\rm GW}) &=& \max \left\{0,-\tfrac14 [1+p(3-4q)] \right\}=0, \label{H1} \\
  \bar H_2(\rho_{\rm GW}) &=& \max \left\{0,p^2q\bar q-\tfrac14 (1+p-2pq)^2 \right\}, \label{H2}
\end{eqnarray}
where $\bar q=1-q$. It can be seen that $\bar H_1(\rho_{\rm GW})$
is useless in detecting the entanglement of the GWSs, however
$\bar H_2(\rho_{\rm GW})$ can be nonzero. Thus, it detects
entanglement for the GWSs corresponding to the blue-line-filled
area in Fig.~\ref{fig:HZ}(a). The threshold (border) curve, as a
function of the superposition parameter $q$ in $\rho_{\rm
GW}(p,q)$, corresponds to the smallest allowed values of the
mixing parameter $p,$ for which the entanglement of the GWSs can
be detected. This threshold is shown by the blue dashed curve in
this figure, and is given by
\begin{equation}
  p_{H_2}(q)\;=\; 1/[2(q+\sqrt{q\bar q})-1],
  \label{p_H2}
\end{equation}
for $q\in[1/2,1]$. Let us now apply the Pauli operator $\sigma_1$
(the NOT gate) to the second qubit in the GWS, which results in
the state $\rho_{X}=(I\otimes \sigma_1)\rho_{\rm GW} (I\otimes
\sigma_1)$. Note that any local unitary operation does not change
entanglement measures, but, of course, it can change entanglement
witnesses, which is the case for the HZ criteria. Indeed, this
local transformation results in the HZ witnesses:
\begin{eqnarray}
  \bar H_1(\rho_{X}) &=& \max \left\{0,p^2 q\bar q-\tfrac14 \bar p \right\}, \label{H1x} \\
  \bar H_2(\rho_{X}) &=& \max \left\{0,-\tfrac14 (1-p^2)-p^2q\bar q \right\}=0, \label{H2x}
\end{eqnarray}
where $\bar p=1-p$. It can be seen that the sensitivities of the
HZ witnesses are exchanged for $\rho_{X}$ compared to $\rho_{\rm
GW}$. The second criterion cannot detect entanglement, while the
first reveals entanglement of some GWSs corresponding to those
shown in the blue-line-filled area in Fig.~\ref{fig:HZ}(b).
Analogously to Eq.~(\ref{p_H2}), the threshold curve for the first
HZ witness for $\rho_{X}(p,q)$ is given by
\begin{equation}
  p_{H_1X}(q)\;=\; 2/[1+\sqrt{1+16 q\bar q}],
  \label{p_H1}
\end{equation}
for $q\in[0,1]$. Now let us apply an arbitrary rotation along the
$y$-axis of the second qubit in the GWSs. Thus, we transform
$\rho_{\rm GW}$ into $\rho_{\phi}=[I\otimes R_{Y}(\phi)]\rho_{\rm
GW} [I\otimes R^\dagger_Y(\phi)]$, where the rotation is described
by $R_{Y}(\phi)=[c,-s;s,c]$ with  $c=\cos(\phi/2)$ and
$s=\sin(\phi/2)$. The HZ witnesses for the locally rotated GWSs
read:
\begin{eqnarray}
  \bar H_1(\rho_{\phi}) &=& \max \left\{0,-
   \tfrac14 \big[c^2[1 + p(3 - 4q)] \right. \nonumber\\
    && \hspace{1.5cm}\left. + s^2(\bar p - 4s^2p^2q\bar q)\big]
   \right\}, \label{H1y} \\
  \bar H_2(\rho_{\phi}) &=& \max \left\{0,c^4p^2q\bar q
   -\tfrac14 f_+(c^2 f_++s^2f_-)\right\},\quad \label{H2y}
\end{eqnarray}
where $f_{\pm}=1\pm p (1-2q)$. The threshold curves for the HZ
witnesses applied to $\rho_{\phi}(p,q)$ are given by
\begin{eqnarray}
  p_{H_1}(q,\phi) &=& \Big(f_1+\sqrt{f_1^2+2f_2}\Big)f^{-1}_2,  \label{p_H1phi} \\
  p_{H_2}(q,\phi) &=& 2/[\sqrt{f} + 2(1 + C_1)q - C_1 -1] ,\label{p_H2phi}
\end{eqnarray}
which are physically meaningful only in the regions of $q$ for a
given $\phi$, such that $p_{H_n}(q,\phi)\in[0,1]$ ($n=1,2$). Here
$f=(1 - C_1)^2(1 - 2q)^2 + 2(4C_1 + C_2 + 3)q\bar q$ with
$C_n=\cos(n\phi)$, $f_1=c^2(3-4q)-s^2$ and $f_2=8q\bar q s^4$. As
seen in Fig.~\ref{fig:HZ}, the lowest value of $q$ for which the
entanglement of the GWSs can be detected via the HZ witness $\bar
H_1(\rho_{\phi})$ [$\bar H_2(\rho_{\phi})$] is 0 (1/2) for
$\phi=\pi$ ($\phi=0$). For both HZ witnesses, the largest allowed
value of $q$ is equal to 1.

Figure~\ref{fig:HZ} shows a comparison of the two approaches to
analyze a hierarchy of quantum correlations, i.e.: the criteria
hierarchy, which is based on the HZ witnesses, and the CC
hierarchy, which is based on the discussed quantum correlation
measures. Any good measure of entanglement results in the same CC
hierarchy for the GWSs, while the criterion hierarchy depend on
the applied nonuniversal witnesses and can reveal only a subset of
the entangled GWSs, which correspond to regimes \#2--\#5. This
figure explains our motivation of experimentally demonstrating in
detail only the CC hierarchy instead of the hierarchy based on the
HZ witnesses, or using other either sufficient or necessary
conditions of quantum correlations. Unfortunately, by contrast to
such a hierarchy of criteria, it is experimentally challenging to
reveal such a CC hierarchy for the GWSs \emph{without} QST.

\subsection{Quantum state tomography via moments of annihilation and creation
operators} \label{App:Wunsche}

Here we give an example showing that some very limited additional
measurements on a given state can supplement a partial state
reconstruction into a full QST.

We recall that a  general single-mode density matrix $\rho$ of a
bosonic field can be reconstructed from the following moments of the
annihilation and creation operators via the
formula~\cite{Wunsche1996}:
\begin{equation}
\< m_{1}|\rho |m_{2}\> =\sum_{j=0}^{\infty }{\frac{1}{j!%
\sqrt{m_{1}!m_{2}!}}}\< (a^{\dag })^{m_{2}+j}a^{m_{1}+j}\>.
\label{Wunsche1}
\end{equation}
Note that this formula can be divergent for some states of the
radiation field including thermal field with the mean photon
number $\< n\> \geq 1$. However, for finite-dimensional states,
the above sum becomes finite. In particular, a two-mode version of
Eq.~(\ref{Wunsche1}) leads to the moment-based representation:
\begin{equation}
\left[
\begin{array}{cccc}
f  & \<b^{\dag }\>-\<n_1b^{\dag }\> & \<a^{\dag }\>-\< a^{\dag
}n_2\>
& \<a^{\dag }b^{\dag }\>\\
\<b\>-\<n_1b\> & \<n_2\>-\<n_1n_2\> & \<a^{\dag }b\> &
\<a^{\dag }n_2\>\\
\<a\>-\<an_2\> & \<ab^{\dag }\> & \<n_1\>-\<n_1n_2\> & \<
n_1b^{\dag }\>\\
\<ab\> & \<an_2\> & \<n_1b\> & \<n_1n_2\>
\end{array}
\right] \label{Wunsche2}
\end{equation}
of a general two-qubit state $\rho$, where
$f=1-\<n_1\>-\<n_2\>+\<n_1n_2\>$, and the annihilation operator
$a=a_{1}$ (and analogously $b=a_{2}$) is simply $a=\sigma_-=[0,
1;0, 0]$, i.e., the qubit lowering operator. Thus, an arbitrary
two-qubit state can be completely reconstructed by measuring only
the moments: $\<n_i\>$, $\<n_1 n_2\>$, $\<a_i\>$, $\<n_i
a_{2-i}\>$, $\<a_1 a_2\>,$ and $\<a_1 a_2^\dagger\>$ for $i=1,2$.

Note that experimental implementations of the HZ witnesses require
measuring $\<n_i\>$, $\<n_1 n_2\>$, $\<a_1 a_2\>,$ and $\<a_1
a_2^\dagger\>$. Thus, by measuring additionally only the moments
$\<a_i\>$ and $\<n_i a_{2-i}\>$, one can collect all the
information required for a complete QST, which one can thus
calculate any properties of an experimentally-reconstructed
two-qubit state.

%


\begin{thebibliography}{102}%
\makeatletter
\providecommand \@ifxundefined [1]{%
 \@ifx{#1\undefined}
}%
\providecommand \@ifnum [1]{%
 \ifnum #1\expandafter \@firstoftwo
 \else \expandafter \@secondoftwo
 \fi
}%
\providecommand \@ifx [1]{%
 \ifx #1\expandafter \@firstoftwo
 \else \expandafter \@secondoftwo
 \fi
}%
\providecommand \natexlab [1]{#1}%
\providecommand \enquote  [1]{``#1''}%
\providecommand \bibnamefont  [1]{#1}%
\providecommand \bibfnamefont [1]{#1}%
\providecommand \citenamefont [1]{#1}%
\providecommand \href@noop [0]{\@secondoftwo}%
\providecommand \href [0]{\begingroup \@sanitize@url \@href}%
\providecommand \@href[1]{\@@startlink{#1}\@@href}%
\providecommand \@@href[1]{\endgroup#1\@@endlink}%
\providecommand \@sanitize@url [0]{\catcode `\\12\catcode
`\$12\catcode
  `\&12\catcode `\#12\catcode `\^12\catcode `\_12\catcode `\%12\relax}%
\providecommand \@@startlink[1]{}%
\providecommand \@@endlink[0]{}%
\providecommand \url  [0]{\begingroup\@sanitize@url \@url }%
\providecommand \@url [1]{\endgroup\@href {#1}{\urlprefix }}%
\providecommand \urlprefix  [0]{URL }%
\providecommand \Eprint [0]{\href }%
\providecommand \doibase [0]{http://dx.doi.org/}%
\providecommand \selectlanguage [0]{\@gobble}%
\providecommand \bibinfo  [0]{\@secondoftwo}%
\providecommand \bibfield  [0]{\@secondoftwo}%
\providecommand \translation [1]{[#1]}%
\providecommand \BibitemOpen [0]{}%
\providecommand \bibitemStop [0]{}%
\providecommand \bibitemNoStop [0]{.\EOS\space}%
\providecommand \EOS [0]{\spacefactor3000\relax}%
\providecommand \BibitemShut  [1]{\csname bibitem#1\endcsname}%
\let\auto@bib@innerbib\@empty
\bibitem [{\citenamefont {Einstein}\ \emph {et~al.}(1935)\citenamefont
  {Einstein}, \citenamefont {Podolsky},\ and\ \citenamefont
  {Rosen}}]{Einstein1935}%
  \BibitemOpen
  \bibinfo {author} {A.~Einstein}, \bibinfo {author} {B.~Podolsky},\ and\
  \bibinfo {author} {N.~Rosen},\ \emph {\bibinfo {title} {Can
  Quantum-Mechanical Description of Physical Reality Be Considered
  Complete?}},\ \href {https://doi.org/10.1103/PhysRev.47.777} {\bibfield
  {journal} {\bibinfo  {journal} {Phys. Rev.}\ }\textbf {\bibinfo {volume}
  {47}},\ \bibinfo {pages} {777} (\bibinfo {year} {1935})}\BibitemShut
  {NoStop}%
\bibitem [{\citenamefont {Schr\"{o}dinger}(1935)}]{Schrodinger1935}%
  \BibitemOpen
  \bibinfo {author} {E.~Schr\"{o}dinger},\ \emph {\bibinfo {title} {Discussion
  of Probability Relations between Separated Systems}},\ \href
  {https://doi.org/10.1017/s0305004100013554} {\bibfield  {journal} {\bibinfo
  {journal} {Math. Proc. Camb. Phil. Soc.}\ }\textbf {\bibinfo {volume} {31}},\
  \bibinfo {pages} {555} (\bibinfo {year} {1935})}\BibitemShut {NoStop}%
\bibitem [{\citenamefont {Schr\"{o}dinger}(1936)}]{Schrodinger1936}%
  \BibitemOpen
  \bibinfo {author} {E.~Schr\"{o}dinger},\ \emph {\bibinfo {title} {Probability
  relations between separated systems}},\ \href
  {https://doi.org/10.1017/s0305004100019137} {\bibfield  {journal} {\bibinfo
  {journal} {Math. Proc. Camb. Phil. Soc.}\ }\textbf {\bibinfo {volume} {32}},\
  \bibinfo {pages} {446} (\bibinfo {year} {1936})}\BibitemShut {NoStop}%
\bibitem [{\citenamefont {Bell}(1964)}]{Bell1964}%
  \BibitemOpen
  \bibinfo {author} {J.~S. Bell},\ \emph {\bibinfo {title} {On the {E}instein
  {P}odolsky {R}osen paradox}},\ \href@noop {} {\bibfield  {journal} {\bibinfo
  {journal} {Physics (Long Island City, New York)}\ }\textbf {\bibinfo {volume}
  {1}},\ \bibinfo {pages} {195} (\bibinfo {year} {1964})}\BibitemShut {NoStop}%
\bibitem [{\citenamefont {Horodecki}\ \emph {et~al.}(2009)\citenamefont
  {Horodecki}, \citenamefont {Horodecki}, \citenamefont {Horodecki},\ and\
  \citenamefont {Horodecki}}]{HorodeckiReview}%
  \BibitemOpen
  \bibinfo {author} {R.~Horodecki}, \bibinfo {author} {P.~Horodecki}, \bibinfo
  {author} {M.~Horodecki},\ and\ \bibinfo {author} {K.~Horodecki},\ \emph
  {\bibinfo {title} {Quantum entanglement}},\ \href
  {http://link.aps.org/doi/10.1103/RevModPhys.81.865} {\bibfield  {journal}
  {\bibinfo  {journal} {Rev. Mod. Phys.}\ }\textbf {\bibinfo {volume} {81}},\
  \bibinfo {pages} {865} (\bibinfo {year} {2009})}\BibitemShut {NoStop}%
\bibitem [{\citenamefont {Brunner}\ \emph {et~al.}(2014)\citenamefont
  {Brunner}, \citenamefont {Cavalcanti}, \citenamefont {Pironio}, \citenamefont
  {Scarani},\ and\ \citenamefont {Wehner}}]{BrunnerReview}%
  \BibitemOpen
  \bibinfo {author} {N.~Brunner}, \bibinfo {author} {D.~Cavalcanti}, \bibinfo
  {author} {S.~Pironio}, \bibinfo {author} {V.~Scarani},\ and\ \bibinfo
  {author} {S.~Wehner},\ \emph {\bibinfo {title} {{B}ell nonlocality}},\ \href
  {http://link.aps.org/doi/10.1103/RevModPhys.86.419} {\bibfield  {journal}
  {\bibinfo  {journal} {Rev. Mod. Phys.}\ }\textbf {\bibinfo {volume} {86}},\
  \bibinfo {pages} {419} (\bibinfo {year} {2014})}\BibitemShut {NoStop}%
\bibitem [{\citenamefont {Cavalcanti}\ and\ \citenamefont
  {Skrzypczyk}(2017)}]{CavalcantiReview}%
  \BibitemOpen
  \bibinfo {author} {D.~Cavalcanti}\ and\ \bibinfo {author} {P.~Skrzypczyk},\
  \emph {\bibinfo {title} {Quantum steering: a review with focus on
  semidefinite programming}},\ \href
  {https://doi.org/10.1088/1361-6633/80/2/024001} {\bibfield  {journal}
  {\bibinfo  {journal} {Rep. Prog. Phys.}\ }\textbf {\bibinfo {volume} {80}},\
  \bibinfo {pages} {024001} (\bibinfo {year} {2017})}\BibitemShut {NoStop}%
\bibitem [{\citenamefont {Uola}\ \emph {et~al.}(2020)\citenamefont {Uola},
  \citenamefont {Costa}, \citenamefont {Nguyen},\ and\ \citenamefont
  {G\"{u}hne}}]{UolaReview}%
  \BibitemOpen
  \bibinfo {author} {R.~Uola}, \bibinfo {author} {A.~C.~S. Costa}, \bibinfo
  {author} {H.~C. Nguyen},\ and\ \bibinfo {author} {O.~G\"{u}hne},\ \emph
  {\bibinfo {title} {Quantum steering}},\ \href
  {https://doi.org/10.1103/revmodphys.92.015001} {\bibfield  {journal}
  {\bibinfo  {journal} {Rev. Mod. Phys.}\ }\textbf {\bibinfo {volume} {92}}, {015001}
  (\bibinfo {year} {2020})}\BibitemShut {NoStop}%
\bibitem [{\citenamefont {Clauser}\ \emph {et~al.}(1969)\citenamefont
  {Clauser}, \citenamefont {Horne}, \citenamefont {Shimony},\ and\
  \citenamefont {Holt}}]{Clauser1969}%
  \BibitemOpen
  \bibinfo {author} {J.~F. Clauser}, \bibinfo {author} {M.~A. Horne}, \bibinfo
  {author} {A.~Shimony},\ and\ \bibinfo {author} {R.~A. Holt},\ \emph {\bibinfo
  {title} {Proposed Experiment to Test Local Hidden-Variable Theories}},\ \href
  {http://link.aps.org/doi/10.1103/PhysRevLett.23.880} {\bibfield  {journal}
  {\bibinfo  {journal} {Phys. Rev. Lett.}\ }\textbf {\bibinfo {volume} {23}},\
  \bibinfo {pages} {880} (\bibinfo {year} {1969})}\BibitemShut {NoStop}%
\bibitem [{\citenamefont {Wiseman}\ \emph {et~al.}(2007)\citenamefont
  {Wiseman}, \citenamefont {Jones},\ and\ \citenamefont
  {Doherty}}]{Wiseman2007}%
  \BibitemOpen
  \bibinfo {author} {H.~M. Wiseman}, \bibinfo {author} {S.~J. Jones},\ and\
  \bibinfo {author} {A.~C. Doherty},\ \emph {\bibinfo {title} {Steering,
  Entanglement, Nonlocality, and the {E}instein-{P}odolsky-{R}osen Paradox}},\
  \href {https://doi.org/10.1103/physrevlett.98.140402} {\bibfield  {journal}
  {\bibinfo  {journal} {Phys. Rev. Lett.}\ }\textbf {\bibinfo {volume} {98}},\
  \bibinfo {pages} {140402} (\bibinfo {year} {2007})}\BibitemShut {NoStop}%
\bibitem [{\citenamefont {Werner}(1989)}]{Werner1989}%
  \BibitemOpen
  \bibinfo {author} {R.~F. Werner},\ \emph {\bibinfo {title} {Quantum states
  with {E}instein-{P}odolsky-{R}osen correlations admitting a hidden-variable
  model}},\ \href {https://doi.org/10.1103/physreva.40.4277} {\bibfield
  {journal} {\bibinfo  {journal} {Phys. Rev. A}\ }\textbf {\bibinfo {volume}
  {40}},\ \bibinfo {pages} {4277} (\bibinfo {year} {1989})}\BibitemShut
  {NoStop}%
\bibitem [{\citenamefont {Zhao}\ \emph {et~al.}(2020)\citenamefont {Zhao},
  \citenamefont {Ku}, \citenamefont {Chen}, \citenamefont {Chen}, \citenamefont
  {Nori}, \citenamefont {Xiang}, \citenamefont {Li}, \citenamefont {Guo},\ and\
  \citenamefont {Chen}}]{Zhao2020}%
  \BibitemOpen
  \bibinfo {author} {Y.-Y. Zhao}, \bibinfo {author} {H.-Y. Ku}, \bibinfo
  {author} {S.-L. Chen}, \bibinfo {author} {H.-B. Chen}, \bibinfo {author}
  {F.~Nori}, \bibinfo {author} {G.-Y. Xiang}, \bibinfo {author} {C.-F. Li},
  \bibinfo {author} {G.-C. Guo},\ and\ \bibinfo {author} {Y.-N. Chen},\ \emph
  {\bibinfo {title} {Experimental demonstration of
  measurement-device-independent measure of quantum steering}},\ \href
  {https://doi.org/10.1038/s41534-020-00307-9} {\bibfield  {journal} {\bibinfo
  {journal} {npj Quantum Inf.}\ }\textbf {\bibinfo {volume} {6}},\ \bibinfo
  {pages} {77} (\bibinfo {year} {2020})}\BibitemShut {NoStop}%
\bibitem [{\citenamefont {Ku}\ \emph {et~al.}(2018{\natexlab{a}})\citenamefont
  {Ku}, \citenamefont {Chen}, \citenamefont {Lambert}, \citenamefont {Chen},\
  and\ \citenamefont {Nori}}]{Ku2018hierarchy}%
  \BibitemOpen
  \bibinfo {author} {H.-Y. Ku}, \bibinfo {author} {S.-L. Chen}, \bibinfo
  {author} {N.~Lambert}, \bibinfo {author} {Y.-N. Chen},\ and\ \bibinfo
  {author} {F.~Nori},\ \emph {\bibinfo {title} {Hierarchy in temporal quantum
  correlations}},\ \href {https://doi.org/10.1103/physreva.98.022104}
  {\bibfield  {journal} {\bibinfo  {journal} {Phys. Rev. A}\ }\textbf {\bibinfo
  {volume} {98}},\ \bibinfo {pages} {022104} (\bibinfo {year}
  {2018}{\natexlab{a}})}\BibitemShut {NoStop}%
\bibitem [{\citenamefont {Fitzsimons}\ \emph {et~al.}(2015)\citenamefont
  {Fitzsimons}, \citenamefont {Jones},\ and\ \citenamefont
  {Vedral}}]{Fitzsimons2015}%
  \BibitemOpen
  \bibinfo {author} {J.~F. Fitzsimons}, \bibinfo {author} {J.~A. Jones},\ and\
  \bibinfo {author} {V.~Vedral},\ \emph {\bibinfo {title} {Quantum correlations
  which imply causation}},\ \href {https://doi.org/10.1038/srep18281}
  {\bibfield  {journal} {\bibinfo  {journal} {Sci. Rep.}\ }\textbf {\bibinfo
  {volume} {5}},\ \bibinfo {pages} {18281} (\bibinfo {year}
  {2015})}\BibitemShut {NoStop}%
\bibitem [{\citenamefont {Fritz}(2010)}]{Fritz2010}%
  \BibitemOpen
  \bibinfo {author} {T.~Fritz},\ \emph {\bibinfo {title} {Quantum correlations
  in the temporal {C}lauser-{H}orne-{S}himony-{H}olt ({CHSH}) scenario}},\
  \href {https://doi.org/10.1088/1367-2630/12/8/083055} {\bibfield  {journal}
  {\bibinfo  {journal} {New J. Phys.}\ }\textbf {\bibinfo {volume} {12}},\
  \bibinfo {pages} {083055} (\bibinfo {year} {2010})}\BibitemShut {NoStop}%
\bibitem [{\citenamefont {Chen}\ \emph {et~al.}(2014)\citenamefont {Chen},
  \citenamefont {Li}, \citenamefont {Lambert}, \citenamefont {Chen},
  \citenamefont {Ota}, \citenamefont {Chen},\ and\ \citenamefont
  {Nori}}]{Chen2014}%
  \BibitemOpen
  \bibinfo {author} {Y.-N. Chen}, \bibinfo {author} {C.-M. Li}, \bibinfo
  {author} {N.~Lambert}, \bibinfo {author} {S.-L. Chen}, \bibinfo {author}
  {Y.~Ota}, \bibinfo {author} {G.-Y. Chen},\ and\ \bibinfo {author} {F.~Nori},\
  \emph {\bibinfo {title} {Temporal steering inequality}},\ \href
  {https://doi.org/10.1103/physreva.89.032112} {\bibfield  {journal} {\bibinfo
  {journal} {Phys. Rev. A}\ }\textbf {\bibinfo {volume} {89}},\ \bibinfo
  {pages} {032112} (\bibinfo {year} {2014})}\BibitemShut {NoStop}%
\bibitem [{\citenamefont {Bartkiewicz}\ \emph {et~al.}(2016)\citenamefont
  {Bartkiewicz}, \citenamefont {{\v{C}}ernoch}, \citenamefont {Lemr},
  \citenamefont {Miranowicz},\ and\ \citenamefont {Nori}}]{Bartkiewicz2016}%
  \BibitemOpen
  \bibinfo {author} {K.~Bartkiewicz}, \bibinfo {author} {A.~{\v{C}}ernoch},
  \bibinfo {author} {K.~Lemr}, \bibinfo {author} {A.~Miranowicz},\ and\
  \bibinfo {author} {F.~Nori},\ \emph {\bibinfo {title} {Temporal steering and
  security of quantum key distribution with mutually unbiased bases against
  individual attacks}},\ \href {https://doi.org/10.1103/physreva.93.062345}
  {\bibfield  {journal} {\bibinfo  {journal} {Phys. Rev. A}\ }\textbf {\bibinfo
  {volume} {93}},\ \bibinfo {pages} {062345} (\bibinfo {year}
  {2016})}\BibitemShut {NoStop}%
\bibitem [{\citenamefont {Thew}\ and\ \citenamefont
  {Munro}(2001)}]{Thew2001PhysRevA}%
  \BibitemOpen
  \bibinfo {author} {R.~T. Thew}\ and\ \bibinfo {author} {W.~J. Munro},\ \emph
  {\bibinfo {title} {Mixed state entanglement: Manipulating
  polarization-entangled photons}},\ \href
  {https://link.aps.org/doi/10.1103/PhysRevA.64.022320} {\bibfield  {journal}
  {\bibinfo  {journal} {Phys. Rev. A}\ }\textbf {\bibinfo {volume} {64}},\
  \bibinfo {pages} {022320} (\bibinfo {year} {2001})}\BibitemShut {NoStop}%
\bibitem [{\citenamefont {Zhang}(2004)}]{Zhang2004PhysRevA}%
  \BibitemOpen
  \bibinfo {author} {C.~Zhang},\ \emph {\bibinfo {title} {Preparation of
  polarization-entangled mixed states of two photons}},\ \href
  {https://link.aps.org/doi/10.1103/PhysRevA.69.014304} {\bibfield  {journal}
  {\bibinfo  {journal} {Phys. Rev. A}\ }\textbf {\bibinfo {volume} {69}},\
  \bibinfo {pages} {014304} (\bibinfo {year} {2004})}\BibitemShut {NoStop}%
\bibitem [{\citenamefont {Wei}\ \emph {et~al.}(2005)\citenamefont {Wei},
  \citenamefont {Altepeter}, \citenamefont {Branning}, \citenamefont
  {Goldbart}, \citenamefont {James}, \citenamefont {Jeffrey}, \citenamefont
  {Kwiat}, \citenamefont {Mukhopadhyay},\ and\ \citenamefont
  {Peters}}]{Wei2005PhysRevA}%
  \BibitemOpen
  \bibinfo {author} {T.-C. Wei}, \bibinfo {author} {J.~B. Altepeter}, \bibinfo
  {author} {D.~Branning}, \bibinfo {author} {P.~M. Goldbart}, \bibinfo {author}
  {D.~F.~V. James}, \bibinfo {author} {E.~Jeffrey}, \bibinfo {author} {P.~G.
  Kwiat}, \bibinfo {author} {S.~Mukhopadhyay},\ and\ \bibinfo {author} {N.~A.
  Peters},\ \emph {\bibinfo {title} {Synthesizing arbitrary two-photon
  polarization mixed states}},\ \href
  {https://link.aps.org/doi/10.1103/PhysRevA.71.032329} {\bibfield  {journal}
  {\bibinfo  {journal} {Phys. Rev. A}\ }\textbf {\bibinfo {volume} {71}},\
  \bibinfo {pages} {032329} (\bibinfo {year} {2005})}\BibitemShut {NoStop}%
\bibitem [{\citenamefont {Lima}\ \emph {et~al.}(2008)\citenamefont {Lima},
  \citenamefont {Torres-Ruiz}, \citenamefont {Neves}, \citenamefont {Delgado},
  \citenamefont {Saavedra},\ and\ \citenamefont {Padua}}]{LIMA2008OptComm}%
  \BibitemOpen
  \bibinfo {author} {G.~Lima}, \bibinfo {author} {F.~Torres-Ruiz}, \bibinfo
  {author} {L.~Neves}, \bibinfo {author} {A.~Delgado}, \bibinfo {author}
  {C.~Saavedra},\ and\ \bibinfo {author} {S.~Padua},\ \emph {\bibinfo {title}
  {Generating mixtures of spatial qubits}},\ \href
  {http://www.sciencedirect.com/science/article/pii/S0030401808005944}
  {\bibfield  {journal} {\bibinfo  {journal} {Opt. Commun.}\ }\textbf {\bibinfo
  {volume} {281}},\ \bibinfo {pages} {5058 } (\bibinfo {year}
  {2008})}\BibitemShut {NoStop}%
\bibitem [{\citenamefont {Ling}\ \emph {et~al.}(2006)\citenamefont {Ling},
  \citenamefont {Han}, \citenamefont {Lamas-Linares},\ and\ \citenamefont
  {Kurtsiefer}}]{Ling2006LasPhys}%
  \BibitemOpen
  \bibinfo {author} {A.~Ling}, \bibinfo {author} {P.~Y. Han}, \bibinfo {author}
  {A.~Lamas-Linares},\ and\ \bibinfo {author} {C.~Kurtsiefer},\ \emph {\bibinfo
  {title} {Preparation of {B}ell states with controlled white noise}},\ \href
  {https://doi.org/10.1134/S1054660X06070206} {\bibfield  {journal} {\bibinfo
  {journal} {Laser Phys.}\ }\textbf {\bibinfo {volume} {16}},\ \bibinfo {pages}
  {1140} (\bibinfo {year} {2006})}\BibitemShut {NoStop}%
\bibitem [{\citenamefont {Liu}\ \emph {et~al.}(2017)\citenamefont {Liu},
  \citenamefont {Wang}, \citenamefont {Li},\ and\ \citenamefont
  {Wang}}]{Liu2017EurLett}%
  \BibitemOpen
  \bibinfo {author} {T.-J. Liu}, \bibinfo {author} {C.-Y. Wang}, \bibinfo
  {author} {J.~Li},\ and\ \bibinfo {author} {Q.~Wang},\ \emph {\bibinfo {title}
  {Experimental preparation of an arbitrary tunable Werner state}},\ \href
  {https://doi.org/10.1209/0295-5075/119/14002} {\bibfield  {journal} {\bibinfo
   {journal} {{EPL} (Europhys. Lett.)}\ }\textbf {\bibinfo {volume} {119}},\
  \bibinfo {pages} {14002} (\bibinfo {year} {2017})}\BibitemShut {NoStop}%
\bibitem [{\citenamefont {Puentes}\ \emph {et~al.}(2006)\citenamefont
  {Puentes}, \citenamefont {Voigt}, \citenamefont {Aiello},\ and\ \citenamefont
  {Woerdman}}]{Puentes2006OptLett}%
  \BibitemOpen
  \bibinfo {author} {G.~Puentes}, \bibinfo {author} {D.~Voigt}, \bibinfo
  {author} {A.~Aiello},\ and\ \bibinfo {author} {J.~P. Woerdman},\ \emph
  {\bibinfo {title} {Tunable spatial decoherers for polarization-entangled
  photons}},\ \href {http://ol.osa.org/abstract.cfm?URI=ol-31-13-2057}
  {\bibfield  {journal} {\bibinfo  {journal} {Opt. Lett.}\ }\textbf {\bibinfo
  {volume} {31}},\ \bibinfo {pages} {2057} (\bibinfo {year}
  {2006})}\BibitemShut {NoStop}%
\bibitem [{\citenamefont {White}\ \emph {et~al.}(2001)\citenamefont {White},
  \citenamefont {James}, \citenamefont {Munro},\ and\ \citenamefont
  {Kwiat}}]{White2001PhysRevA}%
  \BibitemOpen
  \bibinfo {author} {A.~G. White}, \bibinfo {author} {D.~F.~V. James}, \bibinfo
  {author} {W.~J. Munro},\ and\ \bibinfo {author} {P.~G. Kwiat},\ \emph
  {\bibinfo {title} {Exploring Hilbert space: Accurate characterization of
  quantum information}},\ \href
  {https://link.aps.org/doi/10.1103/PhysRevA.65.012301} {\bibfield  {journal}
  {\bibinfo  {journal} {Phys. Rev. A}\ }\textbf {\bibinfo {volume} {65}},\
  \bibinfo {pages} {012301} (\bibinfo {year} {2001})}\BibitemShut {NoStop}%
\bibitem [{\citenamefont {Zhang}\ \emph {et~al.}(2002)\citenamefont {Zhang},
  \citenamefont {Huang}, \citenamefont {Li},\ and\ \citenamefont
  {Guo}}]{Zhang2002PhysRevA}%
  \BibitemOpen
  \bibinfo {author} {Y.-S. Zhang}, \bibinfo {author} {Y.-F. Huang}, \bibinfo
  {author} {C.-F. Li},\ and\ \bibinfo {author} {G.-C. Guo},\ \emph {\bibinfo
  {title} {Experimental preparation of the Werner state via spontaneous
  parametric down-conversion}},\ \href
  {https://link.aps.org/doi/10.1103/PhysRevA.66.062315} {\bibfield  {journal}
  {\bibinfo  {journal} {Phys. Rev. A}\ }\textbf {\bibinfo {volume} {66}},\
  \bibinfo {pages} {062315} (\bibinfo {year} {2002})}\BibitemShut {NoStop}%
\bibitem [{\citenamefont {Cinelli}\ \emph {et~al.}(2004)\citenamefont
  {Cinelli}, \citenamefont {Di~Nepi}, \citenamefont {De~Martini}, \citenamefont
  {Barbieri},\ and\ \citenamefont {Mataloni}}]{Cinelli2004PhysRevA}%
  \BibitemOpen
  \bibinfo {author} {C.~Cinelli}, \bibinfo {author} {G.~Di~Nepi}, \bibinfo
  {author} {F.~De~Martini}, \bibinfo {author} {M.~Barbieri},\ and\ \bibinfo
  {author} {P.~Mataloni},\ \emph {\bibinfo {title} {Parametric source of
  two-photon states with a tunable degree of entanglement and mixing:
  Experimental preparation of Werner states and maximally entangled mixed
  states}},\ \href {https://link.aps.org/doi/10.1103/PhysRevA.70.022321}
  {\bibfield  {journal} {\bibinfo  {journal} {Phys. Rev. A}\ }\textbf {\bibinfo
  {volume} {70}},\ \bibinfo {pages} {022321} (\bibinfo {year}
  {2004})}\BibitemShut {NoStop}%
\bibitem [{\citenamefont {Caminati}\ \emph {et~al.}(2006)\citenamefont
  {Caminati}, \citenamefont {De~Martini}, \citenamefont {Perris}, \citenamefont
  {Sciarrino},\ and\ \citenamefont {Secondi}}]{Caminati2006PhysRevA}%
  \BibitemOpen
  \bibinfo {author} {M.~Caminati}, \bibinfo {author} {F.~De~Martini}, \bibinfo
  {author} {R.~Perris}, \bibinfo {author} {F.~Sciarrino},\ and\ \bibinfo
  {author} {V.~Secondi},\ \emph {\bibinfo {title} {Nonseparable Werner states
  in spontaneous parametric down-conversion}},\ \href
  {https://link.aps.org/doi/10.1103/PhysRevA.73.032312} {\bibfield  {journal}
  {\bibinfo  {journal} {Phys. Rev. A}\ }\textbf {\bibinfo {volume} {73}},\
  \bibinfo {pages} {032312} (\bibinfo {year} {2006})}\BibitemShut {NoStop}%
\bibitem [{\citenamefont {Puentes}\ \emph {et~al.}(2007)\citenamefont
  {Puentes}, \citenamefont {Aiello}, \citenamefont {Voigt},\ and\ \citenamefont
  {Woerdman}}]{Puentes2007PhysRevA}%
  \BibitemOpen
  \bibinfo {author} {G.~Puentes}, \bibinfo {author} {A.~Aiello}, \bibinfo
  {author} {D.~Voigt},\ and\ \bibinfo {author} {J.~P. Woerdman},\ \emph
  {\bibinfo {title} {Entangled mixed-state generation by twin-photon
  scattering}},\ \href {https://link.aps.org/doi/10.1103/PhysRevA.75.032319}
  {\bibfield  {journal} {\bibinfo  {journal} {Phys. Rev. A}\ }\textbf {\bibinfo
  {volume} {75}},\ \bibinfo {pages} {032319} (\bibinfo {year}
  {2007})}\BibitemShut {NoStop}%
\bibitem [{\citenamefont {Aiello}\ \emph {et~al.}(2007)\citenamefont {Aiello},
  \citenamefont {Puentes}, \citenamefont {Voigt},\ and\ \citenamefont
  {Woerdman}}]{Aiello2007PhysRevA}%
  \BibitemOpen
  \bibinfo {author} {A.~Aiello}, \bibinfo {author} {G.~Puentes}, \bibinfo
  {author} {D.~Voigt},\ and\ \bibinfo {author} {J.~P. Woerdman},\ \emph
  {\bibinfo {title} {Maximally entangled mixed-state generation via local
  operations}},\ \href {https://link.aps.org/doi/10.1103/PhysRevA.75.062118}
  {\bibfield  {journal} {\bibinfo  {journal} {Phys. Rev. A}\ }\textbf {\bibinfo
  {volume} {75}},\ \bibinfo {pages} {062118} (\bibinfo {year}
  {2007})}\BibitemShut {NoStop}%
\bibitem [{\citenamefont {Brida}\ \emph {et~al.}(2008)\citenamefont {Brida},
  \citenamefont {Genovese}, \citenamefont {Chekhova},\ and\ \citenamefont
  {Krivitsky}}]{Brida2008PhysRevA}%
  \BibitemOpen
  \bibinfo {author} {G.~Brida}, \bibinfo {author} {M.~Genovese}, \bibinfo
  {author} {M.~V. Chekhova},\ and\ \bibinfo {author} {L.~A. Krivitsky},\ \emph
  {\bibinfo {title} {Tailoring polarization entanglement in
  anisotropy-compensated spontaneous parametric down-conversion}},\ \href
  {https://link.aps.org/doi/10.1103/PhysRevA.77.015805} {\bibfield  {journal}
  {\bibinfo  {journal} {Phys. Rev. A}\ }\textbf {\bibinfo {volume} {77}},\
  \bibinfo {pages} {015805} (\bibinfo {year} {2008})}\BibitemShut {NoStop}%
\bibitem [{\citenamefont {Peters}\ \emph {et~al.}(2004)\citenamefont {Peters},
  \citenamefont {Altepeter}, \citenamefont {Branning}, \citenamefont {Jeffrey},
  \citenamefont {Wei},\ and\ \citenamefont {Kwiat}}]{Peters2004PhysRevLett}%
  \BibitemOpen
  \bibinfo {author} {N.~A. Peters}, \bibinfo {author} {J.~B. Altepeter},
  \bibinfo {author} {D.~Branning}, \bibinfo {author} {E.~R. Jeffrey}, \bibinfo
  {author} {T.-C. Wei},\ and\ \bibinfo {author} {P.~G. Kwiat},\ \emph {\bibinfo
  {title} {Maximally Entangled Mixed States: Creation and Concentration}},\
  \href {https://link.aps.org/doi/10.1103/PhysRevLett.92.133601} {\bibfield
  {journal} {\bibinfo  {journal} {Phys. Rev. Lett.}\ }\textbf {\bibinfo
  {volume} {92}},\ \bibinfo {pages} {133601} (\bibinfo {year}
  {2004})}\BibitemShut {NoStop}%
\bibitem [{\citenamefont {Barbieri}\ \emph {et~al.}(2004)\citenamefont
  {Barbieri}, \citenamefont {De~Martini}, \citenamefont {Di~Nepi},\ and\
  \citenamefont {Mataloni}}]{Barbieri2004PhysRevLett}%
  \BibitemOpen
  \bibinfo {author} {M.~Barbieri}, \bibinfo {author} {F.~De~Martini}, \bibinfo
  {author} {G.~Di~Nepi},\ and\ \bibinfo {author} {P.~Mataloni},\ \emph
  {\bibinfo {title} {Generation and Characterization of Werner States and
  Maximally Entangled Mixed States by a Universal Source of Entanglement}},\
  \href {https://link.aps.org/doi/10.1103/PhysRevLett.92.177901} {\bibfield
  {journal} {\bibinfo  {journal} {Phys. Rev. Lett.}\ }\textbf {\bibinfo
  {volume} {92}},\ \bibinfo {pages} {177901} (\bibinfo {year}
  {2004})}\BibitemShut {NoStop}%
\bibitem [{\citenamefont {Kwiat}\ \emph {et~al.}(2000)\citenamefont {Kwiat},
  \citenamefont {Berglund}, \citenamefont {Altepeter},\ and\ \citenamefont
  {White}}]{Kwiat2000Science}%
  \BibitemOpen
  \bibinfo {author} {P.~G. Kwiat}, \bibinfo {author} {A.~J. Berglund}, \bibinfo
  {author} {J.~B. Altepeter},\ and\ \bibinfo {author} {A.~G. White},\ \emph
  {\bibinfo {title} {Experimental Verification of Decoherence-Free
  Subspaces}},\ \href {https://science.sciencemag.org/content/290/5491/498}
  {\bibfield  {journal} {\bibinfo  {journal} {Science}\ }\textbf {\bibinfo
  {volume} {290}},\ \bibinfo {pages} {498} (\bibinfo {year}
  {2000})}\BibitemShut {NoStop}%
\bibitem [{\citenamefont {Munro}\ \emph {et~al.}(2001)\citenamefont {Munro},
  \citenamefont {James}, \citenamefont {White},\ and\ \citenamefont
  {Kwiat}}]{Munro2001PhysRevA}%
  \BibitemOpen
  \bibinfo {author} {W.~J. Munro}, \bibinfo {author} {D.~F.~V. James}, \bibinfo
  {author} {A.~G. White},\ and\ \bibinfo {author} {P.~G. Kwiat},\ \emph
  {\bibinfo {title} {Maximizing the entanglement of two mixed qubits}},\ \href
  {https://link.aps.org/doi/10.1103/PhysRevA.64.030302} {\bibfield  {journal}
  {\bibinfo  {journal} {Phys. Rev. A}\ }\textbf {\bibinfo {volume} {64}},\
  \bibinfo {pages} {030302} (\bibinfo {year} {2001})}\BibitemShut {NoStop}%
\bibitem [{\citenamefont {Jeong}\ \emph {et~al.}(2013)\citenamefont {Jeong},
  \citenamefont {Lee},\ and\ \citenamefont {Kim}}]{Jeong2013PhysRevA}%
  \BibitemOpen
  \bibinfo {author} {Y.-C. Jeong}, \bibinfo {author} {J.-C. Lee},\ and\
  \bibinfo {author} {Y.-H. Kim},\ \emph {\bibinfo {title} {Experimental
  implementation of a fully controllable depolarizing quantum operation}},\
  \href {https://link.aps.org/doi/10.1103/PhysRevA.87.014301} {\bibfield
  {journal} {\bibinfo  {journal} {Phys. Rev. A}\ }\textbf {\bibinfo {volume}
  {87}},\ \bibinfo {pages} {014301} (\bibinfo {year} {2013})}\BibitemShut
  {NoStop}%
\bibitem [{\citenamefont {Lemr}\ \emph {et~al.}(2016)\citenamefont {Lemr},
  \citenamefont {Bartkiewicz},\ and\ \citenamefont {\ifmmode~\check{C}\else
  \v{C}\fi{}ernoch}}]{Lemr2016PhysRevA}%
  \BibitemOpen
  \bibinfo {author} {K.~Lemr}, \bibinfo {author} {K.~Bartkiewicz},\ and\
  \bibinfo {author} {A.~\ifmmode~\check{C}\else \v{C}\fi{}ernoch},\ \emph
  {\bibinfo {title} {Experimental measurement of collective nonlinear
  entanglement witness for two qubits}},\ \href
  {https://link.aps.org/doi/10.1103/PhysRevA.94.052334} {\bibfield  {journal}
  {\bibinfo  {journal} {Phys. Rev. A}\ }\textbf {\bibinfo {volume} {94}},\
  \bibinfo {pages} {052334} (\bibinfo {year} {2016})}\BibitemShut {NoStop}%
\bibitem [{\citenamefont {Gavenda}\ \emph {et~al.}(2005)\citenamefont
  {Gavenda}, \citenamefont {\v{C}ernoch}, \citenamefont {Soubusta},
  \citenamefont {Du\v{S}ek},\ and\ \citenamefont
  {Filip}}]{GAVENDA2005ModPhysLett}%
  \BibitemOpen
  \bibinfo {author} {M.~Gavenda}, \bibinfo {author} {A.~\v{C}ernoch}, \bibinfo
  {author} {J.~Soubusta}, \bibinfo {author} {M.~Du\v{S}ek},\ and\ \bibinfo
  {author} {R.~Filip},\ \emph {\bibinfo {title} {Knowledge excess duality and
  violation of {B}ell inequalities: theory and experiment}},\ \href
  {https://doi.org/10.1142/S021798490500827X} {\bibfield  {journal} {\bibinfo
  {journal} {Mod. Phys. Lett. B}\ }\textbf {\bibinfo {volume} {19}},\ \bibinfo
  {pages} {195} (\bibinfo {year} {2005})}\BibitemShut {NoStop}%
\bibitem [{\citenamefont {Aspect}\ \emph {et~al.}(1982)\citenamefont {Aspect},
  \citenamefont {Dalibard},\ and\ \citenamefont {Roger}}]{Aspect1982}%
  \BibitemOpen
  \bibinfo {author} {A.~Aspect}, \bibinfo {author} {J.~Dalibard},\ and\
  \bibinfo {author} {G.~Roger},\ \emph {\bibinfo {title} {Experimental Test of
  {B}ell's Inequalities Using Time-Varying Analyzers}},\ \href
  {https://doi.org/10.1103/physrevlett.49.1804} {\bibfield  {journal} {\bibinfo
   {journal} {Phys. Rev. Lett.}\ }\textbf {\bibinfo {volume} {49}},\ \bibinfo
  {pages} {1804} (\bibinfo {year} {1982})}\BibitemShut {NoStop}%
\bibitem [{\citenamefont
  {{Christensen~\emph{et~al.}}}(2013)}]{Christensen2013}%
  \BibitemOpen
  \bibinfo {author} {B.~G. {Christensen~\emph{et~al.}}},\ \emph {\bibinfo
  {title} {Detection-Loophole-Free Test of Quantum Nonlocality, and
  Applications}},\ \href {https://doi.org/10.1103/physrevlett.111.130406}
  {\bibfield  {journal} {\bibinfo  {journal} {Phys. Rev. Lett.}\ }\textbf
  {\bibinfo {volume} {111}},\ \bibinfo {pages} {130406} (\bibinfo {year}
  {2013})}\BibitemShut {NoStop}%
\bibitem [{\citenamefont {{Giustina~\emph{et~al.}}}(2015)}]{Giustina2015}%
  \BibitemOpen
  \bibinfo {author} {M.~{Giustina~\emph{et~al.}}},\ \emph {\bibinfo {title}
  {Significant-Loophole-Free Test of {B}ell's Theorem with Entangled
  Photons}},\ \href {https://doi.org/10.1103/physrevlett.115.250401} {\bibfield
   {journal} {\bibinfo  {journal} {Phys. Rev. Lett.}\ }\textbf {\bibinfo
  {volume} {115}},\ \bibinfo {pages} {250401} (\bibinfo {year}
  {2015})}\BibitemShut {NoStop}%
\bibitem [{\citenamefont {{Hensen \emph{et~al.}}}(2015)}]{Hensen2015}%
  \BibitemOpen
  \bibinfo {author} {B.~{Hensen \emph{et~al.}}},\ \emph {\bibinfo {title}
  {Loophole-free {B}ell inequality violation using electron spins separated by
  1.3 kilometres}},\ \href {https://doi.org/10.1038/nature15759} {\bibfield
  {journal} {\bibinfo  {journal} {Nature (London)}\ }\textbf {\bibinfo {volume}
  {526}},\ \bibinfo {pages} {682} (\bibinfo {year} {2015})}\BibitemShut
  {NoStop}%
\bibitem [{\citenamefont {Shchukin}\ and\ \citenamefont
  {Vogel}(2005{\natexlab{a}})}]{Shchukin2005}%
  \BibitemOpen
  \bibinfo {author} {E.~Shchukin}\ and\ \bibinfo {author} {W.~Vogel},\ \emph
  {\bibinfo {title} {Inseparability Criteria for Continuous Bipartite Quantum
  States}},\ \href {https://doi.org/10.1103/physrevlett.95.230502} {\bibfield
  {journal} {\bibinfo  {journal} {Phys, Rev. Lett.}\ }\textbf {\bibinfo
  {volume} {95}},\ \bibinfo {pages} {230502} (\bibinfo {year}
  {2005}{\natexlab{a}})}\BibitemShut {NoStop}%
\bibitem [{\citenamefont {Miranowicz}\ and\ \citenamefont
  {Piani}(2006)}]{Miranowicz2006}%
  \BibitemOpen
  \bibinfo {author} {A.~Miranowicz}\ and\ \bibinfo {author} {M.~Piani},\ \emph
  {\bibinfo {title} {Comment on {\textquotedblleft}Inseparability Criteria for
  Continuous Bipartite Quantum States{\textquotedblright}}},\ \href
  {https://doi.org/10.1103/physrevlett.97.058901} {\bibfield  {journal}
  {\bibinfo  {journal} {Phys, Rev. Lett.}\ }\textbf {\bibinfo {volume} {97}},\
  \bibinfo {pages} {058901} (\bibinfo {year} {2006})}\BibitemShut {NoStop}%
\bibitem [{\citenamefont {Miranowicz}\ \emph {et~al.}(2009)\citenamefont
  {Miranowicz}, \citenamefont {Piani}, \citenamefont {Horodecki},\ and\
  \citenamefont {Horodecki}}]{Miranowicz2009}%
  \BibitemOpen
  \bibinfo {author} {A.~Miranowicz}, \bibinfo {author} {M.~Piani}, \bibinfo
  {author} {P.~Horodecki},\ and\ \bibinfo {author} {R.~Horodecki},\ \emph
  {\bibinfo {title} {Inseparability criteria based on matrices of moments}},\
  \href {https://doi.org/10.1103/physreva.80.052303} {\bibfield  {journal}
  {\bibinfo  {journal} {Phys. Rev. A}\ }\textbf {\bibinfo {volume} {80}},\
  \bibinfo {pages} {052303} (\bibinfo {year} {2009})}\BibitemShut {NoStop}%
\bibitem [{\citenamefont {Kogias}\ \emph {et~al.}(2015)\citenamefont {Kogias},
  \citenamefont {Skrzypczyk}, \citenamefont {Cavalcanti}, \citenamefont
  {Ac{\'{\i}}n},\ and\ \citenamefont {Adesso}}]{Kogias2015}%
  \BibitemOpen
  \bibinfo {author} {I.~Kogias}, \bibinfo {author} {P.~Skrzypczyk}, \bibinfo
  {author} {D.~Cavalcanti}, \bibinfo {author} {A.~Ac{\'{\i}}n},\ and\ \bibinfo
  {author} {G.~Adesso},\ \emph {\bibinfo {title} {Hierarchy of Steering
  Criteria Based on Moments for All Bipartite Quantum Systems}},\ \href
  {https://doi.org/10.1103/physrevlett.115.210401} {\bibfield  {journal}
  {\bibinfo  {journal} {Phys. Rev. Lett.}\ }\textbf {\bibinfo {volume} {115}},\
  \bibinfo {pages} {210401} (\bibinfo {year} {2015})}\BibitemShut {NoStop}%
\bibitem [{\citenamefont {Navascu{\'{e}}s}\ \emph {et~al.}(2007)\citenamefont
  {Navascu{\'{e}}s}, \citenamefont {Pironio},\ and\ \citenamefont
  {Ac{\'{\i}}n}}]{Navascus2007}%
  \BibitemOpen
  \bibinfo {author} {M.~Navascu{\'{e}}s}, \bibinfo {author} {S.~Pironio},\ and\
  \bibinfo {author} {A.~Ac{\'{\i}}n},\ \emph {\bibinfo {title} {Bounding the
  Set of Quantum Correlations}},\ \href
  {https://doi.org/10.1103/physrevlett.98.010401} {\bibfield  {journal}
  {\bibinfo  {journal} {Phys. Rev. Lett.}\ }\textbf {\bibinfo {volume} {98}},\
  \bibinfo {pages} {010401} (\bibinfo {year} {2007})}\BibitemShut {NoStop}%
\bibitem [{\citenamefont {Richter}\ and\ \citenamefont
  {Vogel}(2002)}]{Richter2002}%
  \BibitemOpen
  \bibinfo {author} {T.~Richter}\ and\ \bibinfo {author} {W.~Vogel},\ \emph
  {\bibinfo {title} {Nonclassicality of Quantum States: A Hierarchy of
  Observable Conditions}},\ \href
  {https://doi.org/10.1103/PhysRevLett.89.283601} {\bibfield  {journal}
  {\bibinfo  {journal} {Phys. Rev. Lett.}\ }\textbf {\bibinfo {volume} {89}},\
  \bibinfo {pages} {283601} (\bibinfo {year} {2002})}\BibitemShut {NoStop}%
\bibitem [{\citenamefont {Vogel}(2008)}]{Vogel2008}%
  \BibitemOpen
  \bibinfo {author} {W.~Vogel},\ \emph {\bibinfo {title} {Nonclassical
  Correlation Properties of Radiation Fields}},\ \href
  {https://doi.org/10.1103/physrevlett.100.013605} {\bibfield  {journal}
  {\bibinfo  {journal} {Phys. Rev. Lett.}\ }\textbf {\bibinfo {volume} {100}},\
  \bibinfo {pages} {013605} (\bibinfo {year} {2008})}\BibitemShut {NoStop}%
\bibitem [{\citenamefont {Miranowicz}\ \emph {et~al.}(2010)\citenamefont
  {Miranowicz}, \citenamefont {Bartkowiak}, \citenamefont {Wang}, \citenamefont
  {xi~Liu},\ and\ \citenamefont {Nori}}]{Miranowicz2010}%
  \BibitemOpen
  \bibinfo {author} {A.~Miranowicz}, \bibinfo {author} {M.~Bartkowiak},
  \bibinfo {author} {X.~Wang}, \bibinfo {author} {Y.~xi~Liu},\ and\ \bibinfo
  {author} {F.~Nori},\ \emph {\bibinfo {title} {Testing nonclassicality in
  multimode fields: A unified derivation of classical inequalities}},\ \href
  {https://doi.org/10.1103/physreva.82.013824} {\bibfield  {journal} {\bibinfo
  {journal} {Phys. Rev. A}\ }\textbf {\bibinfo {volume} {82}},\ \bibinfo
  {pages} {013824} (\bibinfo {year} {2010})}\BibitemShut {NoStop}%
\bibitem [{\citenamefont {{\.{Z}}yczkowski}\ \emph {et~al.}(1998)\citenamefont
  {{\.{Z}}yczkowski}, \citenamefont {Horodecki}, \citenamefont {Sanpera},\ and\
  \citenamefont {Lewenstein}}]{Zyczkowski1998}%
  \BibitemOpen
  \bibinfo {author} {K.~{\.{Z}}yczkowski}, \bibinfo {author} {P.~Horodecki},
  \bibinfo {author} {A.~Sanpera},\ and\ \bibinfo {author} {M.~Lewenstein},\
  \emph {\bibinfo {title} {Volume of the set of separable states}},\ \href
  {https://doi.org/10.1103/physreva.58.883} {\bibfield  {journal} {\bibinfo
  {journal} {Phys. Rev. A}\ }\textbf {\bibinfo {volume} {58}},\ \bibinfo
  {pages} {883} (\bibinfo {year} {1998})}\BibitemShut {NoStop}%
\bibitem [{\citenamefont {Horodecki}\ \emph {et~al.}(1996)\citenamefont
  {Horodecki}, \citenamefont {Horodecki},\ and\ \citenamefont
  {Horodecki}}]{Horodecki1996negativity}%
  \BibitemOpen
  \bibinfo {author} {M.~Horodecki}, \bibinfo {author} {P.~Horodecki},\ and\
  \bibinfo {author} {R.~Horodecki},\ \emph {\bibinfo {title} {Separability of
  mixed states: necessary and sufficient conditions}},\ \href
  {https://doi.org/10.1016/s0375-9601(96)00706-2} {\bibfield  {journal}
  {\bibinfo  {journal} {Phys. Lett. A}\ }\textbf {\bibinfo {volume} {223}},\
  \bibinfo {pages} {1} (\bibinfo {year} {1996})}\BibitemShut {NoStop}%
\bibitem [{\citenamefont {Audenaert}\ \emph {et~al.}(2003)\citenamefont
  {Audenaert}, \citenamefont {Plenio},\ and\ \citenamefont
  {Eisert}}]{Audenaert2003}%
  \BibitemOpen
  \bibinfo {author} {K.~Audenaert}, \bibinfo {author} {M.~B. Plenio},\ and\
  \bibinfo {author} {J.~Eisert},\ \emph {\bibinfo {title} {Entanglement Cost
  under Positive-Partial-Transpose-Preserving Operations}},\ \href
  {https://doi.org/10.1103/physrevlett.90.027901} {\bibfield  {journal}
  {\bibinfo  {journal} {Phys. Rev. Lett.}\ }\textbf {\bibinfo {volume} {90}},\
  \bibinfo {pages} {027901} (\bibinfo {year} {2003})}\BibitemShut {NoStop}%
\bibitem [{\citenamefont {Ishizaka}(2004)}]{Ishizaka2004}%
  \BibitemOpen
  \bibinfo {author} {S.~Ishizaka},\ \emph {\bibinfo {title} {Binegativity and
  geometry of entangled states in two qubits}},\ \href
  {https://doi.org/10.1103/physreva.69.020301} {\bibfield  {journal} {\bibinfo
  {journal} {Phys. Rev. A}\ }\textbf {\bibinfo {volume} {69}},\ \bibinfo
  {pages} {020301} (\bibinfo {year} {2004})}\BibitemShut {NoStop}%
\bibitem [{\citenamefont {{C. Eltschka and J. Siewert}}(2013)}]{Eltschka2013}%
  \BibitemOpen
  \bibinfo {author} {{C. Eltschka and J. Siewert}},\ \emph {\bibinfo {title}
  {Negativity as an Estimator of Entanglement Dimension}},\ \href
  {https://doi.org/10.1103/PhysRevLett.111.100503} {\bibfield  {journal}
  {\bibinfo  {journal} {Phys. Rev. Lett.}\ }\textbf {\bibinfo {volume} {111}},\
  \bibinfo {pages} {100503} (\bibinfo {year} {2013})}\BibitemShut {NoStop}%
\bibitem [{\citenamefont {Wootters}(1998)}]{Wootters1998}%
  \BibitemOpen
  \bibinfo {author} {W.~K. Wootters},\ \emph {\bibinfo {title} {Entanglement of
  Formation of an Arbitrary State of Two Qubits}},\ \href
  {https://doi.org/10.1103/physrevlett.80.2245} {\bibfield  {journal} {\bibinfo
   {journal} {Phys. Rev. Lett.}\ }\textbf {\bibinfo {volume} {80}},\ \bibinfo
  {pages} {2245} (\bibinfo {year} {1998})}\BibitemShut {NoStop}%
\bibitem [{\citenamefont {Asb\'oth}\ \emph {et~al.}(2005)\citenamefont
  {Asb\'oth}, \citenamefont {Calsamiglia},\ and\ \citenamefont
  {Ritsch}}]{Asboth2005}%
  \BibitemOpen
  \bibinfo {author} {J.~K. Asb\'oth}, \bibinfo {author} {J.~Calsamiglia},\ and\
  \bibinfo {author} {H.~Ritsch},\ \emph {\bibinfo {title} {Computable Measure
  of Nonclassicality for Light}},\ \href
  {https://link.aps.org/10.1103/PhysRevLett.94.173602} {\bibfield  {journal}
  {\bibinfo  {journal} {Phys. Rev. Lett.}\ }\textbf {\bibinfo {volume} {94}},\
  \bibinfo {pages} {173602} (\bibinfo {year} {2005})}\BibitemShut {NoStop}%
\bibitem [{\citenamefont {Miranowicz}\ \emph
  {et~al.}(2015{\natexlab{a}})\citenamefont {Miranowicz}, \citenamefont
  {Bartkiewicz}, \citenamefont {Pathak}, \citenamefont {Pe\v{r}ina~Jr.},
  \citenamefont {Chen},\ and\ \citenamefont {Nori}}]{Adam2015a}%
  \BibitemOpen
  \bibinfo {author} {A.~Miranowicz}, \bibinfo {author} {K.~Bartkiewicz},
  \bibinfo {author} {A.~Pathak}, \bibinfo {author} {J.~Pe\v{r}ina~Jr.},
  \bibinfo {author} {Y.~Chen},\ and\ \bibinfo {author} {F.~Nori},\ \emph
  {\bibinfo {title} {Statistical mixtures of states can be more quantum than
  their superpositions: {C}omparison of nonclassicality measures for
  single-qubit states}},\ \href
  {http://link.aps.org/doi/10.1103/PhysRevA.91.042309} {\bibfield  {journal}
  {\bibinfo  {journal} {Phys. Rev. A}\ }\textbf {\bibinfo {volume} {91}},\
  \bibinfo {pages} {042309} (\bibinfo {year} {2015}{\natexlab{a}})}\BibitemShut
  {NoStop}%
\bibitem [{\citenamefont {Miranowicz}\ \emph
  {et~al.}(2015{\natexlab{b}})\citenamefont {Miranowicz}, \citenamefont
  {Bartkiewicz}, \citenamefont {Lambert}, \citenamefont {Chen},\ and\
  \citenamefont {Nori}}]{Adam2015b}%
  \BibitemOpen
  \bibinfo {author} {A.~Miranowicz}, \bibinfo {author} {K.~Bartkiewicz},
  \bibinfo {author} {N.~Lambert}, \bibinfo {author} {Y.~Chen},\ and\ \bibinfo
  {author} {F.~Nori},\ \emph {\bibinfo {title} {Increasing relative
  nonclassicality quantified by standard entanglement potentials by dissipation
  and unbalanced beam splitting}},\ \href
  {http://link.aps.org/doi/10.1103/PhysRevA.92.062314} {\bibfield  {journal}
  {\bibinfo  {journal} {Phys. Rev. A}\ }\textbf {\bibinfo {volume} {92}},\
  \bibinfo {pages} {062314} (\bibinfo {year} {2015}{\natexlab{b}})}\BibitemShut
  {NoStop}%
\bibitem [{\citenamefont {Bartkiewicz}\ \emph
  {et~al.}(2015{\natexlab{a}})\citenamefont {Bartkiewicz}, \citenamefont
  {Horodecki}, \citenamefont {Lemr}, \citenamefont {Miranowicz},\ and\
  \citenamefont {{\.{Z}}yczkowski}}]{Bartkiewicz2015}%
  \BibitemOpen
  \bibinfo {author} {K.~Bartkiewicz}, \bibinfo {author} {P.~Horodecki},
  \bibinfo {author} {K.~Lemr}, \bibinfo {author} {A.~Miranowicz},\ and\
  \bibinfo {author} {K.~{\.{Z}}yczkowski},\ \emph {\bibinfo {title} {Method for
  universal detection of two-photon polarization entanglement}},\ \href
  {http://dx.doi.org/10.1103/PhysRevA.91.032315} {\bibfield  {journal}
  {\bibinfo  {journal} {Phys. Rev. A}\ }\textbf {\bibinfo {volume} {91}},\
  \bibinfo {pages} {032315} (\bibinfo {year} {2015}{\natexlab{a}})}\BibitemShut
  {NoStop}%
\bibitem [{\citenamefont {Augusiak}\ \emph {et~al.}(2008)\citenamefont
  {Augusiak}, \citenamefont {Demianowicz},\ and\ \citenamefont
  {Horodecki}}]{Augusiak2008}%
  \BibitemOpen
  \bibinfo {author} {R.~Augusiak}, \bibinfo {author} {M.~Demianowicz},\ and\
  \bibinfo {author} {P.~Horodecki},\ \emph {\bibinfo {title} {Universal
  observable detecting all two-qubit entanglement and determinant-based
  separability tests}},\ \href {https://doi.org/10.1103/physreva.77.030301}
  {\bibfield  {journal} {\bibinfo  {journal} {Phys. Rev. A}\ }\textbf {\bibinfo
  {volume} {77}},\ \bibinfo {pages} {030301} (\bibinfo {year}
  {2008})}\BibitemShut {NoStop}%
\bibitem [{\citenamefont {Skrzypczyk}\ \emph {et~al.}(2014)\citenamefont
  {Skrzypczyk}, \citenamefont {Navascu\'es},\ and\ \citenamefont
  {Cavalcanti}}]{Skrzypczyk2014}%
  \BibitemOpen
  \bibinfo {author} {P.~Skrzypczyk}, \bibinfo {author} {M.~Navascu\'es},\ and\
  \bibinfo {author} {D.~Cavalcanti},\ \emph {\bibinfo {title} {Quantifying
  {E}instein-{P}odolsky-{R}osen Steering}},\ \href
  {http://link.aps.org/doi/10.1103/PhysRevLett.112.180404} {\bibfield
  {journal} {\bibinfo  {journal} {Phys. Rev. Lett.}\ }\textbf {\bibinfo
  {volume} {112}},\ \bibinfo {pages} {180404} (\bibinfo {year}
  {2014})}\BibitemShut {NoStop}%
\bibitem [{\citenamefont {Piani}\ and\ \citenamefont
  {Watrous}(2015)}]{Piani2015}%
  \BibitemOpen
  \bibinfo {author} {M.~Piani}\ and\ \bibinfo {author} {J.~Watrous},\ \emph
  {\bibinfo {title} {Necessary and Sufficient Quantum Information
  Characterization of Einstein-Podolsky-Rosen Steering}},\ \href
  {https://doi.org/10.1103/physrevlett.114.060404} {\bibfield  {journal}
  {\bibinfo  {journal} {Phys. Rev. Lett.}\ }\textbf {\bibinfo {volume} {114}},\
  \bibinfo {pages} {060404} (\bibinfo {year} {2015})}\BibitemShut {NoStop}%
\bibitem [{\citenamefont {Ku}\ \emph {et~al.}(2018{\natexlab{b}})\citenamefont
  {Ku}, \citenamefont {Chen}, \citenamefont {Budroni}, \citenamefont
  {Miranowicz}, \citenamefont {Chen},\ and\ \citenamefont {Nori}}]{Ku2018}%
  \BibitemOpen
  \bibinfo {author} {H.-Y. Ku}, \bibinfo {author} {S.-L. Chen}, \bibinfo
  {author} {C.~Budroni}, \bibinfo {author} {A.~Miranowicz}, \bibinfo {author}
  {Y.-N. Chen},\ and\ \bibinfo {author} {F.~Nori},\ \emph {\bibinfo {title}
  {Einstein-{P}odolsky-{R}osen steering: {I}ts geometric quantification and
  witness}},\ \href {https://doi.org/10.1103/physreva.97.022338} {\bibfield
  {journal} {\bibinfo  {journal} {Phys. Rev. A}\ }\textbf {\bibinfo {volume}
  {97}},\ \bibinfo {pages} {022338} (\bibinfo {year}
  {2018}{\natexlab{b}})}\BibitemShut {NoStop}%
\bibitem [{\citenamefont {Chen}\ \emph {et~al.}(2016)\citenamefont {Chen},
  \citenamefont {Lambert}, \citenamefont {Li}, \citenamefont {Miranowicz},
  \citenamefont {Chen},\ and\ \citenamefont {Nori}}]{Chen2016}%
  \BibitemOpen
  \bibinfo {author} {S.-L. Chen}, \bibinfo {author} {N.~Lambert}, \bibinfo
  {author} {C.-M. Li}, \bibinfo {author} {A.~Miranowicz}, \bibinfo {author}
  {Y.-N. Chen},\ and\ \bibinfo {author} {F.~Nori},\ \emph {\bibinfo {title}
  {Quantifying Non-{M}arkovianity with Temporal Steering}},\ \href
  {https://doi.org/10.1103/PhysRevLett.116.020503} {\bibfield  {journal}
  {\bibinfo  {journal} {Phys. Rev. Lett.}\ }\textbf {\bibinfo {volume} {116}},\
  \bibinfo {pages} {020503} (\bibinfo {year} {2016})}\BibitemShut {NoStop}%
\bibitem [{\citenamefont {Ku}\ \emph {et~al.}(2016)\citenamefont {Ku},
  \citenamefont {Chen}, \citenamefont {Chen}, \citenamefont {Lambert},
  \citenamefont {Chen},\ and\ \citenamefont {Nori}}]{Ku2016}%
  \BibitemOpen
  \bibinfo {author} {H.-Y. Ku}, \bibinfo {author} {S.-L. Chen}, \bibinfo
  {author} {H.-B. Chen}, \bibinfo {author} {N.~Lambert}, \bibinfo {author}
  {Y.-N. Chen},\ and\ \bibinfo {author} {F.~Nori},\ \emph {\bibinfo {title}
  {Temporal steering in four dimensions with applications to coupled qubits and
  magnetoreception}},\ \href {https://doi.org/10.1103/physreva.94.062126}
  {\bibfield  {journal} {\bibinfo  {journal} {Phys. Rev. A}\ }\textbf {\bibinfo
  {volume} {94}},\ \bibinfo {pages} {062126} (\bibinfo {year}
  {2016})}\BibitemShut {NoStop}%
\bibitem [{\citenamefont {Chen}\ \emph {et~al.}(2017)\citenamefont {Chen},
  \citenamefont {Lambert}, \citenamefont {Li}, \citenamefont {Chen},
  \citenamefont {Chen}, \citenamefont {Miranowicz},\ and\ \citenamefont
  {Nori}}]{Chen2017}%
  \BibitemOpen
  \bibinfo {author} {S.-L. Chen}, \bibinfo {author} {N.~Lambert}, \bibinfo
  {author} {C.-M. Li}, \bibinfo {author} {G.-Y. Chen}, \bibinfo {author} {Y.-N.
  Chen}, \bibinfo {author} {A.~Miranowicz},\ and\ \bibinfo {author} {F.~Nori},\
  \emph {\bibinfo {title} {Spatio-Temporal Steering for Testing Nonclassical
  Correlations in Quantum Networks}},\ \href
  {https://doi.org/10.1038/s41598-017-03789-4} {\bibfield  {journal} {\bibinfo
  {journal} {Sci. Rep.}\ }\textbf {\bibinfo {volume} {7}},\ \bibinfo {pages}
  {3728} (\bibinfo {year} {2017})}\BibitemShut {NoStop}%
\bibitem [{\citenamefont {Hirsch}\ \emph {et~al.}(2016)\citenamefont {Hirsch},
  \citenamefont {Quintino}, \citenamefont {V{\'{e}}rtesi}, \citenamefont
  {Pusey},\ and\ \citenamefont {Brunner}}]{Hirsch2016}%
  \BibitemOpen
  \bibinfo {author} {F.~Hirsch}, \bibinfo {author} {M.~T. Quintino}, \bibinfo
  {author} {T.~V{\'{e}}rtesi}, \bibinfo {author} {M.~F. Pusey},\ and\ \bibinfo
  {author} {N.~Brunner},\ \emph {\bibinfo {title} {Algorithmic Construction of
  Local Hidden Variable Models for Entangled Quantum States}},\ \href
  {https://doi.org/10.1103/physrevlett.117.190402} {\bibfield  {journal}
  {\bibinfo  {journal} {Phys. Rev. Lett.}\ }\textbf {\bibinfo {volume} {117}},\
  \bibinfo {pages} {190402} (\bibinfo {year} {2016})}\BibitemShut {NoStop}%
\bibitem [{\citenamefont {Fillettaz}\ \emph {et~al.}(2018)\citenamefont
  {Fillettaz}, \citenamefont {Hirsch}, \citenamefont {Designolle},\ and\
  \citenamefont {Brunner}}]{Fillettaz2018}%
  \BibitemOpen
  \bibinfo {author} {M.~Fillettaz}, \bibinfo {author} {F.~Hirsch}, \bibinfo
  {author} {S.~Designolle},\ and\ \bibinfo {author} {N.~Brunner},\ \emph
  {\bibinfo {title} {Algorithmic construction of local models for entangled
  quantum states: {O}ptimization for two-qubit states}},\ \href
  {https://doi.org/10.1103/physreva.98.022115} {\bibfield  {journal} {\bibinfo
  {journal} {Phys. Rev. A}\ }\textbf {\bibinfo {volume} {98}},\ \bibinfo
  {pages} {022115} (\bibinfo {year} {2018})}\BibitemShut {NoStop}%
\bibitem [{\citenamefont {Horodecki}\ \emph {et~al.}(1995)\citenamefont
  {Horodecki}, \citenamefont {Horodecki},\ and\ \citenamefont
  {Horodecki}}]{Horodecki1995}%
  \BibitemOpen
  \bibinfo {author} {R.~Horodecki}, \bibinfo {author} {P.~Horodecki},\ and\
  \bibinfo {author} {M.~Horodecki},\ \emph {\bibinfo {title} {Violating {B}ell
  inequality by mixed states: necessary and sufficient condition}},\ \href
  {https://doi.org/10.1016/0375-9601(95)00214-n} {\bibfield  {journal}
  {\bibinfo  {journal} {Phys. Lett. A}\ }\textbf {\bibinfo {volume} {200}},\
  \bibinfo {pages} {340} (\bibinfo {year} {1995})}\BibitemShut {NoStop}%
\bibitem [{\citenamefont {Horodecki}(1996)}]{Horodecki1996}%
  \BibitemOpen
  \bibinfo {author} {R.~Horodecki},\ \emph {\bibinfo {title} {Two-spin-1/2
  mixtures and {B}ell's inequalities}},\ \href
  {https://doi.org/10.1016/0375-9601(95)00904-3} {\bibfield  {journal}
  {\bibinfo  {journal} {Phys. Lett. A}\ }\textbf {\bibinfo {volume} {210}},\
  \bibinfo {pages} {223} (\bibinfo {year} {1996})}\BibitemShut {NoStop}%
\bibitem [{\citenamefont {Bennett}\ \emph {et~al.}(1999)\citenamefont
  {Bennett}, \citenamefont {DiVincenzo}, \citenamefont {Fuchs}, \citenamefont
  {Mor}, \citenamefont {Rains}, \citenamefont {Shor}, \citenamefont {Smolin},\
  and\ \citenamefont {Wootters}}]{Bennett1999}%
  \BibitemOpen
  \bibinfo {author} {C.~H. Bennett}, \bibinfo {author} {D.~P. DiVincenzo},
  \bibinfo {author} {C.~A. Fuchs}, \bibinfo {author} {T.~Mor}, \bibinfo
  {author} {E.~Rains}, \bibinfo {author} {P.~W. Shor}, \bibinfo {author} {J.~A.
  Smolin},\ and\ \bibinfo {author} {W.~K. Wootters},\ \emph {\bibinfo {title}
  {Quantum nonlocality without entanglement}},\ \href
  {https://doi.org/10.1103/physreva.59.1070} {\bibfield  {journal} {\bibinfo
  {journal} {Phys. Rev. A}\ }\textbf {\bibinfo {volume} {59}},\ \bibinfo
  {pages} {1070} (\bibinfo {year} {1999})}\BibitemShut {NoStop}%
\bibitem [{\citenamefont {Popescu}(1994)}]{Popescu1994}%
  \BibitemOpen
  \bibinfo {author} {S.~Popescu},\ \emph {\bibinfo {title} {Bell's inequalities
  versus teleportation: What is nonlocality?}},\ \href
  {https://doi.org/10.1103/physrevlett.72.797} {\bibfield  {journal} {\bibinfo
  {journal} {Phys. Rev. Lett.}\ }\textbf {\bibinfo {volume} {72}},\ \bibinfo
  {pages} {797} (\bibinfo {year} {1994})}\BibitemShut {NoStop}%
\bibitem [{\citenamefont {Miranowicz}(2004)}]{Adam2004}%
  \BibitemOpen
  \bibinfo {author} {A.~Miranowicz},\ \emph {\bibinfo {title} {Violation of
  {B}ell inequality and entanglement of decaying {W}erner~states}},\ \href
  {http://dx.doi.org/10.1016/j.physleta.2004.05.001} {\bibfield  {journal}
  {\bibinfo  {journal} {Phys. Lett. A}\ }\textbf {\bibinfo {volume} {327}},\
  \bibinfo {pages} {272} (\bibinfo {year} {2004})}\BibitemShut {NoStop}%
\bibitem [{\citenamefont {Bartkiewicz}\ \emph
  {et~al.}(2017{\natexlab{a}})\citenamefont {Bartkiewicz}, \citenamefont
  {Lemr}, \citenamefont {{\v{C}}ernoch},\ and\ \citenamefont
  {Miranowicz}}]{Bartkiewicz2017}%
  \BibitemOpen
  \bibinfo {author} {K.~Bartkiewicz}, \bibinfo {author} {K.~Lemr}, \bibinfo
  {author} {A.~{\v{C}}ernoch},\ and\ \bibinfo {author} {A.~Miranowicz},\ \emph
  {\bibinfo {title} {Bell nonlocality and fully entangled fraction measured in
  an entanglement-swapping device without quantum state tomography}},\ \href
  {https://doi.org/10.1103/physreva.95.030102} {\bibfield  {journal} {\bibinfo
  {journal} {Phys. Rev. A}\ }\textbf {\bibinfo {volume} {95}},\ \bibinfo
  {pages} {030102} (\bibinfo {year} {2017}{\natexlab{a}})}\BibitemShut
  {NoStop}%
\bibitem [{\citenamefont {Elitzur}\ \emph {et~al.}(1992)\citenamefont
  {Elitzur}, \citenamefont {Popescu},\ and\ \citenamefont
  {Rohrlich}}]{Elitzur1992}%
  \BibitemOpen
  \bibinfo {author} {A.~C. Elitzur}, \bibinfo {author} {S.~Popescu},\ and\
  \bibinfo {author} {D.~Rohrlich},\ \emph {\bibinfo {title} {Quantum
  nonlocality for each pair in an ensemble}},\ \href
  {https://doi.org/10.1016/0375-9601(92)90952-i} {\bibfield  {journal}
  {\bibinfo  {journal} {Phys. Lett. A}\ }\textbf {\bibinfo {volume} {162}},\
  \bibinfo {pages} {25} (\bibinfo {year} {1992})}\BibitemShut {NoStop}%
\bibitem [{\citenamefont {Barrett}\ \emph {et~al.}(2006)\citenamefont
  {Barrett}, \citenamefont {Kent},\ and\ \citenamefont
  {Pironio}}]{Barrett2006}%
  \BibitemOpen
  \bibinfo {author} {J.~Barrett}, \bibinfo {author} {A.~Kent},\ and\ \bibinfo
  {author} {S.~Pironio},\ \emph {\bibinfo {title} {Maximally Nonlocal and
  Monogamous Quantum Correlations}},\ \href
  {https://link.aps.org/doi/10.1103/PhysRevLett.97.170409} {\bibfield
  {journal} {\bibinfo  {journal} {Phys. Rev. Lett.}\ }\textbf {\bibinfo
  {volume} {97}},\ \bibinfo {pages} {170409} (\bibinfo {year}
  {2006})}\BibitemShut {NoStop}%
\bibitem [{\citenamefont {Aolita}\ \emph {et~al.}(2012)\citenamefont {Aolita},
  \citenamefont {Gallego}, \citenamefont {Ac\'{\i}n}, \citenamefont {Chiuri},
  \citenamefont {Vallone}, \citenamefont {Mataloni},\ and\ \citenamefont
  {Cabello}}]{Aolita2012}%
  \BibitemOpen
  \bibinfo {author} {L.~Aolita}, \bibinfo {author} {R.~Gallego}, \bibinfo
  {author} {A.~Ac\'{\i}n}, \bibinfo {author} {A.~Chiuri}, \bibinfo {author}
  {G.~Vallone}, \bibinfo {author} {P.~Mataloni},\ and\ \bibinfo {author}
  {A.~Cabello},\ \emph {\bibinfo {title} {Fully nonlocal quantum
  correlations}},\ \href {https://link.aps.org/doi/10.1103/PhysRevA.85.032107}
  {\bibfield  {journal} {\bibinfo  {journal} {Phys. Rev. A}\ }\textbf {\bibinfo
  {volume} {85}},\ \bibinfo {pages} {032107} (\bibinfo {year}
  {2012})}\BibitemShut {NoStop}%
\bibitem [{\citenamefont {Ghosh}\ \emph {et~al.}(2001)\citenamefont {Ghosh},
  \citenamefont {Kar}, \citenamefont {Sen(De)},\ and\ \citenamefont
  {Sen}}]{Ghosh2001}%
  \BibitemOpen
  \bibinfo {author} {S.~Ghosh}, \bibinfo {author} {G.~Kar}, \bibinfo {author}
  {A.~Sen(De)},\ and\ \bibinfo {author} {U.~Sen},\ \emph {\bibinfo {title}
  {Mixedness in the Bell violation versus entanglement of formation}},\ \href
  {https://doi.org/10.1103/physreva.64.044301} {\bibfield  {journal} {\bibinfo
  {journal} {Phys. Rev. A}\ }\textbf {\bibinfo {volume} {64}},\ \bibinfo
  {pages} {044301} (\bibinfo {year} {2001})}\BibitemShut {NoStop}%
\bibitem [{\citenamefont {Wei}\ \emph {et~al.}(2003)\citenamefont {Wei},
  \citenamefont {Nemoto}, \citenamefont {Goldbart}, \citenamefont {Kwiat},
  \citenamefont {Munro},\ and\ \citenamefont {Verstraete}}]{Wei2003}%
  \BibitemOpen
  \bibinfo {author} {T.-C. Wei}, \bibinfo {author} {K.~Nemoto}, \bibinfo
  {author} {P.~M. Goldbart}, \bibinfo {author} {P.~G. Kwiat}, \bibinfo {author}
  {W.~J. Munro},\ and\ \bibinfo {author} {F.~Verstraete},\ \emph {\bibinfo
  {title} {Maximal entanglement versus entropy for mixed quantum states}},\
  \href {https://doi.org/10.1103/physreva.67.022110} {\bibfield  {journal}
  {\bibinfo  {journal} {Phys. Rev. A}\ }\textbf {\bibinfo {volume} {67}},\
  \bibinfo {pages} {022110} (\bibinfo {year} {2003})}\BibitemShut {NoStop}%
\bibitem [{\citenamefont {Kwiat}\ \emph {et~al.}(1999)\citenamefont {Kwiat},
  \citenamefont {Waks}, \citenamefont {White}, \citenamefont {Appelbaum},\ and\
  \citenamefont {Eberhard}}]{Kwiat1999PRA}%
  \BibitemOpen
  \bibinfo {author} {P.~G. Kwiat}, \bibinfo {author} {E.~Waks}, \bibinfo
  {author} {A.~G. White}, \bibinfo {author} {I.~Appelbaum},\ and\ \bibinfo
  {author} {P.~H. Eberhard},\ \emph {\bibinfo {title} {Ultrabright source of
  polarization-entangled photons}},\ \href
  {https://link.aps.org/doi/10.1103/PhysRevA.60.R773} {\bibfield  {journal}
  {\bibinfo  {journal} {Phys. Rev. A}\ }\textbf {\bibinfo {volume} {60}},\
  \bibinfo {pages} {R773} (\bibinfo {year} {1999})}\BibitemShut {NoStop}%
\bibitem [{\citenamefont {Halenkov\'{a}}\ \emph {et~al.}(2012)\citenamefont
  {Halenkov\'{a}}, \citenamefont {\v{C}ernoch}, \citenamefont {Lemr},
  \citenamefont {Soubusta},\ and\ \citenamefont
  {Drusov\'{a}}}]{Halenkova2012ApplOpt}%
  \BibitemOpen
  \bibinfo {author} {E.~Halenkov\'{a}}, \bibinfo {author} {A.~\v{C}ernoch},
  \bibinfo {author} {K.~Lemr}, \bibinfo {author} {J.~Soubusta},\ and\ \bibinfo
  {author} {S.~Drusov\'{a}},\ \emph {\bibinfo {title} {Experimental
  implementation of the multifunctional compact two-photon state analyzer}},\
  \href {http://ao.osa.org/abstract.cfm?URI=ao-51-4-474} {\bibfield  {journal}
  {\bibinfo  {journal} {Appl. Opt.}\ }\textbf {\bibinfo {volume} {51}},\
  \bibinfo {pages} {474} (\bibinfo {year} {2012})}\BibitemShut {NoStop}%
\bibitem [{\citenamefont {Banaszek}\ \emph {et~al.}(1999)\citenamefont
  {Banaszek}, \citenamefont {D'Ariano}, \citenamefont {Paris},\ and\
  \citenamefont {Sacchi}}]{Banaszek1999}%
  \BibitemOpen
  \bibinfo {author} {K.~Banaszek}, \bibinfo {author} {G.~M. D'Ariano}, \bibinfo
  {author} {M.~G.~A. Paris},\ and\ \bibinfo {author} {M.~F. Sacchi},\ \emph
  {\bibinfo {title} {Maximum-likelihood estimation of the density matrix}},\
  \href {https://link.aps.org/doi/10.1103/PhysRevA.61.010304} {\bibfield
  {journal} {\bibinfo  {journal} {Phys. Rev. A}\ }\textbf {\bibinfo {volume}
  {61}},\ \bibinfo {pages} {010304} (\bibinfo {year} {1999})}\BibitemShut
  {NoStop}%
\bibitem [{\citenamefont {James}\ \emph {et~al.}(2001)\citenamefont {James},
  \citenamefont {Kwiat}, \citenamefont {Munro},\ and\ \citenamefont
  {White}}]{James2001}%
  \BibitemOpen
  \bibinfo {author} {D.~F.~V. James}, \bibinfo {author} {P.~G. Kwiat}, \bibinfo
  {author} {W.~J. Munro},\ and\ \bibinfo {author} {A.~G. White},\ \emph
  {\bibinfo {title} {Measurement of qubits}},\ \href
  {https://link.aps.org/doi/10.1103/PhysRevA.64.052312} {\bibfield  {journal}
  {\bibinfo  {journal} {Phys. Rev. A}\ }\textbf {\bibinfo {volume} {64}},\
  \bibinfo {pages} {052312} (\bibinfo {year} {2001})}\BibitemShut {NoStop}%
\bibitem [{\citenamefont {Hradil}\ \emph {et~al.}(2004)\citenamefont {Hradil},
  \citenamefont {\v{R}eh\'{a}\v{c}ek}, \citenamefont {Fiur\'{a}\v{s}ek},\ and\
  \citenamefont {Je\v{z}ek}}]{Hradil2004}%
  \BibitemOpen
  \bibinfo {author} {Z.~Hradil}, \bibinfo {author} {J.~\v{R}eh\'{a}\v{c}ek},
  \bibinfo {author} {J.~Fiur\'{a}\v{s}ek},\ and\ \bibinfo {author}
  {M.~Je\v{z}ek},\ in\ \href {https://doi.org/10.1007/978-3-540-44481-7_3}
  {\emph {\bibinfo {booktitle} {(eds) Quantum State Estimation. Lecture Notes
  in Physics, vol 649}}},\ \bibinfo {editor} {edited by\ \bibinfo {editor}
  {P.~M.}\ and\ \bibinfo {editor} {\v{R}eh\'{a}\v{c}ek J.}}\ (\bibinfo
  {publisher} {Springer, Berlin, Heidelberg},\ \bibinfo {address} {Oxford},\
  \bibinfo {year} {2004})\ Chap.~\bibinfo {chapter} {3}, pp.\ \bibinfo {pages}
  {266--290}\BibitemShut {NoStop}%
\bibitem [{\citenamefont {Sanderson}\ and\ \citenamefont
  {Curtin}(2016)}]{Sanderson2016}%
  \BibitemOpen
  \bibinfo {author} {C.~Sanderson}\ and\ \bibinfo {author} {R.~Curtin},\ \emph
  {\bibinfo {title} {Armadillo: a template-based {C}$++$ library for linear
  algebra}},\ \href {https://doi.org/10.21105/joss.00026} {\bibfield  {journal}
  {\bibinfo  {journal} {J. Open Source Software}\ }\textbf {\bibinfo {volume}
  {1}},\ \bibinfo {pages} {26} (\bibinfo {year} {2016})}\BibitemShut {NoStop}%
\bibitem [{\citenamefont {Sanderson}\ and\ \citenamefont
  {Curtin}(2018)}]{Sanderson2018}%
  \BibitemOpen
  \bibinfo {author} {C.~Sanderson}\ and\ \bibinfo {author} {R.~Curtin},\ in\
  \href {https://doi.org/10.1007/978-3-319-96418-8_50} {\emph {\bibinfo
  {booktitle} {Mathematical Software {\textendash} {ICMS} 2018}}},\ \bibinfo
  {series} {Lecture Notes in Computer Science}, Vol.\ \bibinfo {volume}
  {10931}\ (\bibinfo  {publisher} {Springer Int. Pub.},\ \bibinfo {year}
  {2018})\ pp.\ \bibinfo {pages} {422--430}\BibitemShut {NoStop}%
\bibitem [{sup()}]{supplement}%
  \BibitemOpen
  \href@noop {} {}\bibinfo {note} {See the Supplementary Material at
  http://xxx, which includes the density matrices for our experimental GWSs and
  Werner states.}\BibitemShut {Stop}%
\bibitem [{\citenamefont {Girdhar}\ and\ \citenamefont
  {Cavalcanti}(2016)}]{Girdhar2016}%
  \BibitemOpen
  \bibinfo {author} {P.~Girdhar}\ and\ \bibinfo {author} {E.~G. Cavalcanti},\
  \emph {\bibinfo {title} {All two-qubit states that are steerable via
  {C}lauser-{H}orne-{S}himony-{H}olt-type correlations are {B}ell nonlocal}},\
  \href {https://doi.org/10.1103/physreva.94.032317} {\bibfield  {journal}
  {\bibinfo  {journal} {Phys. Rev. A}\ }\textbf {\bibinfo {volume} {94}},\
  \bibinfo {pages} {032317} (\bibinfo {year} {2016})}\BibitemShut {NoStop}%
\bibitem [{\citenamefont {Cavalcanti}\ \emph {et~al.}(2015)\citenamefont
  {Cavalcanti}, \citenamefont {Foster}, \citenamefont {Fuwa},\ and\
  \citenamefont {Wiseman}}]{Cavalcanti2015}%
  \BibitemOpen
  \bibinfo {author} {E.~G. Cavalcanti}, \bibinfo {author} {C.~J. Foster},
  \bibinfo {author} {M.~Fuwa},\ and\ \bibinfo {author} {H.~M. Wiseman},\ \emph
  {\bibinfo {title} {Analog of the {C}lauser-{H}orne-{S}himony-{H}olt
  inequality for steering}},\ \href {https://doi.org/10.1364/josab.32.000a74}
  {\bibfield  {journal} {\bibinfo  {journal} {J. Opt. Soc. Am. B}\ }\textbf
  {\bibinfo {volume} {32}},\ \bibinfo {pages} {A74} (\bibinfo {year}
  {2015})}\BibitemShut {NoStop}%
\bibitem [{\citenamefont {Costa}\ and\ \citenamefont
  {Angelo}(2016)}]{Costa2016}%
  \BibitemOpen
  \bibinfo {author} {A.~C.~S. Costa}\ and\ \bibinfo {author} {R.~M. Angelo},\
  \emph {\bibinfo {title} {Quantification of {E}instein-{P}odolski-{R}osen
  steering for two-qubit states}},\ \href
  {https://doi.org/10.1103/physreva.93.020103} {\bibfield  {journal} {\bibinfo
  {journal} {Phys. Rev. A}\ }\textbf {\bibinfo {volume} {93}},\ \bibinfo
  {pages} {020103} (\bibinfo {year} {2016})}\BibitemShut {NoStop}%
\bibitem [{\citenamefont {Brukner}(2014)}]{Brukner2014}%
  \BibitemOpen
  \bibinfo {author} {{\v{C}}.~Brukner},\ \emph {\bibinfo {title} {Quantum
  causality}},\ \href {https://doi.org/10.1038/nphys2930} {\bibfield  {journal}
  {\bibinfo  {journal} {Nat. Phys.}\ }\textbf {\bibinfo {volume} {10}},\
  \bibinfo {pages} {259} (\bibinfo {year} {2014})}\BibitemShut {NoStop}%
\bibitem [{\citenamefont {Bartkiewicz}\ \emph
  {et~al.}(2015{\natexlab{b}})\citenamefont {Bartkiewicz}, \citenamefont
  {Beran}, \citenamefont {Lemr}, \citenamefont {Norek},\ and\ \citenamefont
  {Miranowicz}}]{Bartkiewicz2015b}%
  \BibitemOpen
  \bibinfo {author} {K.~Bartkiewicz}, \bibinfo {author} {J.~Beran}, \bibinfo
  {author} {K.~Lemr}, \bibinfo {author} {M.~Norek},\ and\ \bibinfo {author}
  {A.~Miranowicz},\ \emph {\bibinfo {title} {Quantifying entanglement of a
  two-qubit system via measurable and invariant moments of its partially
  transposed density matrix}},\ \href
  {https://doi.org/10.1103/physreva.91.022323} {\bibfield  {journal} {\bibinfo
  {journal} {Phys. Rev. A}\ }\textbf {\bibinfo {volume} {91}},\ \bibinfo
  {pages} {022323} (\bibinfo {year} {2015}{\natexlab{b}})}\BibitemShut
  {NoStop}%
\bibitem [{\citenamefont {Bartkiewicz}\ \emph
  {et~al.}(2017{\natexlab{b}})\citenamefont {Bartkiewicz}, \citenamefont
  {Chimczak},\ and\ \citenamefont {Lemr}}]{Bartkiewicz2017b}%
  \BibitemOpen
  \bibinfo {author} {K.~Bartkiewicz}, \bibinfo {author} {G.~Chimczak},\ and\
  \bibinfo {author} {K.~Lemr},\ \emph {\bibinfo {title} {Direct method for
  measuring and witnessing quantum entanglement of arbitrary two-qubit states
  through Hong-Ou-Mandel interference}},\ \href
  {https://link.aps.org/doi/10.1103/PhysRevA.95.022331} {\bibfield  {journal}
  {\bibinfo  {journal} {Phys. Rev. A}\ }\textbf {\bibinfo {volume} {95}},\
  \bibinfo {pages} {022331} (\bibinfo {year} {2017}{\natexlab{b}})}\BibitemShut
  {NoStop}%
\bibitem [{\citenamefont {Bennett}\ \emph {et~al.}(1996)\citenamefont
  {Bennett}, \citenamefont {DiVincenzo}, \citenamefont {Smolin},\ and\
  \citenamefont {Wootters}}]{Bennett1996}%
  \BibitemOpen
  \bibinfo {author} {C.~H. Bennett}, \bibinfo {author} {D.~P. DiVincenzo},
  \bibinfo {author} {J.~A. Smolin},\ and\ \bibinfo {author} {W.~K. Wootters},\
  \emph {\bibinfo {title} {Mixed-state entanglement and quantum error
  correction}},\ \href {https://doi.org/10.1103/physreva.54.3824} {\bibfield
  {journal} {\bibinfo  {journal} {Phys. Rev. A}\ }\textbf {\bibinfo {volume}
  {54}},\ \bibinfo {pages} {3824} (\bibinfo {year} {1996})}\BibitemShut
  {NoStop}%
\bibitem [{\citenamefont {Cavalcanti}\ \emph {et~al.}(2009)\citenamefont
  {Cavalcanti}, \citenamefont {Jones}, \citenamefont {Wiseman},\ and\
  \citenamefont {Reid}}]{Cavalcanti2009}%
  \BibitemOpen
  \bibinfo {author} {E.~G. Cavalcanti}, \bibinfo {author} {S.~J. Jones},
  \bibinfo {author} {H.~M. Wiseman},\ and\ \bibinfo {author} {M.~D. Reid},\
  \emph {\bibinfo {title} {Experimental criteria for steering and the
  {E}instein-{P}odolsky-{R}osen paradox}},\ \href
  {https://doi.org/10.1103/physreva.80.032112} {\bibfield  {journal} {\bibinfo
  {journal} {Phys. Rev. A}\ }\textbf {\bibinfo {volume} {80}},\ \bibinfo
  {pages} {032112} (\bibinfo {year} {2009})}\BibitemShut {NoStop}%
\bibitem [{\citenamefont {Grant}\ and\ \citenamefont {Boyd}(2012)}]{CVX}%
  \BibitemOpen
  \bibinfo {author} {M.~Grant}\ and\ \bibinfo {author} {S.~Boyd},\ \href
  {http://cvxr.com/cvx/} {\emph {\bibinfo {title} {{CVX}: {M}atlab software for
  disciplined convex programming, {\rm http://cvxr.com/cvx}}}} (\bibinfo {year}
  {2012})\BibitemShut {NoStop}%
\bibitem [{\citenamefont {Cavalcanti}\ \emph {et~al.}(2016)\citenamefont
  {Cavalcanti}, \citenamefont {Guerini}, \citenamefont {Rabelo},\ and\
  \citenamefont {Skrzypczyk}}]{Cavalcanti2016}%
  \BibitemOpen
  \bibinfo {author} {D.~Cavalcanti}, \bibinfo {author} {L.~Guerini}, \bibinfo
  {author} {R.~Rabelo},\ and\ \bibinfo {author} {P.~Skrzypczyk},\ \emph
  {\bibinfo {title} {General Method for Constructing Local Hidden Variable
  Models for Entangled Quantum States}},\ \href
  {https://doi.org/10.1103/physrevlett.117.190401} {\bibfield  {journal}
  {\bibinfo  {journal} {Phys. Rev. Lett.}\ }\textbf {\bibinfo {volume} {117}},\
  \bibinfo {pages} {190401} (\bibinfo {year} {2016})}\BibitemShut {NoStop}%
\bibitem [{\citenamefont {Shchukin}\ and\ \citenamefont
  {Vogel}(2005{\natexlab{b}})}]{Shchukin2005exp}%
  \BibitemOpen
  \bibinfo {author} {E.~V. Shchukin}\ and\ \bibinfo {author} {W.~Vogel},\ \emph
  {\bibinfo {title} {Nonclassical moments and their measurement}},\ \href
  {https://doi.org/10.1103/physreva.72.043808} {\bibfield  {journal} {\bibinfo
  {journal} {Phys. Rev. A}\ }\textbf {\bibinfo {volume} {72}},\ \bibinfo
  {pages} {043808} (\bibinfo {year} {2005}{\natexlab{b}})}\BibitemShut
  {NoStop}%
\bibitem [{\citenamefont {Hillery}\ and\ \citenamefont
  {Zubairy}(2006)}]{Hillery2006}%
  \BibitemOpen
  \bibinfo {author} {M.~Hillery}\ and\ \bibinfo {author} {M.~S. Zubairy},\
  \emph {\bibinfo {title} {Entanglement Conditions for Two-Mode States}},\
  \href {https://doi.org/10.1103/physrevlett.96.050503} {\bibfield  {journal}
  {\bibinfo  {journal} {Phys. Rev. Lett.}\ }\textbf {\bibinfo {volume} {96}},\
  \bibinfo {pages} {050503} (\bibinfo {year} {2006})}\BibitemShut {NoStop}%
\bibitem [{\citenamefont {Gomes}\ \emph {et~al.}(2009)\citenamefont {Gomes},
  \citenamefont {Salles}, \citenamefont {Toscano}, \citenamefont {Ribeiro},\
  and\ \citenamefont {Walborn}}]{Gomes2009}%
  \BibitemOpen
  \bibinfo {author} {R.~M. Gomes}, \bibinfo {author} {A.~Salles}, \bibinfo
  {author} {F.~Toscano}, \bibinfo {author} {P.~H.~S. Ribeiro},\ and\ \bibinfo
  {author} {S.~P. Walborn},\ \emph {\bibinfo {title} {Quantum entanglement
  beyond {G}aussian criteria}},\ \href
  {https://doi.org/10.1073/pnas.0908329106} {\bibfield  {journal} {\bibinfo
  {journal} {PNAS}\ }\textbf {\bibinfo {volume} {106}},\ \bibinfo {pages}
  {21517} (\bibinfo {year} {2009})}\BibitemShut {NoStop}%
\bibitem [{\citenamefont {W\"{u}nsche}(1996)}]{Wunsche1996}%
  \BibitemOpen
  \bibinfo {author} {A.~W\"{u}nsche},\ \emph {\bibinfo {title} {Tomographic
  reconstruction of the density operator from its normally ordered moments}},\
  \href {https://doi.org/10.1103/physreva.54.5291} {\bibfield  {journal}
  {\bibinfo  {journal} {Phys. Rev. A}\ }\textbf {\bibinfo {volume} {54}},\
  \bibinfo {pages} {5291} (\bibinfo {year} {1996})}\BibitemShut {NoStop}%
\end{thebibliography}
\end{document}